\documentclass[fleqn,usenatbib]{mnras}

\usepackage{newtxtext,newtxmath}
\usepackage{longtable}
\usepackage{supertabular,booktabs}

\usepackage[T1]{fontenc}

\DeclareRobustCommand{\VAN}[3]{#2}
\let\VANthebibliography\thebibliography
\def\thebibliography{\DeclareRobustCommand{\VAN}[3]{##3}\VANthebibliography}

\usepackage{graphicx}	
\usepackage{amsmath}	

\newcommand{\HII}{H\,{\sc ii}\rm\,}
\newcommand{\NII}{{[N\,{\sc ii}]}}

\newcommand{\NIIl}{{[N\,{\sc ii}]\,$\lambda$}}
\newcommand{\NIII}{{N\,{\sc iii}]}}

\newcommand{\NIV}{{N\,{\sc iv}]}}

\newcommand{\SII}{{[S\,{\sc ii}]}}

\newcommand{\OIII}{{[O\,{\sc iii}]}}
\newcommand{\NeIII}{{[Ne\,{\sc iii}]}}

\newcommand{\OIIIl}{{[O\,{\sc iii}]\,$\lambda$}}
\newcommand{\OII}{{[O\,{\sc ii}]}}

\newcommand{\OIIll}{{[O\,{\sc ii}]\,$\lambda\lambda$}}

\newcommand{\Ha}{H$\alpha$}

\newcommand{\lOH}{$12+\log({\rm O}/{\rm H})$}
\newcommand{\logNO}{$\log({\rm N}/{\rm O})$}
\newcommand{\SDmass}{$\Sigma_{M_*}$}
\newcommand{\SDsfr}{$\Sigma_{\rm SFR}$}

\newcommand{\gonem}{G140M/F070LP}
\newcommand{\gtwom}{G235M/F170LP}
\newcommand{\gthreem}{G395M/F290LP}

\newcommand{\Nall}{588}
\newcommand{\NallTe}{40}
\newcommand{\NallSL}{548}

\newcommand{\Nmass}{484}

\newcommand{\NmassSL}{453}

\defcitealias{Cataldi2025}{C25}
\defcitealias{Pilyugin2009}{P09}

\title[N/O at $z\sim1.5-7$]{JADES: Evolution of nitrogen abundances in star-forming galaxies from $z\sim1.5-7$}

\author[A. J. Cameron et al.]{
Alex J. Cameron,$^{1,2,3}$\thanks{E-mail: alex.cameron@nbi.ku.dk}
Courtney Carreira,$^{4}$
Charlotte Simmonds,$^{5}$
Andrew J.\ Bunker,$^{3}$
Aayush Saxena,$^{3,6}$
\newauthor
Stefano Carniani,$^{7}$
Stéphane Charlot,$^{8}$
Jacopo Chevallard,$^{3}$
Emma Curtis-Lake,$^{9}$
Kevin Hainline,$^{10}$
\newauthor
Ryan Hausen,$^{11}$
Xihan Ji,$^{12,13}$
Zhiyuan Ji,$^{10}$
Benjamin D.\ Johnson,$^{14}$
Pierluigi Rinaldi,$^{15}$
Brant Robertson,$^{4}$
\newauthor
Jan Scholtz,$^{12,13}$
Maddie S. Silcock,$^{9}$
Sandro Tacchella,$^{12,13}$
James A. A. Trussler,$^{14}$
Hannah \"Ubler,$^{16}$
\newauthor
Christina C. Williams,$^{17}$
Christopher N. A. Willmer,$^{10}$
Chris Willott,$^{18}$
Joris Witstok,$^{1,2}$
\\
$^{1}$Cosmic Dawn Center (DAWN), Copenhagen, Denmark\\
$^{2}$Niels Bohr Institute, University of Copenhagen, Jagtvej 128, DK-2200, Copenhagen, Denmark\\
$^{3}$Department of Physics, University of Oxford, Denys Wilkinson Building, Keble Road, Oxford, OX1 3RH, UK\\
$^{4}$Department of Astronomy and Astrophysics University of California, Santa Cruz, 1156 High Street, Santa Cruz CA 96054, USA \\
$^{5}$Departamento de Astronomía, Universidad de Chile, Camino El Observatorio 1515, Las Condes, Santiago, Chile\\
$^{6}$Department of Physics and Astronomy, University College London, Gower Street, London WC1E 6BT, UK\\
$^{7}$Scuola Normale Superiore, Piazza dei Cavalieri 7, I-56126 Pisa, Italy\\
$^{8}$Sorbonne Universit\'e, CNRS, UMR 7095, Institut d'Astrophysique de Paris, 98 bis bd Arago, 75014 Paris, France\\
$^{9}$Centre for Astrophysics Research, Department of Physics, Astronomy and Mathematics, University of Hertfordshire, Hatfield AL10 9AB, UK\\
$^{10}$Steward Observatory, University of Arizona, 933 N. Cherry Avenue, Tucson, AZ 85721, USA\\
$^{11}$Department of Physics and Astronomy, The Johns Hopkins University, 3400 N. Charles St., Baltimore, MD 21218\\
$^{12}$Kavli Institute for Cosmology, University of Cambridge, Madingley Road, Cambridge, CB3 OHA, UK\\
$^{13}$Cavendish Laboratory - Astrophysics Group, University of Cambridge, 19 JJ Thomson Avenue, Cambridge, CB3 OHE, UK\\
$^{14}$Center for Astrophysics $|$ Harvard \& Smithsonian, 60 Garden St., Cambridge MA 02138 USA\\
$^{15}$Space Telescope Science Institute, 3700 San Martin Drive, Baltimore, Maryland 21218, USA\\
$^{16}$Max-Planck-Institut f\"ur extraterrestrische Physik (MPE), Gie{\ss}enbachstra{\ss}e 1, 85748 Garching, Germany\\
$^{17}$NSF National Optical-Infrared Astronomy Research Laboratory, 950 North Cherry Avenue, Tucson, AZ 85719, USA\\
$^{18}$NRC Herzberg, 5071 West Saanich Rd, Victoria, BC V9E 2E7, Canada\\
\vspace{-1 cm}
}

\date{Accepted XXX. Received YYY; in original form ZZZ}

\pubyear{2026}

\begin{document}
\label{firstpage}
\pagerange{\pageref{firstpage}--\pageref{lastpage}}
\maketitle

\begin{abstract}
We present nitrogen abundance measurements based on the low-ionisation \NIIl6583 emission line for \Nall\ galaxies between $1.5<z<7.0$ from the JWST Advanced Deep Extragalactic Survey (JADES).
We detect the temperature-sensitive \OIIIl4363 auroral line in \NallTe\ galaxies in our sample, affording $T_e$-based abundances for this subset.
We find that the average N/O abundance ratio in our low-metallicity sample is at least 0.1 dex higher than $z\sim0$ samples. 
In particular, we find significant scatter toward high N/O, with five galaxies being identified with enhanced nitrogen abundances ($\log({\rm N}/{\rm O})>-1.1$) at low-metallicity ($12+\log({\rm O}/{\rm H})<8.0$) from $T_e$-based measurements. Meanwhile, applying strong-line abundance measurements to the remainder of our sample reveals a further 14 candidate galaxies passing these abundance cuts, 
implying that around 13~\% of $12+\log({\rm O}/{\rm H})<8.0$ galaxies at these redshifts are nitrogen-enhanced at this level. 
We find that N/O abundance in low-metallicity systems correlates with SFR, \SDsfr, and \SDmass at high redshift, while only in high-metallicity systems does a correlation with $M_*$ emerge.
Despite healthy representation of these `moderately nitrogen-enhanced' galaxies ($-1.1<\log({\rm N}/{\rm O})\leq-0.6$), no galaxies in our low-metallicity sample are identified as having $\log({\rm N}/{\rm O})>-0.6$, abundances that are typical of high-redshift \NIII- and \NIV-emitters. 
This demonstrates that the extreme nitrogen enhancements seen in some \NIII- and \NIV-emitters are only attained during the most extreme starbursts. This suggests that these elevated abundances are caused by enrichment from young massive stars in extreme environments and that the impact of this enrichment pathway is milder, though still important, for high-redshift systems on the star-forming main sequence.

\end{abstract}



\section{Introduction} \label{sec:intro}

Chemical abundances within the interstellar medium (ISM) of galaxies depend on many processes, including nucleosynthesis, gas mixing, outflows, depletion and dilution \citep[e.g.,][]{Matteucci2012,MaiolinoMannucci2019}.
In particular, since different sources of nucleosynthesis operate over different timescales with different chemical yields, ratios of abundances of different metals are a powerful constraint on star formation histories and the properties of current and previous generations of stellar populations \citep{Izotov1999, Berg2019_CNO_Dwarf, Kobayashi2020, KobayashiTaylor2023}.

Core-collapse supernovae (CCSNe), arising from stars with initial masses  $\sim8-25~M_\odot$, are an important chemical enrichment mechanism, producing large quantities of $\alpha$-elements on relatively short timescales after the onset of star formation \citep[e.g.][]{Nomoto2013}. Indeed, oxygen, the Universe's most abundant metal, is primarily synthesised via this channel \citep{KobayashiTaylor2023}.

Nitrogen is an interesting element since, as an odd-numbered element, it is only produced in modest amounts by CCSNe \citep{Nomoto2013, LimongiChieffi2018}.
Among $z\sim0$ galaxies, the nitrogen-to-oxygen abundance ratio (N/O) has been shown to have no correlation with gas-phase oxygen abundance (O/H) at low oxygen abundances ($12+\log({\rm O}/{\rm H})\lesssim8.0$; \citealt{Izotov1999, PerezMontero2009, Berg2019_CNO_Dwarf, ArellanoCordova2025_CLASSY, Scholte2026}), but rises steeply with higher oxygen abundances \citep{VilaCostas1993, Pilyugin2012, Berg2020}, greatly exceeding the N/O ratio from pure CCSNe yields. 
At high-metallicity, this is typically attributed to secondary enrichment during the asymptotic giant branch (AGB) phase of intermediate mass ($\sim4-7~M_\odot$) stars \citep{Kobayashi2011, Romano2022}, which occurs on timescales of $>100$~Myr \citep{Vincenzo2018}.
Meanwhile the level of (and scatter about) the plateau at low oxygen abundance is thought to comprise contributions from both CCSNe yields and pre-supernovae winds of massive stars \citep[e.g.][]{LopezSanchez2010, Roy2021}.

Prior to \emph{JWST}, ground-based studies exploring the N/O -- O/H relation out to $z\sim2$ were largely limited to relatively metal-enriched galaxies, finding no significant evolution in this relation \citep{Strom2018, HaydenPawson2022}. 
However, \emph{JWST}/NIRSpec spectroscopy of low metallicity galaxies at higher redshifts has begun to change this picture.

In \citet{Cameron2023_GNz11}, it was shown that the strong rest-frame ultraviolet \NIII$\lambda\lambda$1750 and \NIV$\lambda\lambda$1483, 1486 emission lines in the spectrum of the $z=10.6$ galaxy GN-z11 \citep{Bunker2023_GNz11} implied that the N/O abundance was at least a factor of $\sim4$ higher than the solar N/O, despite the sub-solar O/H abundance. Independent studies have since confirmed this super-solar N/O measurement \citep[e.g.,][]{Senchyna2023, Martinez2025}.
Subsequent studies have identified many more of these highly nitrogen enhanced systems feature strong \NIII\ and/or \NIV\ emission implying highly enhanced N/O at low O/H 
\citep{Isobe2023, MarquesChaves2023, Castellano2024, Ji2024, Ji2026, Topping2024, Schaerer2024, Berg2025_WN, Naidu2025, Napolitano2025, Morel2025_Nemitters}.
In addition to high-ionisation emission lines with high-equivalent widths, these systems are typically observed to be compact, with high electron densities ($n_e\gtrsim10^5$~cm$^{-3}$; \citealt{Senchyna2023, Maiolino2024_GNz11, Topping2025}) and high star-formation rates (SFR; \citealt{MarquesChaves2023, Schaerer2024}) commonly inferred.
The strong deviation of these systems from what is observed in $z\sim0$ samples suggests differences either in the enrichment mechanisms or the metal-mixing in these systems.

Many studies have subsequently explored models that can reconcile such high nitrogen abundances at low O/H. 
On the one hand, enrichment from winds of Wolf-Rayet stars  \citep{KobayashiFerrara2023, Bekki2023, Watanabe2024, Berg2025_WN} and very massive stars ($100-500~M_\odot$; \citealt{Vink2023, Higgins2025}) have been invoked as viable sources of strong nitrogen enrichment, as have as yet unobserved supermassive stars ($\geq1000~M_\odot$) formed in dense environments \citep{Cameron2023_GNz11, Charbonnel2023, NageleUmeda2023, Ji2024, Nandal2025}.
On the other hand, other studies have shown that galactic winds with strong oxygen-loading \citep{Rizzuti2025} or the formation of stars from pre-processed nitrogen-rich molecular gas \citep{McClymont2025_NO} can result in the required abundance patterns without invoking a strong contribution from massive star winds.

Although these are clearly interesting systems, at the depths available in most \emph{JWST}/NIRSpec $z\gtrsim3$ spectroscopic programs, nitrogen abundances measured from these \NIII\ and \NIV\ features can rarely be obtained for systems with sub-solar N/O \citep{Zhu2025, Martinez2025}. As a result, measurements of the full distribution of nitrogen abundances across the high-redshift population are limited.
Conversely, nitrogen abundances measured from the optical \NIIl6583 / \OIIll3726, 3729 ratio are sensitive down to low N/O ratios in systems on the star-forming main sequence, and are a mainstay of previously mentioned $z\sim0$ studies.
Despite this, relatively few \NII-based abundance studies have been performed at high-redshift \citep{ArellanoCordova2025_EXCELS, Stiavelli2025, Zhang2025_WR, Cataldi2025, Schaerer2026}, with no consensus yet on the behaviour of N/O at low O/H in galaxies more representative of the star-forming main sequence.

In this work, we perform a systematic survey of nitrogen abundances based on this low-ionisation \NII\ line using data from the JWST Advanced Deep Extragalactic Surey (JADES; \citealt{Eisenstein2023_JADES}). This allows us to be sensitive to much lower N/O abundances (\logNO $\lesssim-1.6$), across galaxies with a more diverse range of ionisation conditions.
The paper is structured as follows: the data sets and analysis techniques used are outlined in Section~\ref{sec:data}. Section~\ref{sec:abundances} then describes our chemical abundances measurements and associated uncertainties. We present the key results in Section~\ref{sec:results} before presenting a discussion of the implications of these in Section~\ref{sec:discussion}. We provide a summary of the work in Section~\ref{sec:conclusion}.

% ...

\section{Data \& Methods} \label{sec:data}

\subsection{Observations \& sample assembly} 
\label{sub:observations}

We draw our sample from the spectroscopic component of the JWST Advanced Deep Extragalactic Survey (JADES; \citealt{Eisenstein2023_JADES}).
A detailed description of the planning and execution of JADES spectroscopic observations can be found in the JADES NIRSpec data release papers \citep{Bunker2023_DR, DEugenio2025_DR3, CurtisLake2025_DR4, Scholtz2025_DR4}, however, here we briefly review the main aspects as relevant to this work.

The JADES spectroscopic sample was compiled from over 30 different pointings observed with the NIRSpec/MOS mode \citep{Jakobsen2022_NS_Overview, Ferruit2022_NS_MOS} across the GOODS-South and GOODS-North legacy fields.
All pointings are observed in the low-resolution Prism/CLEAR mode and all three medium resolution ($R\sim1000$) gratings: \gonem, \gtwom, and \gthreem, which have observed wavelength ranges of $0.70-2.20$~$\mu$m, $1.66-4.00$~$\mu$m, and $2.87-5.48$~$\mu$m, respectively.

Since this work hinges on measurements of the \NIIl6583 emission line, which is significantly blended with \Ha\ at Prism resolution, our analysis focuses on the $R\sim1000$ spectroscopy.
We note that during target assignment, low-priority targets are removed from grating observations if their spectra overlap with protected higher priority targets, meaning around 8~\% of the full JADES sample is only observed with Prism.

JADES observations were split into two main tiers: `deep' and `medium'\footnote{JADES also included an `ultradeep' tier, but we exclude that from this study as it did not observe in the \gtwom\ grating.}, with the former obtaining deeper observations for a smaller sample. In practice, the difference in depth between the two primarily relates to Prism observations, since for `deep', the Prism integrations were four times longer than the grating integrations, while for `medium' these were approximately equal. 
In fact, because individual targets in the medium tiers could be observed in multiple overlapping pointings, there is no great difference in depth between these tiers in the $R\sim1000$ spectroscopy relevant to this study.
All targets in our sample are observed with at least $t_{\rm exp}=0.57$ h per grating, but the median depth is $t_{\rm exp}=2.3$ h, and the deepest spectra in our sample reach $t_{\rm exp}=6.9$ h.

The two main sub-samples within JADES target prioritisation scheme were: (1) $m_{\rm UV}$-selected galaxies at $z>5.7$ and (2) $m_{\rm F444W}$-selected galaxies at $1.5<z\leq5.7$ (approximating a stellar mass based selection).
Given the  wavelength coverage of the NIRSpec $R\sim1000$ gratings does not extend much past $5~\mu$m, our \NII-based study is limited to galaxies at $z<7$.
We further impose a lower bound of $z>1.5$ for this analysis to align with the cutoff for the JADES $m_{\rm F444W}$-based selection.

\begin{figure}
    \centering
    \includegraphics[width=\columnwidth]{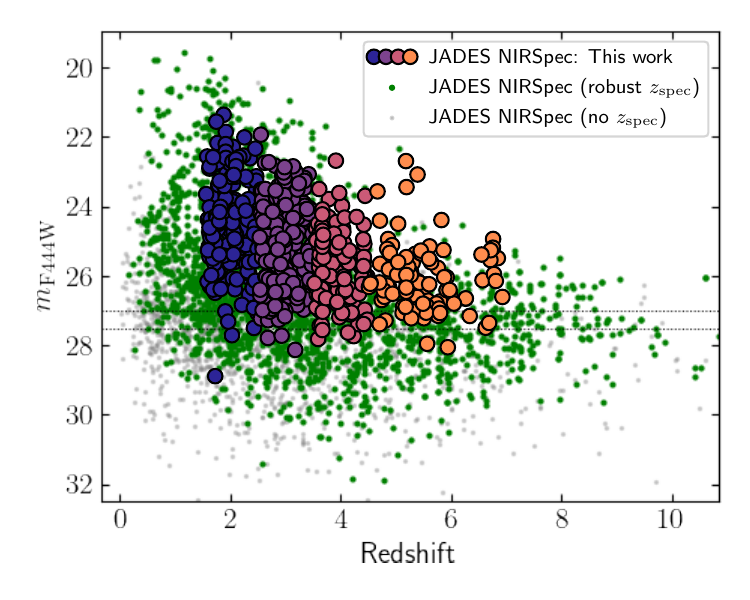}
    \includegraphics[width=\columnwidth]{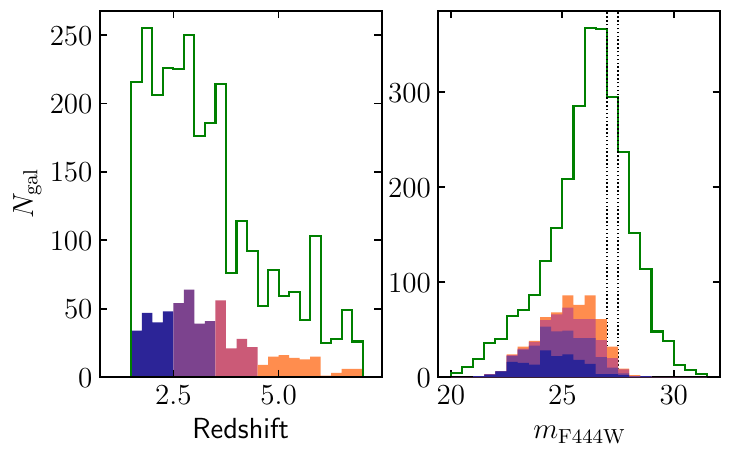}
    \caption{
    \textit{Upper panel:} $m_{\rm F444W}$ vs. redshift for JADES NIRSpec galaxies entering into this sample. Large circles show the galaxies making it into our final sample (see Section~\ref{sec:data}) with colours denoting redshift. Green points show the distribution of all JADES galaxies with robust spectroscopic redshifts, while grey points show the portion of the JADES sample for which no reliable redshift was obtained (with the point location given by the photometric redshift).
    \textit{Lower left:} Redshift distribution of final sample (colours) vs. all JADES galaxies with a robust spectroscopic redshift and $1.5<z\leq 7$. 
    \textit{Lower right:} As for lower left, except showing the $m_{\rm F444W}$ distribution.
    Black dashed lines at $m_{\rm F444W}=27.0$ and $27.5$ show the magnitude cuts for the $m_{\rm F444W}$-limited portion of the JADES NIRSpec selection for Medium/JWST and Deep/JWST respectively (see \citealt{CurtisLake2025_DR4} for details).}
    \label{fig:selection}
\end{figure}

Figure~\ref{fig:selection} shows the distribution of $m_{\rm F444W}$ as a function of redshift for the JADES spectroscopic sample. The parent sample, shown in grey, includes filler targets not selected within our main target allocation scheme, but which were added at a lower priority. The $m_{\rm F444W}$ cuts employed for the main sample are shown by the black dotted lines -- the fainter $m_{\rm F444W}<27.5$ cut corresponds to the `deep' tiers, while $m_{\rm F444W}<27.0$ was employed for the `medium' tiers which dominate the number counts here. 
In the end, many of our filler targets returned robust redshifts (green points). 

We take our initial sample as anything with a confirmed spectroscopic redshift $1.5<z_{\rm spec}<7.0$ for which we have coverage of each of \OIIll3727, H$\beta$, \OIIIl5007, \Ha, and \NIIl6583, meaning that some objects are discarded from our sample due to one or more of these lines falling within the detector chip gap.

A brief note on JADES target nomenclature: given that our sample spans multiple JADES tiers, and ID numbers can be duplicated across tiers, we identify individual galaxies by listing both their tier and ID number. The tier name has three components: [field]-[depth][hst/jwst], which, respectively, indicate (1) whether the pointing is in GOODS-S or GOODS-N, (2) whether the pointing was observed at `medium' or `deep' depth, and (3) whether the target selection was performed from catalogues based on primarily \emph{HST}- or \emph{JWST}-based imaging (see \citealt{CurtisLake2025_DR4} for details). Throughout this paper we interchangeably refer to tier names in their full form, or in shorthand, where the field name is shortened to `gs' or `gn' and single character designations are used for depth (`m' or `d') and selection (`h' or `j'). For example, goods-s-mediumjwst is used interchangeably with `gsmj'.

Throughout this paper we use 1D and 2D spectra reduced by the JADES pipeline, as described in \citet{Scholtz2025_DR4}. 
All our JADES observations employ a 3-point nodding scheme, and our default approach is to perform local background subtraction using this 3-point scheme.
Many of the targets in this paper are spatially extended, and thus we default to using 1D extractions from the full shutter open area.

\begin{figure*}
    \centering
    \includegraphics[width=0.2475\textwidth]{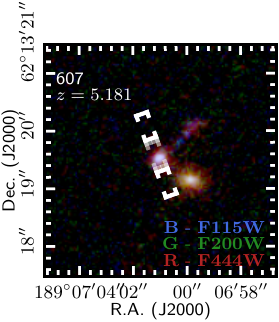}
    \includegraphics[width=0.495\textwidth]{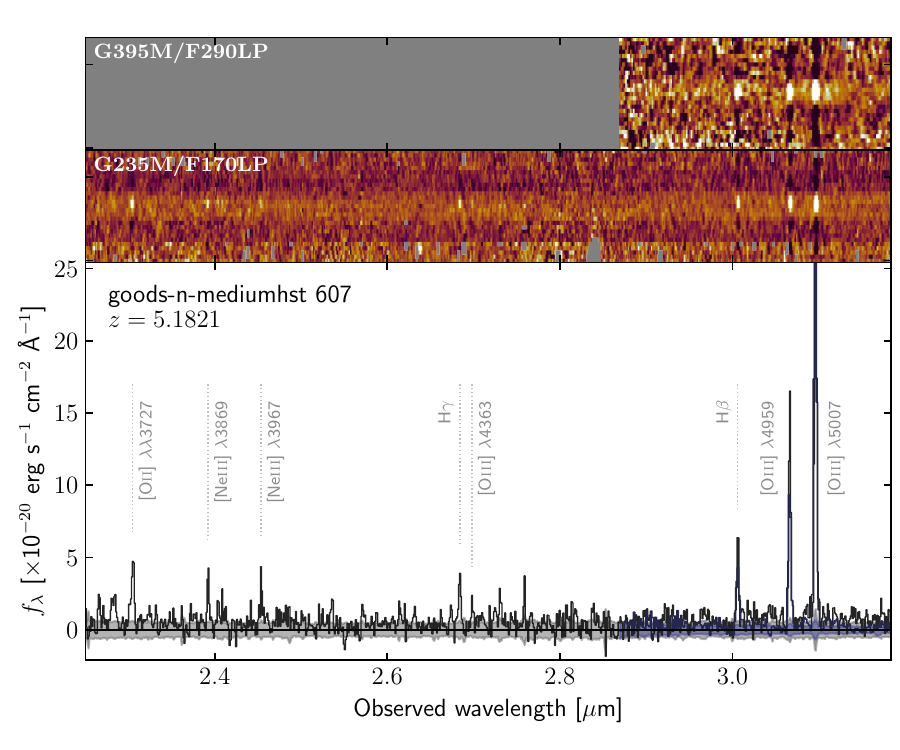}
    \includegraphics[width=0.2475\textwidth]{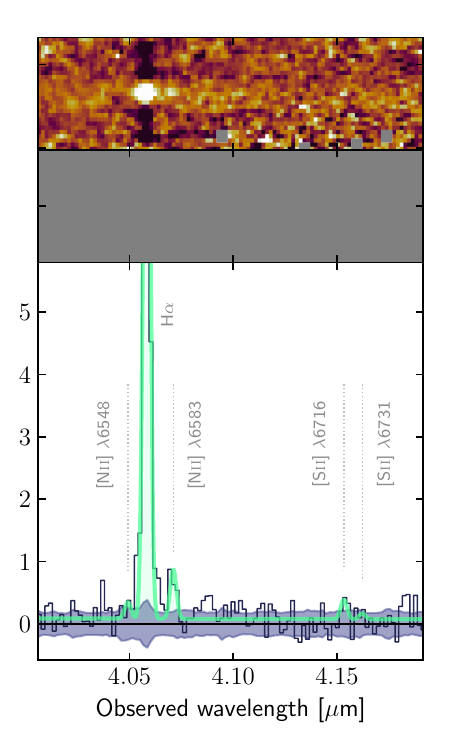}
    \includegraphics[width=0.2475\textwidth]{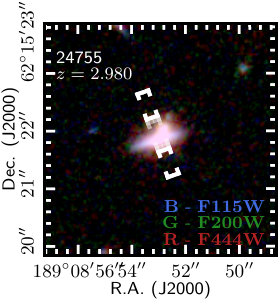}
    \includegraphics[width=0.495\textwidth]{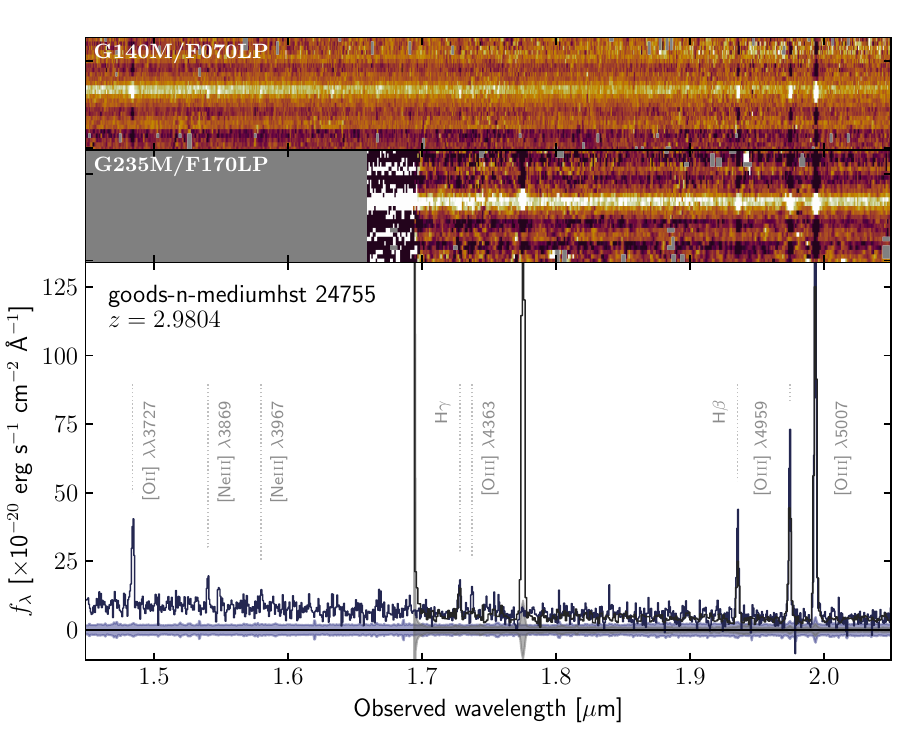}
    \includegraphics[width=0.2475\textwidth]{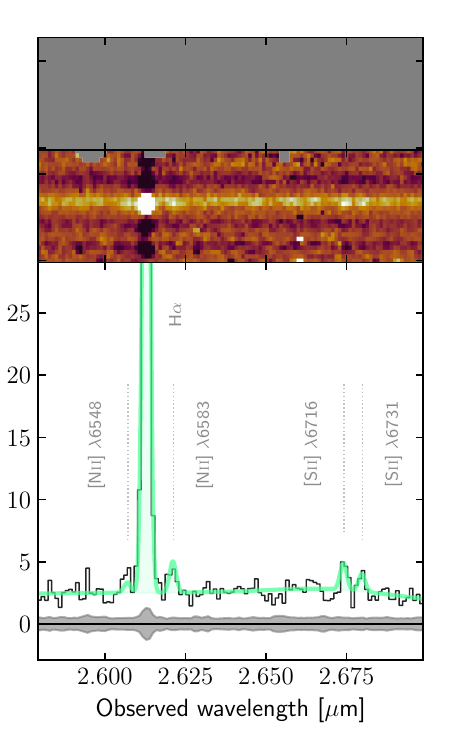}
    \includegraphics[width=0.2475\textwidth]{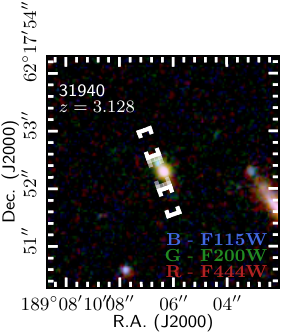}
    \includegraphics[width=0.495\textwidth]{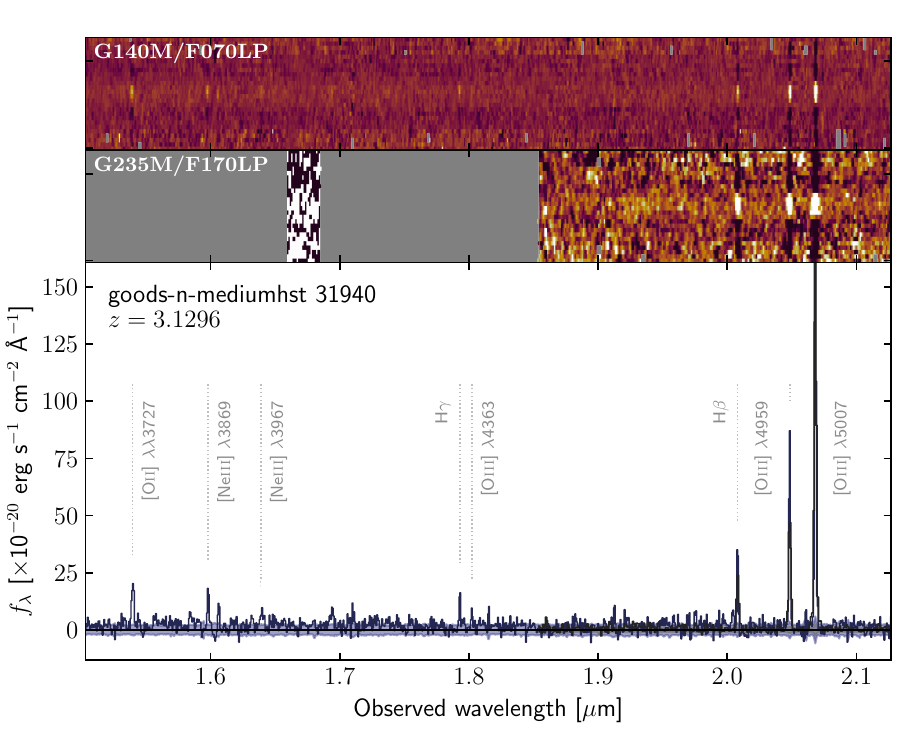}
    \includegraphics[width=0.2475\textwidth]{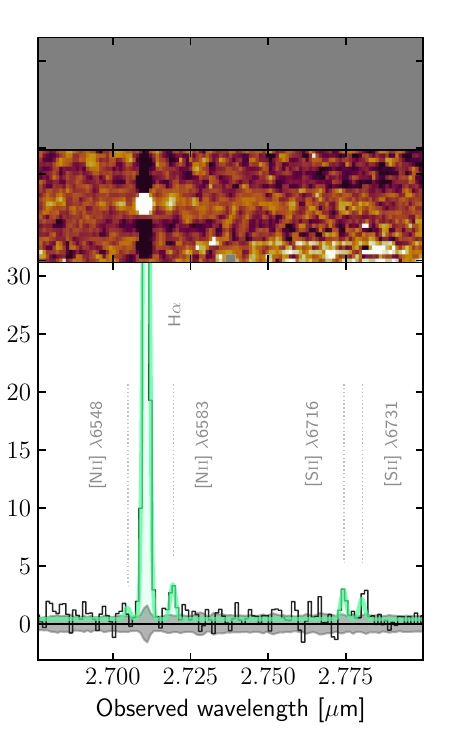}
    \caption{Example spectra of three galaxies in the $T_e$-based sample. 
    \textit{Left:} RGB thumbnail showing the location of the three-shutter slitlet.  
    \textit{Middle:} Section of the NIRSpec spectrum in which key rest-frame optical emission lines from \OIIll3726, 3729 to \OIIIl5007 are observed.
    \textit{Right:} Section of the NIRSpec spectrum surrounding \Ha, \NIIl6583, and \SII$\lambda\lambda$6716, 6731. Our best-fit model to this complex is shown in green.
    There is generally overlapping wavelength coverage between different gratings.
    Across both plot regions, the top panel shows the 2D spectrum from either G140M/F070LP or G395M/F290LP (see label), while the second panel shows the 2D spectrum from G235M/F170LP. In the 1D spectrum, G235M/F170LP is in black and other gratings are in blue. 
    }
    \label{fig:spec_auroral}
\end{figure*}

\begin{figure}
    \centering
    \includegraphics[width=\columnwidth]{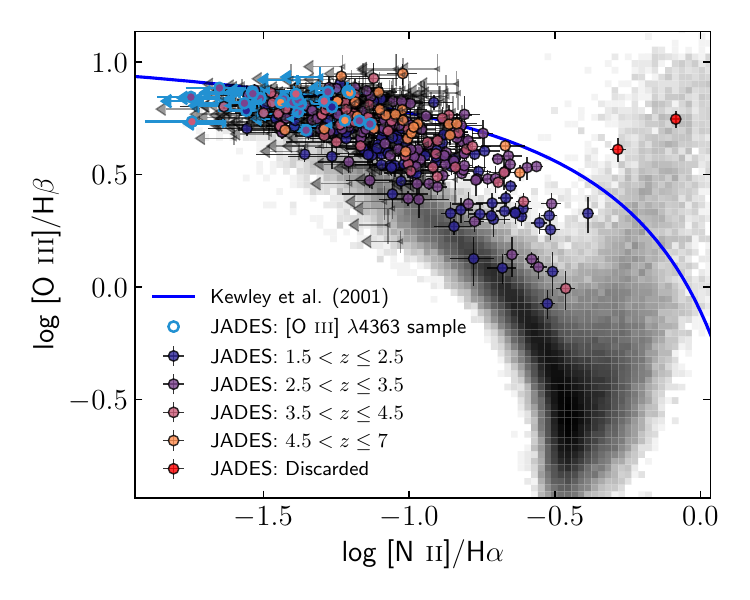}
    \includegraphics[width=\columnwidth]{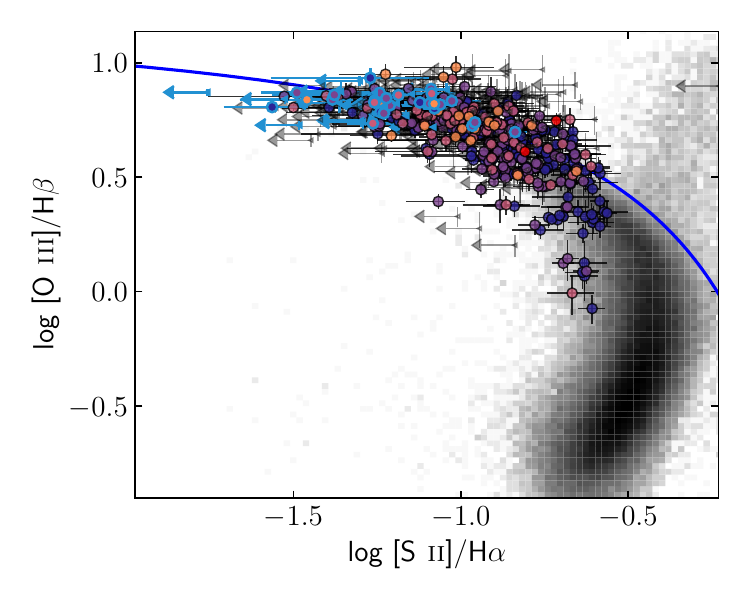}
    \includegraphics[width=\columnwidth]{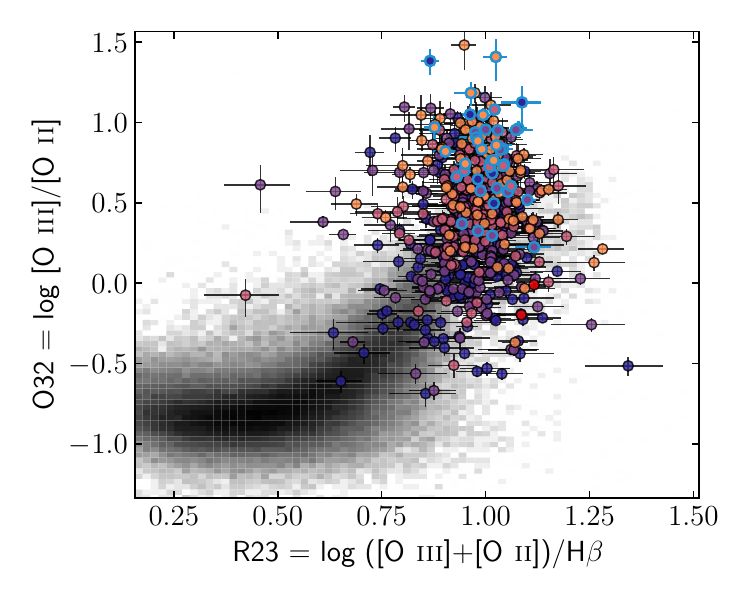}
    \caption{
    Diagnostic diagrams showing the location of galaxies in our sample in emission line ratio space compared to $z\sim0$ galaxies from SDSS \citep[][grey 2D histogram]{Aihara2011}.
    Red points show the galaxies that we discard from our sample due to significant AGN contamination, while the other colours indicate the approximate redshift of each source.
    The \citet{Kewley2001} AGN-SF demarcation line is shown in blue.
    }
    \label{fig:diagnostic_diagrams}
\end{figure}

\subsection{Emission line fitting}
\label{sub:line_fitting}

The continuum is generally at least marginally detected in the $R\sim1000$ spectra. To fit the continuum, we begin by masking out any spectral regions with prominent emission lines. We then drew 1000 perturbed spectra where each spectral pixel was adjusted randomly from a normal distribution centered on the measured flux with a standard deviation of the 1-$\sigma$ uncertainty output by the pipeline. To each realisation, we fit the continuum with a univariate spline. Our best-fit continuum was taken by combining the median at each spectral pixel across all 1000 fits, while we adopt the standard deviation at each spectral pixel as the uncertainty on the continuum fit. We visually confirm that this results in a reasonable fit and then subtract this continuum from the 1D spectrum before performing emission line fits.

We model the emission lines with single-component Gaussian profiles. The \Ha\ + [N~{\sc ii}] $\lambda\lambda$6548, 6583 + \SII\ $\lambda\lambda$6716, 6731 complex was fit simultaneously, with the line widths and redshift being tied. The flux ratio of \NII $\lambda$6583/$\lambda$6548 was fixed to 3.0, but the total flux of this \NII\ doublet was allowed to vary freely, as were the fluxes of \Ha\ and each of the two \SII\ lines. 
Examples of our \Ha\ + \NII\ + \SII\ fitting are shown in the right panels of Figure~\ref{fig:spec_auroral}. Note that, throughout this paper, mentions of `\NII\ flux' refer to the flux of only the \NIIl6583 component, not the combined doublet.
All other emission lines were fit individually with the line width and redshift allowed to vary within a small buffer range.
The fitted lines relevant for this study include \OIIll3727, \NeIII$\lambda$3869, H$\gamma$, \OIIIl4363, H$\beta$, \OIIIl4959, \OIIIl5007, in addition to the \Ha+\NII+\SII\ complex.\footnote{We note that the redshift range of our sample limits our ability to observe the \NIII$\lambda\lambda$1750 complex and the \NIV$\lambda\lambda$1483, 1486 doublet. These are unobservable for most of our sample and even where they are covered by the \gonem\ grating, the relatively shallow observations in that grating mean that these are not viably detected unless nitrogen abundances are extremely enhanced.}

We adopt the integrated flux of the best-fit profile as the line flux and the formal uncertainty from the fit as the uncertainty. If this results in a signal-to-noise ratio of $S/N<3$ then we consider the line undetected and adopt a 3-$\sigma$ upper limit by integrating the variance spectrum across five spectral pixels centered on the expected line centroid.

We correct the emission lines ratios for dust reddening using the measured \Ha/H$\beta$ ratios, assuming an intrinsic ratio of 2.79 (assuming case B recombination with $T_e=15,000$~K and $n_e=100$~cm$^{-3}$; calculated with {\sc pyneb} \citealt{Luridiana2015_pyneb}) and adopting the SMC dust law from \citep{Gordon2003}.
Throughout the rest of this paper we work with these dust-corrected emission line values.

While the vast majority of this paper focuses on these fits to the medium resolution spectra, we do consider the low-resolution prism spectra for the purpose of measuring the equivalent width of \Ha\ since the improved continuum sensitivity makes for a more reliable measurement of this quantity. The prism continuum was fit in the same way as described above for the medium-resolution spectra. Given the lower resolution, the \Ha\ + \NII\ complex is blended, and we simply fit this with a single Gaussian component. For galaxies in which we detect \NIIl6583 in the medium resolution spectrum, we use the \NII/\Ha\ ratio to account for the \NII\ contribution to this component when calculating the \Ha\ equivalent width, while for galaxies without a detection of \NII, we simply adopt the combined flux of this component as the \Ha\ flux, noting that the upper limits on \NII\ generally imply a maximum contribution of less than $5-10$~\% (see top panel of Figure~\ref{fig:diagnostic_diagrams}). The resulting equivalent width values are only used in Section~\ref{sub:correlation_galaxy_properties}.

\subsection{Final sample}
\label{sub:final_sample}

From our sample of JADES galaxies with emission line fits, we applied a signal-to-noise cut on H$\beta$ of $S/N_{H\beta}>5$. 

We use the BPT diagram (Figure~\ref{fig:diagnostic_diagrams}, top panel) to discard galaxies which are clearly contaminated with significant emission powered by an active galactic nucleus (AGN).
We note that it has been widely reported that AGN-SF demarcation lines -- such as the \citet{Kewley2001} line shown in blue -- may not be reliable for separating these populations at high-redshift \citep[e.g.][]{Scholtz2025}
In short, this is partly because the harder ionising fields of lower metallicity stellar populations at $z\gtrsim2$ drive an offset from the $z\sim0$ BPT sequence \citep[e.g.][]{Kewley2013, Steidel2014} -- an effect which is clearly seen for the locus of our sample in Figure~\ref{fig:diagnostic_diagrams}, relative to $z\sim0$ galaxies from SDSS \citep{Aihara2011}.
This effect is compounded by the fact that, at low-metallicity, AGN tend to shift toward lower values of \NII/\Ha\ \citep[e.g.][]{Groves2006}, an effect which has been observed in high-redshift AGN observed with \emph{JWST} \citep{Ubler2023}.
Given that our abundance analysis relies heavily on the \NIIl6583 line, we do not discard galaxies with log \NII/\Ha\ $<-0.5$, even if they exceed the \citet{Kewley2001} demarcation, so as not to bias ourselves against selecting galaxies with enhanced nitrogen abundances.
In the end, we discard only two galaxies as being clearly AGN-contaminated based on the BPT diagram which both were above this demarcation line and had $N2>-0.5$ (shown as red points in Figure~\ref{fig:diagnostic_diagrams}).

In the S2-VO7 diagram (middle panel) and R23-O32 diagram (bottom panel), we find that our sample largely follows the extension of the $z\sim0$ star-forming sequence, consistent with other high-redshift samples reported from \emph{JWST} observations \citep[e.g.][]{Cameron2023_ratios, Cataldi2025}.

We also discard any objects with broad emission lines. This is partly in order to remove broad-line AGN from our sample \citep[e.g.,][]{Maiolino2023_BLsample}. However, even for objects where the broad component also appears in the forbidden lines (i.e. galaxies with strong outflows; \citealt{Carniani2024_outflows}), we remove these from our sample since a significant broad component can confuse the fitting of \NII\ and \Ha.

Applying these extra sample selection constraints, we arrive at a final sample of \Nall\ galaxies, of which \NallTe\ have \OIIIl4363 detections. 
In the next sections, we outline our SED and morphological fitting, which are based on NIRCam imaging. Only \Nmass\ (including 31 with \OIIIl4363) of these \Nall\ galaxies have robust galaxy properties measured from SED and morphological fitting, limiting some of our analyses to this subset.
The coloured points and histograms in Figure~\ref{fig:selection} show the redshift and magnitude distributions of our final sample compared to the distributions of all JADES galaxies with $1.5<z_{\rm spec}<7.0$.
The redshift distribution of our final sample approximately mimics that of the parent sample. In the bottom right panel of Figure~\ref{fig:selection}, we see that the number counts in our final sample level off at $m_{\rm F444W}\approx25.5$, around a magnitude brighter than in the parent sample. However, we do retain some objects even fainter than the $m_{\rm F444W}=27$ and $27.5$ survey cuts.

\subsection{SED fitting}
\label{sub:sed_fitting}

To derive basic galaxy properties, we perform SED fitting to \emph{JWST}/NIRCam photometry with {\sc prospector} \citep{Johnson2019_prospector, Johnson2021_prospector} following the procedure outlined in \citet{Simmonds2024}. We use photometric fluxes from the JADES \citep{Eisenstein2023_JADES, Rieke2023_DR1} catalog. 

For any given galaxy, we fit using all available photometric bands. In the best cases, this amounts to \emph{JWST}/NIRCam photometry in F090W, F115W, F150W, F162M, F182M,
F200W, F210M, F250M, F277W, F300M, F335M, F356W, F410M, F430M, F444W, F460M, and F480M as well as photometry from \emph{HST}/ACS (F435W, F606W, F775W, F814W, and F850LP) and \emph{HST}/WFC3 (F105W, F125W, F140W, and F160W). 
These photometric data are derived from imaging based on numerous programs, including:
JADES (\citealt{Eisenstein2023_JADES}),
JEMS (\citealt{Williams2023_JEMS}),
JOF (\citealt{Eisenstein2023_3215}), and
FRESCO (\citealt{Oesch2023_FRESCO}).

Photometry is extracted using a consistent Kron convolved aperture across all bands, and photometric uncertainties are floored at 5\% to account for residual systematic errors not captured by the data reduction pipelines or the SED modelling. Redshifts are fixed to spectroscopic values from NIRSpec. The stellar population modelling follows \citet{Tacchella2022}, adopting a Chabrier IMF \citep{Chabrier2003}, stellar metallicities spanning $0.01$--$1,Z_{\odot}$, and a two-component dust attenuation model that accounts for differential attenuation of young ($<10$ Myr) stars and nebular emission \citep{Kriek2013}. The spectral energy distributions are generated using the FSPS code \citep{Conroy2010}, with nebular emission computed via \textsc{Cloudy} and MIST stellar evolutionary tracks. We adopt a non-parametric star formation history (continuity SFH; \citealt{Leja2019}), in which the SFH is described by eight time bins whose amplitudes and ratios are allowed to vary following a Student’s $t$-distribution with width 0.3, allowing for bursty behaviour when supported by the data. Intergalactic medium absorption is modelled using a flexible scaling of the \citet{Madau1995} prescription.
For further details of the SED-fitting methodology, we refer the reader to \citet{Simmonds2024}.

\subsection{Morphological fits}
\label{sub:size}

We compute sizes and other structural parameters for our sample using \texttt{pysersic} \citep{pysersic}, a Bayesian inference framework for fitting surface brightness profiles to galaxies with robust uncertainty estimation made possible by implementing Markov chain Monte Carlo (MCMC) methods. We adopt sizes 
from the posterior distributions of the structural parameters for single-component Sérsic fits. 

A detailed description of the procedure for fitting Sérsic profiles to galaxies in JADES, and the resulting structural parameters, can be found in 
Carreira et al. (2026).  
In brief, we obtain cutouts of sources from the JADES \textit{JWST}/NIRCam F200W and F444W imaging mosaics, as described in 
Johnson et al. (2026) and Robertson et al. (2026). 
Additionally, we obtain cutouts of the uncertainty mosaic and segmentation map for the source, the latter of which indicates the assignment of pixels in the mosaic to a particular source. These cutouts, alongside a modeled PSF representative of the observational program in which a given source was primarily imaged, are provided to \texttt{pysersic}. We utilize the MCMC sampling mode, resulting in posterior distributions for 7 different parameters: centroid values, total flux, effective (half-light) radius, S\'ersic index $n$, axis-ratio, and position angle. The median value of each posterior distribution is used as the best fit value of that parameter. 

\section{Abundance measurements}
\label{sec:abundances}

The gold-standard for extragalactic abundance studies is to measure abundances via the so-called `direct method'
which relies on directly constraining the electron temperature ($T_e$) with one or more faint, temperature-sensitive auroral lines.
We detect the \OIIIl4363 auroral line in \NallTe\ of our galaxies, allowing us to make temperature-based abundance measurements for only a fraction of our full sample of \Nall\ galaxies. 
Thus, much of the analysis in this paper will focus on abundances derived from strong-line methods. 
Although strong-line methods have larger systematic uncertainties on a galaxy-by-galaxy basis, 
across a sufficiently large sample, they are still effective for studying trends between chemical abundances and galaxy properties \citep[e.g.][]{Sanders2020_Te}.
However, we begin by exploring temperature-based abundance measurements, making use of our \NallTe\ \OIIIl4363 detections.

\begin{figure*}
    \centering
    \includegraphics[width=\textwidth]{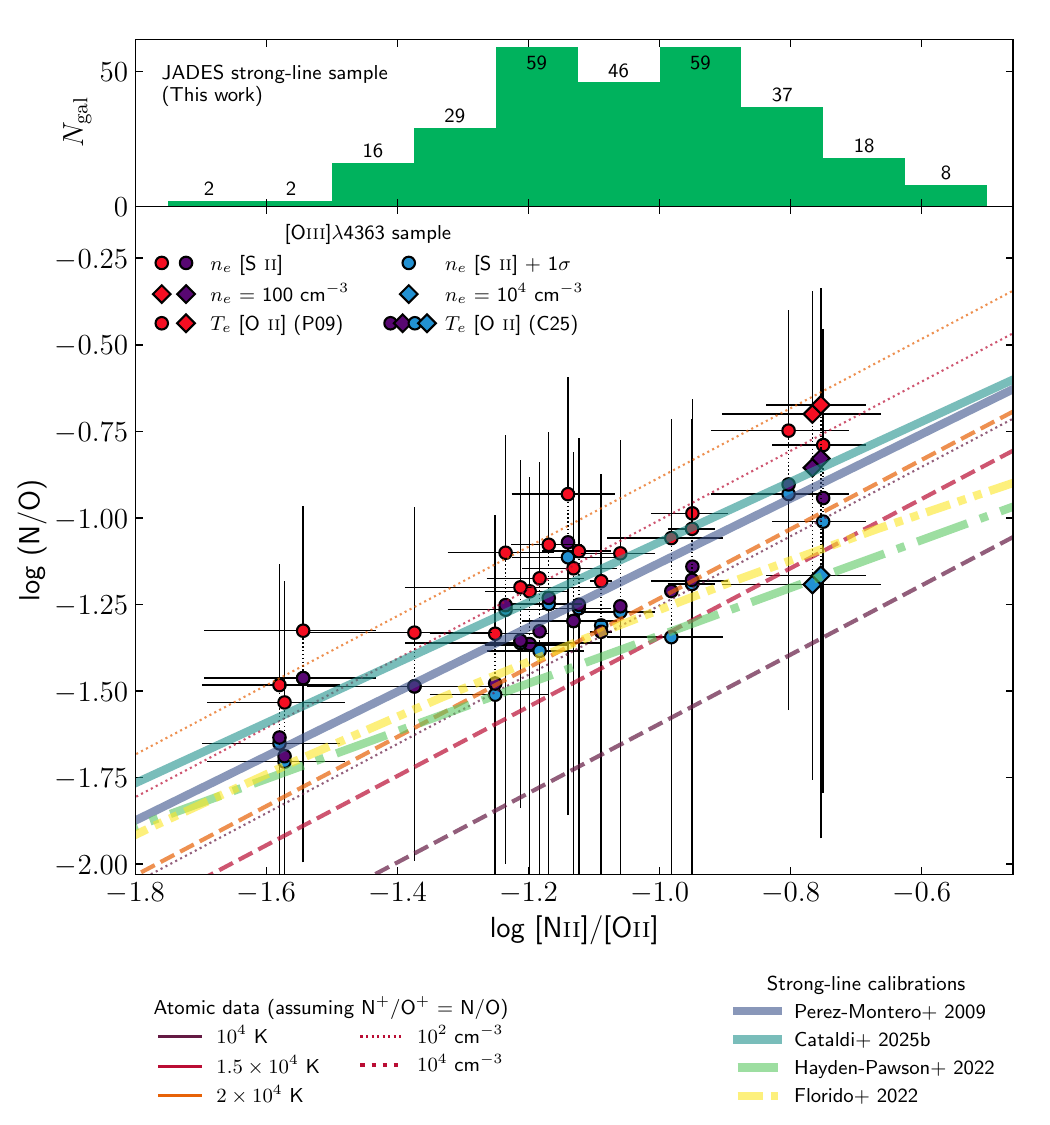}
    \caption{
    \textit{Main panel:} JADES $T_e$-based N/O abundance measurements as a function of measured \NIIl6583/\OIIll3726, 3729 ratio under different $T_e$ and $n_e$ assumptions. All points have $T_e$\OIII\ derived from \OIIIl4363/$\lambda$5007 measurements, comprising a sample of 22 galaxies. Purple points show abundances derived assuming the $T_e$\OII -- $T_e$\OIII\ relation of \citet{Cataldi2025}, while red points show how these change if \citet{Pilyugin2009} is adopted instead. Circles denote galaxies for which $n_e$ is measured from the \SII$\lambda\lambda$ 6716,6731 doublet, and the blue circles show how the purple points would shift if the $+1\sigma$ value was adopted instead of the best-fit $n_e$ value. Diamonds denote galaxies without $n_e$ constraints -- purple and blue diamonds assume $n_e=100$ \& $10^4$~cm$^{-3}$, respectively.
    Dotted and dashed lines show conversions of \NII/\OII to N/O for different $T_e$ and $n_e$ values (assuming ICF(${\rm N}^+/{\rm O}^+)=1$). Dark blue solid, light blue solid, green dot-dash, and yellow double-dot-dash lines show the \citet{PerezMontero2009}, \citet{Cataldi2025_NO}, \citet{HaydenPawson2022}, and \citet{Florido2022} strong-line calibrations respectively.
    \textit{Top panel:} The distribution of \NIIl6583/\OIIll3726, 3729 ratios among galaxies in which both lines are detected with $S/N>3$ in our JADES strong-line sample.
    }
    \label{fig:NO_calib}
\end{figure*}

\subsection{Temperature-based oxygen abundances}
\label{sub:direct_method}

\begin{table*}
    \centering
    \caption{Nebular properties and chemical abundances derived for the \NallTe\ galaxies in our $T_e$-based sample. 
    Columns $1-3$ give unique identifiers and redshift. Columns 4 \& 5  give the $n_e$ derived from the \SII $\lambda$6716/$\lambda$6731 ratio and $T_e$ derived from the \OIII $\lambda$4363/$\lambda$5007 ratio. Column 6 gives the low-ionisation $T_e$\OII\ derived adopting the \citetalias{Cataldi2025} and \citetalias{Pilyugin2009} relations, respectively. Columns 7 \& 9 give \lOH\ and \logNO\ derived using the \citetalias{Cataldi2025} relation, while Columns 8 \& 10 indicate how these values change if the \citetalias{Pilyugin2009} relation is adopted instead.
    }
    \renewcommand{\arraystretch}{1.3} 
    \begin{tabular}{|cc|c|cl|lccccc}
    \hline
    Tier & ID & $z$ & $n_e$ & $T_e$\OIII & $T_e$\OII & $12+\log($O/H)  & $\Delta$log(O/H) & log(N/O)  & $\Delta$log(N/O) \\
         &    &     & cm$^{-3}$ & $10^4$ K & $10^4$ K & ($T_e$\OII: \citetalias{Cataldi2025}, & \citetalias{Pilyugin2009}, & \citetalias{Cataldi2025}, & \citetalias{Pilyugin2009})  \\
    \hline
goods-s-deephst & 18846 & 6.3349 & ... & $1.9_{-0.3}^{+0.3}$ & $1.4-1.9$ & $7.56_{-0.03}^{+0.03}$ & $-0.03$ & $<-0.64$ & $+0.15$ \\
goods-s-mediumhst & 58850 & 6.2633 & ... & $1.7_{-0.3}^{+0.3}$ & $1.3-1.7$ & $7.78_{-0.04}^{+0.03}$ & $-0.01$ & $<-0.41$ & $+0.16$ \\
goods-s-deephst & 22251 & 5.798 & ... & $2.6_{-0.5}^{+0.5}$ & $1.6-2.4$ & $7.39_{-0.03}^{+0.03}$ & $-0.04$ & $<-0.93$ & $+0.14$ \\
goods-s-mediumjwst & 81034 & 5.3904 & ... & $2.6_{-0.4}^{+0.5}$ & $1.7-2.4$ & $7.41_{-0.02}^{+0.02}$ & $-0.02$ & $<-0.97$ & $+0.14$ \\
goods-n-mediumjwst & 79349 & 5.1824 & ... & $2.1_{-0.3}^{+0.3}$ & $1.4-2.0$ & $7.58_{-0.03}^{+0.03}$ & $-0.03$ & $<-0.94$ & $+0.15$ \\
goods-n-mediumhst & 607 & 5.1822 & ... & $1.8_{-0.3}^{+0.3}$ & $1.3-1.8$ & $7.79_{-0.04}^{+0.04}$ & $-0.04$ & $-0.85_{-0.36}^{+0.56}$ & $+0.16$ \\
goods-n-mediumjwst & 70920 & 5.0356 & $<1011$ & $2.1_{-0.2}^{+0.2}$ & $1.4-2.0$ & $7.63_{-0.03}^{+0.03}$ & $-0.04$ & $<-1.27$ & $+0.15$ \\
goods-s-mediumjwst & 206035 & 4.7735 & ... & $1.9_{-0.3}^{+0.3}$ & $1.3-1.8$ & $7.64_{-0.03}^{+0.03}$ & $-0.04$ & $-0.83_{-0.34}^{+0.76}$ & $+0.15$ \\
goods-s-mediumhst & 15325 & 4.7002 & ... & $2.5_{-0.4}^{+0.5}$ & $1.6-2.3$ & $7.42_{-0.04}^{+0.04}$ & $-0.01$ & $<-0.69$ & $+0.15$ \\
goods-n-mediumhst & 3008 & 4.5345 & $1930_{-877}^{+4029}$ & $1.7_{-0.3}^{+0.3}$ & $1.2-1.7$ & $7.85_{-0.02}^{+0.02}$ & $-0.05$ & $-1.21_{-0.34}^{+0.7}$ & $+0.15$ \\
goods-n-mediumjwst & 10000865 & 4.407 & $10_{-10}^{+122}$ & $1.5_{-0.2}^{+0.2}$ & $1.1-1.5$ & $8.06_{-0.02}^{+0.02}$ & $-0.06$ & $-1.36_{-0.33}^{+0.84}$ & $+0.15$ \\
goods-n-mediumjwst & 16553 & 4.3822 & $<100$ & $1.8_{-0.3}^{+0.3}$ & $1.3-1.8$ & $7.82_{-0.03}^{+0.02}$ & $-0.05$ & $<-1.36$ & $+0.16$ \\
goods-s-deephst & 7892 & 4.2286 & ... & $3.0_{-0.7}^{+0.9}$ & $1.8-2.8$ & $7.26_{-0.04}^{+0.03}$ & $-0.04$ & $<-0.93$ & $+0.14$ \\
goods-s-deephst & 4270 & 4.0224 & $<100$ & $1.7_{-0.2}^{+0.2}$ & $1.2-1.7$ & $7.77_{-0.02}^{+0.02}$ & $-0.06$ & $-1.49_{-0.36}^{+0.5}$ & $+0.16$ \\
goods-s-mediumjwst & 191576 & 3.7937 & $284_{-65}^{+140}$ & $1.3_{-0.1}^{+0.1}$ & $1.1-1.3$ & $8.14_{-0.02}^{+0.01}$ & $-0.05$ & $-1.18_{-0.32}^{+1.08}$ & $+0.15$ \\
goods-s-mediumjwst & 88182 & 3.6505 & ... & $1.9_{-0.3}^{+0.3}$ & $1.4-1.9$ & $7.7_{-0.03}^{+0.03}$ & $-0.05$ & $<-1.08$ & $+0.15$ \\
goods-n-mediumjwst & 42939 & 3.4671 & $470_{-56}^{+1208}$ & $1.5_{-0.2}^{+0.2}$ & $1.1-1.5$ & $7.98_{-0.02}^{+0.02}$ & $-0.03$ & $-0.94_{-0.34}^{+0.78}$ & $+0.15$ \\
goods-s-mediumhst & 17366 & 3.4591 & ... & $1.7_{-0.2}^{+0.2}$ & $1.3-1.7$ & $7.84_{-0.03}^{+0.03}$ & $-0.04$ & $<-1.09$ & $+0.16$ \\
goods-s-mediumjwst & 60310684 & 3.3467 & $<165$ & $1.5_{-0.2}^{+0.2}$ & $1.2-1.6$ & $7.96_{-0.02}^{+0.02}$ & $-0.05$ & $-1.35_{-0.37}^{+0.48}$ & $+0.15$ \\
goods-s-mediumjwst & 159438 & 3.2389 & $856_{-343}^{+812}$ & $1.6_{-0.1}^{+0.1}$ & $1.2-1.6$ & $7.88_{-0.01}^{+0.01}$ & $-0.04$ & $-1.14_{-0.33}^{+0.87}$ & $+0.15$ \\
goods-n-mediumhst & 31940 & 3.1295 & $<389$ & $1.7_{-0.3}^{+0.3}$ & $1.3-1.7$ & $7.76_{-0.03}^{+0.03}$ & $-0.03$ & $-0.9_{-0.35}^{+0.62}$ & $+0.16$ \\
goods-s-mediumjwst & 54612 & 3.0833 & $315_{-211}^{+-182}$ & $1.3_{-0.1}^{+0.1}$ & $1.0-1.3$ & $8.19_{-0.01}^{+0.01}$ & $-0.1$ & $-1.33_{-0.31}^{+1.4}$ & $+0.15$ \\
goods-s-mediumhst & 8767 & 3.0633 & $1002_{-408}^{+1198}$ & $1.9_{-0.3}^{+0.3}$ & $1.3-1.8$ & $7.79_{-0.02}^{+0.02}$ & $-0.07$ & $-1.33_{-0.34}^{+0.77}$ & $+0.15$ \\
goods-s-mediumjwst & 192837 & 3.0 & $<100$ & $1.5_{-0.2}^{+0.2}$ & $1.1-1.5$ & $8.02_{-0.02}^{+0.01}$ & $-0.04$ & $-1.3_{-0.34}^{+0.78}$ & $+0.15$ \\
goods-n-mediumhst & 24755 & 2.9803 & $10_{-10}^{+585}$ & $2.9_{-0.5}^{+0.6}$ & $1.8-2.7$ & $7.35_{-0.03}^{+0.03}$ & $-0.05$ & $-1.07_{-0.34}^{+0.75}$ & $+0.14$ \\
goods-n-mediumhst & 33391 & 2.9016 & $475_{-165}^{+233}$ & $1.5_{-0.1}^{+0.1}$ & $1.1-1.5$ & $7.99_{-0.02}^{+0.02}$ & $-0.05$ & $-1.26_{-0.33}^{+0.89}$ & $+0.15$ \\
goods-s-mediumjwst & 10012070 & 2.8105 & $10_{-10}^{+252}$ & $1.7_{-0.2}^{+0.2}$ & $1.2-1.7$ & $7.97_{-0.02}^{+0.02}$ & $-0.07$ & $-1.69_{-0.35}^{+0.62}$ & $+0.16$ \\
goods-s-mediumjwst & 225349 & 2.7851 & $325_{-10}^{+769}$ & $2.7_{-0.6}^{+0.7}$ & $1.7-2.5$ & $7.37_{-0.05}^{+0.05}$ & $-0.02$ & $<-1.0$ & $+0.14$ \\
goods-s-mediumjwst & 164658 & 2.6749 & $<100$ & $3.0_{-0.7}^{+0.9}$ & $1.8-2.8$ & $7.48_{-0.05}^{+0.04}$ & $-0.08$ & $-1.46_{-0.36}^{+0.53}$ & $+0.14$ \\
goods-s-mediumjwst & 176834 & 2.6699 & $369_{-98}^{+225}$ & $2.1_{-0.3}^{+0.3}$ & $1.4-2.0$ & $7.68_{-0.02}^{+0.02}$ & $-0.06$ & $-1.63_{-0.35}^{+0.62}$ & $+0.15$ \\
goods-s-mediumjwst & 202446 & 2.6167 & $233_{-17}^{+201}$ & $2.0_{-0.4}^{+0.4}$ & $1.4-1.9$ & $7.72_{-0.02}^{+0.02}$ & $-0.1$ & $-1.23_{-0.33}^{+0.91}$ & $+0.15$ \\
goods-s-mediumjwst & 167544 & 2.5762 & $76_{-10}^{+151}$ & $2.2_{-0.4}^{+0.5}$ & $1.5-2.1$ & $7.56_{-0.03}^{+0.03}$ & $-0.08$ & $-1.25_{-0.34}^{+0.73}$ & $+0.15$ \\
goods-n-mediumhst & 29455 & 2.5309 & ... & $2.0_{-0.3}^{+0.3}$ & $1.4-1.9$ & $7.71_{-0.05}^{+0.04}$ & $-0.03$ & $<-1.02$ & $+0.15$ \\
goods-s-mediumjwst & 51489 & 2.4561 & $<864$ & $2.8_{-0.6}^{+0.7}$ & $1.7-2.6$ & $7.44_{-0.06}^{+0.05}$ & $-0.01$ & $<-0.61$ & $+0.14$ \\
goods-n-mediumhst & 23638 & 2.4173 & $10_{-10}^{+201}$ & $2.2_{-0.3}^{+0.4}$ & $1.4-2.1$ & $7.62_{-0.03}^{+0.02}$ & $-0.05$ & $<-1.44$ & $+0.15$ \\
goods-s-mediumjwst & 60309998 & 2.1989 & ... & $1.7_{-0.2}^{+0.2}$ & $1.2-1.7$ & $7.67_{-0.03}^{+0.02}$ & $-0.02$ & $<-0.65$ & $+0.16$ \\
goods-n-mediumhst & 26751 & 2.1763 & $10_{-10}^{+193}$ & $1.6_{-0.2}^{+0.2}$ & $1.2-1.6$ & $7.93_{-0.02}^{+0.02}$ & $-0.08$ & $-1.25_{-0.33}^{+0.92}$ & $+0.15$ \\
goods-n-mediumhst & 29849 & 1.8142 & $<149$ & $1.6_{-0.2}^{+0.2}$ & $1.2-1.6$ & $7.8_{-0.02}^{+0.02}$ & $-0.03$ & $<-1.05$ & $+0.16$ \\
goods-n-mediumhst & 25451 & 1.676 & ... & $1.8_{-0.2}^{+0.3}$ & $1.3-1.7$ & $7.85_{-0.03}^{+0.02}$ & $-0.03$ & $<-0.99$ & $+0.16$ \\
goods-n-mediumhst & 25771 & 1.675 & $10_{-10}^{+438}$ & $1.2_{-0.2}^{+0.1}$ & $1.0-1.3$ & $8.17_{-0.02}^{+0.01}$ & $-0.06$ & $-1.48_{-0.34}^{+0.69}$ & $+0.14$ \\
    \hline
    \end{tabular}
    \label{tab:auroral_sample}
\end{table*}

\renewcommand{\arraystretch}{1} 

For our \NallTe\ galaxies with \OIIIl4363 auroral line detections (`$T_e$-based sample' hereafter), we derive the oxygen abundance (O/H) following the methodology outlined in \citet{Cameron2023_GNz11}.
Briefly, we assume the total oxygen abundance is the sum of singly- and doubly-ionised oxygen (${\rm O}/{\rm H}={\rm O}^{+}/{\rm H}^{+} + {\rm O}^{2+}/{\rm H}^{+}$), deriving each ionic abundance from the \OIIIl5007/H$\beta$ and \OIIll3727/H$\beta$ line ratios respectively, using emissivities calculated with \texttt{pyneb} \citep{Luridiana2015_pyneb}.

Where available, we use $n_e$ measured from the \SII\ $\lambda$6716/$\lambda$6731 ratio for the emissivity calculation. For 17 galaxies in our $T_e$-based sample, detections of the \SII\ lines afford a $n_e$ measurement with values ranging up to $n_e\approx2000$~cm$^{-3}$, although the majority are below $n_e\lesssim 500$~cm$^{-3}$ (Table~\ref{tab:auroral_sample}). For a further 9 galaxies we can only place an upper limit on the $n_e$, and for these we adopt a density of $n_e=100$~cm$^{-3}$ for our emissivity calculation.
The remaining 14 galaxies lack a detection of \SII\ altogether. For these, we also adopt $n_e=100$~cm$^{-3}$ as our fiducial assumption, but also consider the effect of assuming a higher density of $n_e=10^4$~cm$^{-3}$ (see discussion below).

We use the \OIIIl4363 / \OIIIl5007 ratio to derive $T_e$\OIII, which we adopt as the temperature of the high-ionisation zone. Given that none of these galaxies exhibits a detection of any low-ionisation auroral lines, such as \OIIll7320, 7330 or \NIIl5755, we must rely on an empirical calibration to derive $T_e$\OII\ from $T_e$\OIII.

Many such $T_e$\OII\ -- $T_e$\OIII\ calibrations exist \citep[e.g.][]{Campbell1986, Garnett1992, Pilyugin2009}, but this approach suffers from significant intrinsic scatter observed around this relation \citep[e.g.][]{Berg2020, Yates2020_TeRelation, MendezDelgado2023}.
Furthermore, little is yet known about the evolution of these relations toward higher redshift and lower metallicity as existing data sets in these regimes are still relatively small \citep[e.g.][]{Strom2023, Rogers2023, HamelBravo2025}, and exhibit significant scatter \citep[e.g.][]{Cataldi2025, Sanders2025_TeCalib}. 
For this analysis, we adopt two temperature calibrations: those of \citet{Pilyugin2009} (\citetalias{Pilyugin2009} hereafter) and \citet{Cataldi2025} (\citetalias{Cataldi2025} hereafter). The \citetalias{Pilyugin2009} relation is fairly steep and predicts $T_e$\OII\ $\approx$ $T_e$\OIII, while the \citetalias{Cataldi2025} relation is much shallower and predicts that $T_e$\OII\ rarely exceeds $\sim$15,000 K.
In this work we do not explore which of these two scenarios better represents our sample, but rather consider both to bracket the range of likely values.

Columns 7 \& 8 of Table~\ref{tab:auroral_sample} give the derived 12+log(O/H) values assuming the \citetalias{Cataldi2025} relation, and then the difference from that value that adopting \citetalias{Pilyugin2009} would lead to. The \citetalias{Pilyugin2009} relation results in oxygen abundances that are systematically lower (owing to the higher \OII\ temperature), but the difference is only mild ($\leq0.1$).

\subsection{Nitrogen abundance measurement}
\label{sub:Te_based_NO}

Because we do not observe higher ionisation states of nitrogen (i.e. N$^{2+}$ or N$^{3+}$) in this sample, in addition to the $T_e$- and $n_e$-dependence, we must also consider the impact of the `ionisation correction factor' (ICF), which corrects for these unseen ionisation states, in our N/O abundance determination.
The ICF of ${\rm N}^{+}/{\rm O}^{+}$ is often assumed to be unity (i.e. ${\rm N}^{+}/{\rm O}^{+}={\rm N}/{\rm O}$) \citep[e.g.][]{PeimbertCostero1969, ArellanoCordova2025_CLASSY} due to the similar ionisation potentials of these species. \citet{Amayo2021} derived an ICF based on photoionisation models which
does suggest a non-zero ICF, albeit with a very modest value of $\log({\rm N}/{\rm O})\approx\log({\rm N}^+/{\rm O^+}) + 0.1$.
Modeling in \citet{Martinez2025} found similar ICFs when low density ($n_e =10^2$~cm$^{-3}$) was assumed, although they found that, at higher densities, ${\rm N}^+/{\rm O^+}$ begins to overestimate N/O, with ICFs of around $\log({\rm N}/{\rm O})\approx\log({\rm N}^+/{\rm O^+}) - 0.1$ at $n_e=10^5$~cm$^{-3}$.
As outlined above, where we are able to measure densities in our sample, we find them to be low, albeit acknowledging that $n_e$(\SII) can underestimate density in the presence of high-density gas \citep{Martinez2025}.
As a result, throughout this work, we adopt ${\rm N}^{+}/{\rm O}^{+}={\rm N}/{\rm O}$, noting that adopting a non-zero ICF would generally push total N/O abundances higher by $\lesssim$0.1 dex in systems with $\sim10^2$~cm$^{-3}$, and would move total N/O abundances down by similar amounts toward higher densities.

Figure~\ref{fig:NO_calib} shows how the $T_e$- and $n_e$-dependence of the \NIIl6583 and \OIIll3726, 3729 emissivities are each responsible for around 0.3 dex of systematic uncertainty in the final N/O abundance if not well constrained. The dotted and dashed lines in Figure~\ref{fig:NO_calib} show the conversion from line ratio to ionic abundance for $n_e= 100 {\rm ~and~ } 10^4$ cm$^{-3}$, respectively.
At higher densities, the same \NII/\OII\ ratio corresponds to an abundance ratio that is around 0.3 dex lower.
Meanwhile, $T_e$ values of $10^4$ K (shown in purple) also result in systematically lower N/O values than if $T_e\geq1.5\times10^4$ K is assumed.

From our $T_e$-based sample, 22 galaxies exhibit detections of both \NIIl6583 and \OIIll3727, allowing us to derive the N/O ratio as 

\begin{equation}
    \frac{\text{N}}{\text{O}} \approx \frac{\text{N}^{+}}{\text{O}^{+}} = \frac{f_{\rm [N\textsc{ii}] 6583}}{f_{\rm [O\textsc{ii}] 3726, 3729}} \times \frac{\epsilon_{\rm [O\textsc{ii}] 3726, 3729}}{\epsilon_{\rm [N\textsc{ii}] 6583}},
\label{eq:no_te_method}
\end{equation}

where $\epsilon$ is the emissivity, dependent on both $T_e$ and $n_e$. The remaining 18 galaxies have detections of \OII, but only an upper-limit on \NII, resulting in an upper limit on the N/O ratio.

Of course, the required temperature for this calculation is the low-ionisation temperature which is not measured in any of our sample. As discussed above, obtaining $T_e$\OII\ via a calibration can result in significantly different values depending on which relation is adopted.
Columns 9 \& 10 of Table~\ref{tab:auroral_sample} give the log(N/O) values derived assuming the two $T_e$\OII--$T_e$\OIII relations.  
The shallower \citetalias{Cataldi2025} relation results in log(N/O) values which are $\sim$0.15 dex lower than the \citetalias{Pilyugin2009} relation. Note that the difference is more pronounced than that seen in the total oxygen abundance.
These differences are also shown in Figure~\ref{fig:NO_calib}, with the red points showing the higher derived N/O at fixed \NII/\OII\ ratio resulting from adopting the \citetalias{Pilyugin2009} relation compared to the \citetalias{Cataldi2025} relation (purple points).

The majority of these galaxies (20/22) have a measurement or an upper limit placed on the density from the \SII$\lambda\lambda$6716, 6731 doublet, indicated by the circle markers in Figure~\ref{fig:NO_calib}. We explore the effect of $n_e$ on the derived N/O abundance by showing (as blue circles) the measured N/O if the $+1\sigma$ upper limit of the measured $n_e$ is adopted instead of the best-fit $n_e$ value. In most cases, this results in negligible difference (median difference of $-0.02$ dex), but the most significant cases can result in abundances that are $0.13$ dex lower.

For the two galaxies in this sample which have a \NII\ detection but no constraint on $n_e$ from \SII, we adopt two values: 100~cm$^{-3}$ (purple diamonds) and 10$^4$~cm$^{-3}$ (blue diamonds). 
Changing the assumed density between these values results in a far more significant change of $0.33$ dex to the  derived N/O. 
This upper density value is higher than any of our measured values in this sample and higher also than the majority of low-ionisation density measurements at high-redshift \citep[e.g.][and references therein]{Harikane2025}, albeit not unprecedented \citep[e.g.][]{Berg2025_WN}. 
However, \citet{Martinez2025} showed that \SII-based $n_e$ measurements are always biased low when the density is not homogeneous, suggesting that some systems with measured $n_e(\text{\SII})\approx10^{3}$~cm$^{-3}$ could still host significant quantity of $n_e\gtrsim10^{4}$~cm$^{-3}$ gas. Thus, while we do not have direct evidence for high-density gas in any of our systems, we cannot rule out its presence, and, as shown by the blue diamonds in Figure~\ref{fig:NO_calib}, this would impact the N/O abundances derived. 
Obtaining density measurements from  density tracers such as  C~{\sc iii}]~$\lambda1907/\lambda1909$ or [Ar~{\sc iv}]~$\lambda4714/\lambda4742$, which are sensitive to higher densities, would help elucidate the presence or absence of high-density gas, but are unfortunately not detected at the depth of this data set.

\subsubsection{Nitrogen-to-oxygen strong-line calibration}

In Figure~\ref{fig:NO_calib} we show several strong-line calibrations for mapping \NII/\OII\ to N/O. Three of these were derived from $T_e$-based abundances from \HII\ regions and star-forming galaxies at $z\sim0$ (\citealt{PerezMontero2009}, \citealt{HaydenPawson2022},  and \citealt{Florido2022}; blue solid, green dot-dashed, yellow dot-dot-dashed lines, respectively).
We also show the recent calibration from \citet[light blue solid line]{Cataldi2025_NO} which was determined based on $z>1$ \emph{JWST}/NIRSpec spectroscopy.
These calibrations all return very similar values for log \NII/\OII\ $\lesssim-1.5$, but \citet{PerezMontero2009} and \citet{Cataldi2025_NO} gradually diverge to yield $\sim$0.3 dex higher N/O at log \NII/\OII\ $\gtrsim-0.8$.

Our sample with density measurements (circles), clearly align better with the \citet{PerezMontero2009} and \citet{Cataldi2025_NO} calibrations. This is true even when adopting the \citetalias{Cataldi2025} temperature relation, while adopting the \citetalias{Pilyugin2009} relation shifts the majority of our sample well above these calibrations. The two galaxies without density constraints have among the highest \NII/\OII\ ratios in our sample. Regardless of the adopted $T_e$ relation, these only yield N/O values consistent with the the \citet{HaydenPawson2022} and \citet{Florido2022} relations if $n_e=10^4$~cm$^{-3}$ is assumed. Even in these cases, they fall right on the relation, and adopting the steeper \citetalias{Pilyugin2009} $T_e$ relation or a positive ICF would push these measurements above these two shallower strong-line calibrations. 
Thus, it is clear that our sample favours the steeper calibrations.

A plausible explanation is that the offset toward high N/O values is driven by higher temperatures, seen, for example, in the atomic data lines in Figure~\ref{fig:NO_calib}.
The \citet{HaydenPawson2022} calibration is based on stacks of SDSS spectra, the vast majority of which have \lOH\ $\gtrsim 8$, and the \citet{Florido2022} calibration is based on \HII\ regions with similarly high metallicities. These metallicities are higher than the bulk of our our JADES $T_e$-based sample. As a result, one might expect that the electron temperature in these calibration samples will be much lower than our sample ($T_e\gtrsim1.5\times10^4$ K for all our individual \OIII\ measurements, in line with previous studies at high redshift; e.g. \citealt{Laseter2023, Sanders2023_TeCalib, Sanders2025_TeCalib}).

Meanwhile, the calibration from \citet{Cataldi2025_NO} is based on galaxies with $7.3\lesssim$ \lOH\ $\lesssim8.3$. Similarly, the calibration sample in 
\citet{PerezMontero2009} has a significant number of objects with \lOH\ $\lesssim8$. In their Figure~10, one can see that the `\HII\ galaxies' (which are generally low metallicity) in their sample prefer a calibration with a higher normalisation, while if the sample had been limited to `Giant Extragalactic \HII\ Regions' (which have metallicities comparable to the \citealt{HaydenPawson2022} and \citealt{Florido2022} calibration samples), a flatter calibration more in line with those works would have been derived.
As a result, we adopt the \citet{PerezMontero2009} calibration as our default N/O strong-line calibration.

\subsection{Strong-line oxygen abundances}

To support our strong-line N/O measurements, outlined above, we make strong-line oxygen abundance measurements across our sample, following the methodology outlined in \citet{Curti2023_MZR}.
In short, we adopt the strong-line calibrations of \citet{Sanders2025_TeCalib} measuring the metallicity from multiple strong-line ratios and performing a weighted average of these metallicity measurements to determine the best-fit $12+\log({\rm O}/{\rm H})$ value for the galaxy in question.
Since we ultimately want to explore how nitrogen abundance correlates with oxygen abundance, we only consider ratios of O, Ne, and H lines in these calculations.
Namely, we consider
$R3=\log($\OIII/H$\beta)$, $R2=\log($\OII/H$\beta)$, $O32=\log($\OIII/\OII$)$, $Ne3O2=\log($\NeIII/\OII$)$, and $R23=\log(($\OIII + \OII)/H$\beta)$.
These strong-line calibrations are quoted by \citet{Sanders2025_TeCalib} as being valid over the range $7.3\leq12+\log({\rm O}/{\rm H})\leq8.6$.
Without strictly enforcing this validity range in our fitting, we find that all but nine of our \NallSL\ strong-line metallicities fall within this range.
Eight galaxies in our sample marginally exceed this, with all returning $12+\log({\rm O}/{\rm H})<8.74$, while one galaxy returns a best-fit abundance of $12+\log({\rm O}/{\rm H})=7.06^{+0.07}_{-0.06}$.
The main function of our O/H measurements throughout this paper is to differentiate between nitrogen enrichment at `low-metallicity' ($12+\log({\rm O}/{\rm H})\lesssim8.0$) and `high-metallicity', since nitrogen abundances at high-metallicity are widely accepted to be driven by different enrichment sources (see Section~\ref{sec:intro}).
Thus, these minor extrapolations should not impact our results, and we leave these extrapolated abundances as derived.

\section{Results} \label{sec:results}

\begin{figure*}
    \centering
    \includegraphics[width=0.99\textwidth]{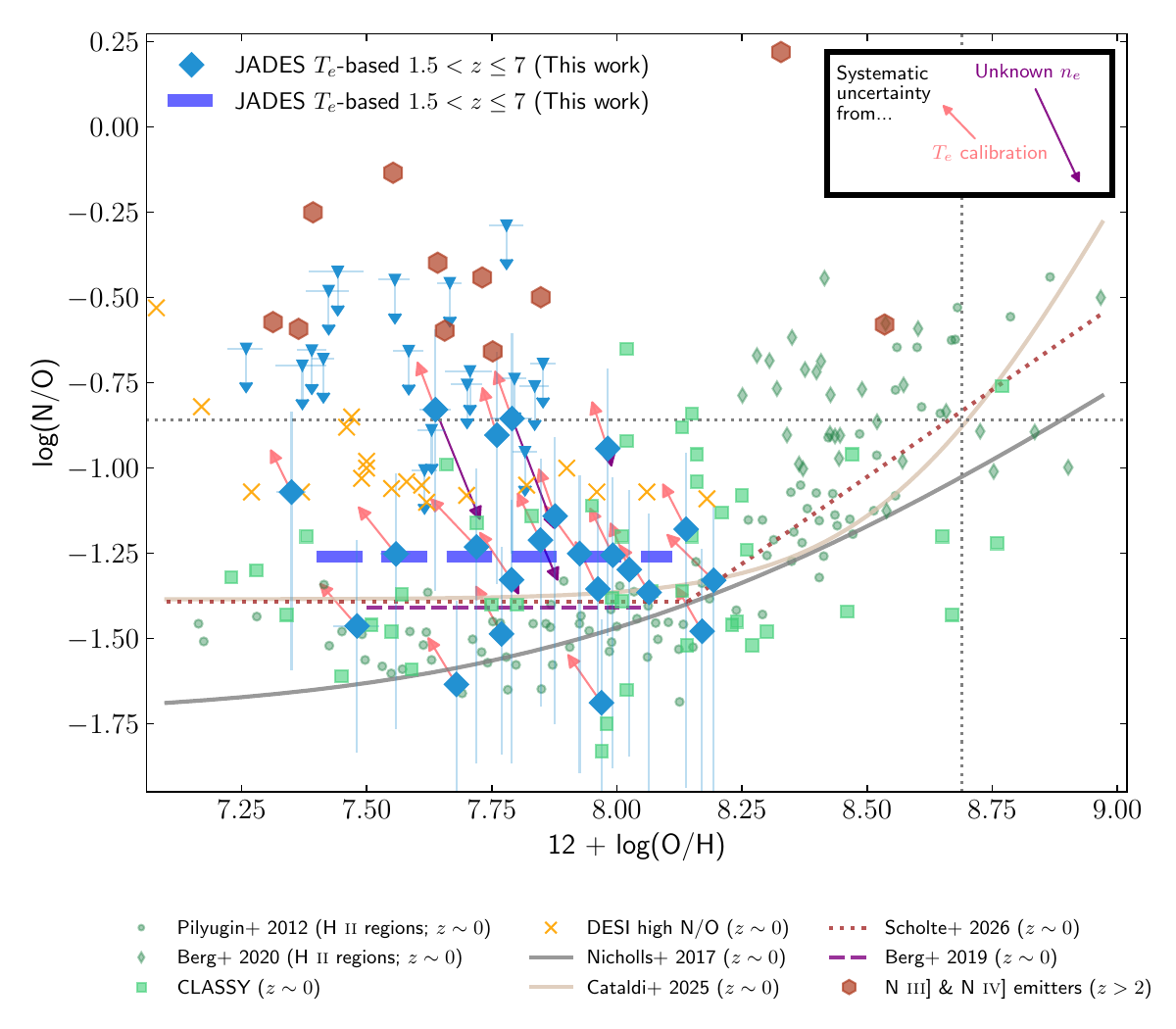}
    \caption{Nitrogen-to-oxygen abundance ratio as a function of oxygen abundance for our JADES $T_e$-based sample (blue diamonds). 
    The blue dashed line shows the weighted average N/O for this \lOH $<8.2$ sample.
    Salmon arrows show how these points would shift if the \citetalias{Pilyugin2009} $T_e$\OII-$T_e$\OIII\ calibration was used instead of the \citetalias{Cataldi2025} adopted for the fiducial values. Purple arrows show the systematic uncertainty arising from a weak or absent $n_e$ constraint (see Section~\ref{sub:Te_based_NO} for details). Green circles and diamonds show $z\sim0$ H~{\sc ii} regions \citep{Pilyugin2012, Berg2020}. Green squares show $z\sim0$ star-forming galaxies from the CLASSY survey \citep{ArellanoCordova2025_CLASSY}. Orange `$\times$' marks show the high-N/O-selected subsample of DESI galaxies from \citet{Bhattacharya2025}.
    High-redshift \NIII- and \NIV-emitters are shown by the brown hexagons, compiled from \citep{Isobe2023, MarquesChaves2023, Castellano2024, Schaerer2024, Martinez2025, Naidu2025}. 
    The grey solid, beige solid, and brown dotted lines show the ${\rm N}/{\rm O}-{\rm O}/{\rm H}$ scaling from \citet{Nicholls2017}, \citet{Cataldi2025_NO}, and \citet{Scholte2026} respectively, while the purple dashed line shows the average N/O value derived for $z\sim0$ metal-poor dwarf galaxies in \citet{Berg2019_CNO_Dwarf}.
    The faint dotted lines show the solar ratios of each abundance.
    }
    \label{fig:NO_OH_diagram}
\end{figure*}

\begin{figure*}
    \centering
    \includegraphics[width=\textwidth]{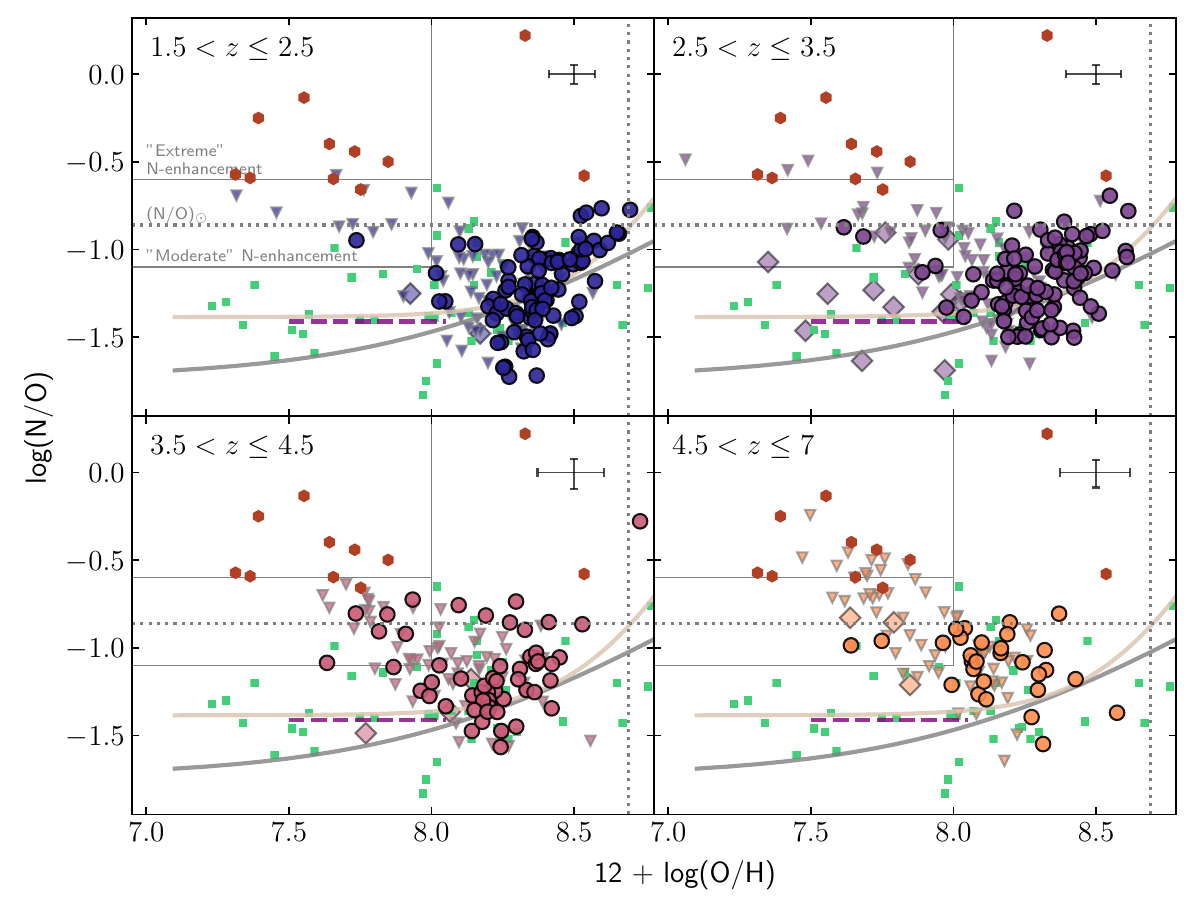}
    \includegraphics[width=\textwidth]{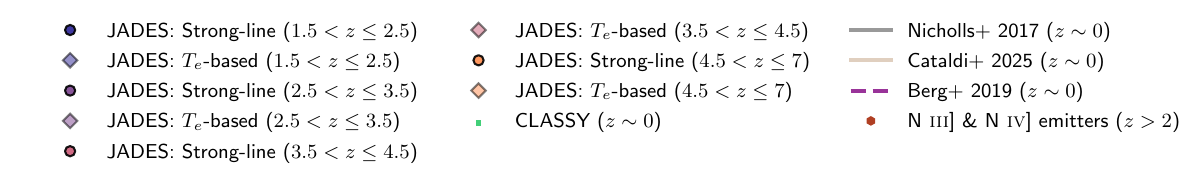}
    \caption{Nitrogen-to-oxygen abundance ratio as a function of oxygen abundance for our JADES strong-line sample, divided into four redshift bins. Circles show our strong-line abundance measurements, while triangles show galaxies for which we only have an upper limit on \logNO. The black errorbars in the upper right corner show the median measurement uncertainties for the strong-line sample.
    Diamond markers show $T_e$-based abundances, carried over from Figure~\ref{fig:NO_OH_diagram}. 
    We include demarcation lines for our `moderate' and `extreme' nitrogen-enhancement selection criteria as a visual aid (see Section~\ref{sub:N_enhanced}).
    Green squares and brown hexagons are as in Figure~\ref{fig:NO_OH_diagram}, as are the grey, beige, and purple lines.
    Faint dotted lines show the solar ratios of each abundance.}
    \label{fig:NO_OH_diagram_zbin}
\end{figure*}

\subsection{Nitrogen abundance variation with oxygen abundance}
\label{sub:NO_OH_diagram}

\subsubsection{Temperature-based abundances}

Figure~\ref{fig:NO_OH_diagram} shows the N/O abundance ratio as a function of O/H from our JADES $T_e$-based sample (diamonds). 
As discussed in Section~\ref{sub:direct_method}, deriving $T_e$\OII\ values from a temperature calibration leads to a $>0.1$ dex uncertainty in the measured N/O, and a somewhat smaller uncertainty in O/H. The salmon arrows in Figure~\ref{fig:NO_OH_diagram} show how these measured abundances shift if the adopted relation is changed from \citetalias{Cataldi2025} (which is shown by the location of the points) to \citetalias{Pilyugin2009} (head of salmon arrows).
For galaxies which lack a $n_e$ constraint from \SII, or has one which has a large uncertainty, the effect of increasing the adopted $n_e$ is shown by the purple arrows.

Comparing to the sample of \HII\ regions at $z\sim0$ (\citealt{Pilyugin2012}, \citealt{Berg2020}; green circles and diamonds respectively) 
and the $z\sim0$ abundance scalings presented in \citet{Nicholls2017}, \citet{Cataldi2025_NO}, and \citet{Scholte2026}, our $T_e$-based sample appears to deviate to higher N/O values. Our sample is limited to low metallicities $12+\log({\rm O}/{\rm H})<8.2$ and those with \NIIl6583 detections have a median value of $\log({\rm N}/{\rm O})=-1.25$ and a weighted mean of $\log({\rm N}/{\rm O})=-1.26\pm0.03$.
This is notably higher than the low-metallicity N/O plateau of $\log({\rm N}/{\rm O})= -1.391 \pm0.003$ derived by \citet{Scholte2026} from a sample of almost 50,000  $T_e$-based abundances in $z\sim0$ galaxies observed with the Dark Energy Spectroscopic Instrument (DESI; \citealt{DESI2023}),
and also than the weighted average of $\log({\rm N}/{\rm O})=-1.41\pm0.09$ found for $z\sim0$ dwarf galaxies in \citet{Berg2019_CNO_Dwarf} (purple dashed line), and for metal-poor galaxies from \citet{Izotov1999} ($\log({\rm N}/{\rm O})= -1.46 \pm 0.14$). 

Green squares show galaxies from the CLASSY survey, which have high SFRs for $z\sim0$ and span a large range of O/H \citep{Berg2022_CLASSY, ArellanoCordova2025_CLASSY}.
These have a median \logNO\ of $-1.36$ which, while somewhat higher than the other $z\sim0$ samples, is still at least 0.10 dex lower than ours. We note that abundances in that work were derived assuming the $T_e$ calibration of \citet{Garnett1992} which lies between the two considered in this work, and results in $T_e$\OII\ values that are $\sim10-20$~\% higher than those from the \citetalias{Cataldi2025} relation, indicating that the true offset between our \logNO\ values and those of \citep{ArellanoCordova2025_CLASSY} would be even higher than the 0.10 dex quoted.

Our result is consistent with a recent analysis of $T_e$-based abundances in a $z\approx 2-3.5$ from \citet{Schaerer2026} which found a mean of $\log({\rm N}/{\rm O})=-1.29^{+0.25}_{-0.21}$, although that sample extends to higher metallicity, ranging up to $12+\log({\rm O}/{\rm H})=8.44$.

Even adopting the `high $n_e$' assumption (purple arrows), the median value of our sample only drops marginally to $\log({\rm N}/{\rm O})=-1.28$, and if we would have adopted the \citetalias{Pilyugin2009} relation as our default, we would recover a median value of $\log({\rm N}/{\rm O})=-1.10$.
We note that a further 10 galaxies in this redshift range did not have \NII\ detected for which we can only place a relatively weak upper limit on the N/O. However, given these upper limits are all well above the median value, there is no suggestion that deeper data to obtain the \NII\ line would push the sample median lower.

\subsubsection{Strong-line abundances}

In Figure~\ref{fig:NO_OH_diagram_zbin}, we expand our focus to show the strong-line abundance measurements, spanning our full sample. We have divided this into four panels showing different redshift slices from $z=1.5 - 7$ for visual clarity. While strong-line oxygen abundance measurements are known to be less reliable than $T_e$-based measurements, the overall trends across a sample this large are likely robust \citep[e.g.][]{Sanders2020_Te}. 

Around 50~\% of our sample has $7.9\lesssim12+\log({\rm O}/{\rm H})\lesssim8.3$, and we see  at least an order of magnitude of spread in the nitrogen abundance distribution within this range, from $-1.72\leq\log({\rm N}/{\rm O})\leq-0.72$, with some upper limits even suggesting values that might be lower still. The median value in this metallicity range is $\log({\rm N}/{\rm O})=-1.2$, comparable to what we found in our auroral line sample extending to lower metallicity.
At higher metallicities, we begin to observe a high fraction of the galaxies with much higher N/O, in line with the N/O -- O/H trend that is observed locally \citep[e.g.,][]{Pilyugin2012, HaydenPawson2022}.

Computing the Kendall's $\tau$ coefficient between \logNO\ and \lOH\ across the full sample (all redshifts and all metallicities), we find a weak correlation ($\tau=0.10$) at high significance ($p=0.002$). Of course, the N/O -- O/H relation is well-established to have little correlation at low metallicity, with a much stronger correlation at higher metallicity, where secondary nitrogen enrichment becomes much more significant. Splitting the sample at the median metallicity, the high metallicity ($12+\log({\rm O}/{\rm H}) > 8.20$) returns a much stronger correlation ($\tau=0.24$, $p<10^{-4}$).
Considering only the low metallicity sample ($12+\log({\rm O}/{\rm H}) \leq 8.0$) in isolation, we find no evidence for any correlation, and instead obtain a \emph{negative} $\tau$ value at low significance ($\tau=-0.07$, $p=0.14$).

Although there is a tentative visual shift towards higher N/O values in the higher redshift panels of Figure~\ref{fig:NO_OH_diagram_zbin}, a $\tau$ test shows that this is not significant, yielding $\tau=0.05$ with $p=0.11$ for the correlation between \logNO\ and $z$ across the full sample, with high- and low-metallicity sub-samples returning very similar values. This suggests that while there does seem to be a general trend of galaxies having higher N/O at fixed O/H toward higher redshift, this is more likely driven by another parameter whose redshift evolution may drive this trend.

\subsection{Nitrogen-enhanced galaxies at low metallicity}
\label{sub:N_enhanced}

To explore the enhancement of N/O at high redshift, we begin by considering the most nitrogen-enhanced galaxies in our sample. 
Throughout this section we consider a galaxy `nitrogen-enhanced' if it has $\log({\rm N}/{\rm O})>-1.1$, corresponding to a factor of two above the $\log({\rm N}/{\rm O})_{\rm plateau}\approx-1.4$ typically found at $z\sim0$ \citep{Berg2019_CNO_Dwarf, Scholte2026}. 
For this section, we limit our focus to galaxies with metallicity below $12+\log({\rm O}/{\rm H})<8.0$. 

In addition to the \NII-based abundances introduced in the previous section, Figures~\ref{fig:NO_OH_diagram}~\&~\ref{fig:NO_OH_diagram_zbin} also show a compilation of abundances derived for high-ionisation \NIII\ and \NIV\ emitters at $z\gtrsim2$. We primarily use values from \citet{Martinez2025}, where available, as these were derived using a custom treatment of the high $n_e$ regions identified in many of these systems \citep[e.g.][]{Senchyna2023, Topping2025, Berg2025_WN}, but the overall finding that these galaxies have highly enhanced N/O abundances has now been demonstrated by numerous studies \citep{Cameron2023_GNz11, Isobe2023, MarquesChaves2023, Castellano2024, Topping2024, Schaerer2024, Naidu2025, Morel2025_Nemitters}.

Immediately we can note that, despite the enhanced average N/O of our sample relative to $z\sim0$ samples, even the highest measured values are still considerably lower than what is measured for the \NIII\ and \NIV\ emitters, which typically have $\log({\rm N}/{\rm O})>-0.6$ (labelled as ``extreme N-enhancement'' in Figure~\ref{fig:NO_OH_diagram_zbin}). This holds for both the $T_e$-based and the strong-line sample (Figures~\ref{fig:NO_OH_diagram}~\&~\ref{fig:NO_OH_diagram_zbin}).
While we identify five nitrogen-enhanced galaxies in our $T_e$-based samples (Table~\ref{tab:highNO_Te}), all of these have abundances of $\log({\rm N}/{\rm O})<-0.8$, well below this ``extreme N-enhancement'' threshold. 

We note that the two most nitrogen-rich galaxies (gnmh\_607 and gsmj\_206035), all also lack $n_e$ constraints from \SII. If these galaxies have densities comparable with the remainder of our sample ($n_e\lesssim10^3$~cm$^{-3}$), they would be confirmed as nitrogen-enhanced systems, but adopting a density of $n_e=10^4$~cm$^{-3}$ would move them below our $\log({\rm N}/{\rm O})>-1.1$ cut. 
That said, gsmj\_206035 was identified by \citet{Morel2025_Nemitters} as a nitrogen-enhanced galaxy based on UV nitrogen lines detected in low-resolution prism spectroscopy, which are far less $n_e$-sensitive when deriving abundances \citep{Cameron2023_GNz11, Martinez2025}.
Nonetheless, this highlights that while $n_e$ uncertainty does not significantly affect our derived value for the low-metallicity plateau, it's clear that our inference on the fraction of galaxies with enhanced N/O abundance ratios can be impacted by the electron density.

We additionally identify 14 candidate nitrogen-enhanced galaxies from our strong-line sample. 
As seen in Figure~\ref{fig:NO_OH_diagram_zbin} and Table~\ref{tab:highNO_SL}, we again find that none of this sample reaches the `extreme nitrogen-enhancement' ($\log({\rm N}/{\rm O})>-0.6$) found in \NIII\ and \NIV\ emitters, with the highest abundance being gsmj\_203994 with \logNO $=-0.72$.
The spectra of all nitrogen-enhanced candidates from the strong-line sample are shown in Figures~\ref{app_fig:highNO_strongline1}~--~\ref{app_fig:highNO_strongline5}.

\begin{figure*}
    \centering
    \includegraphics[width=\textwidth]{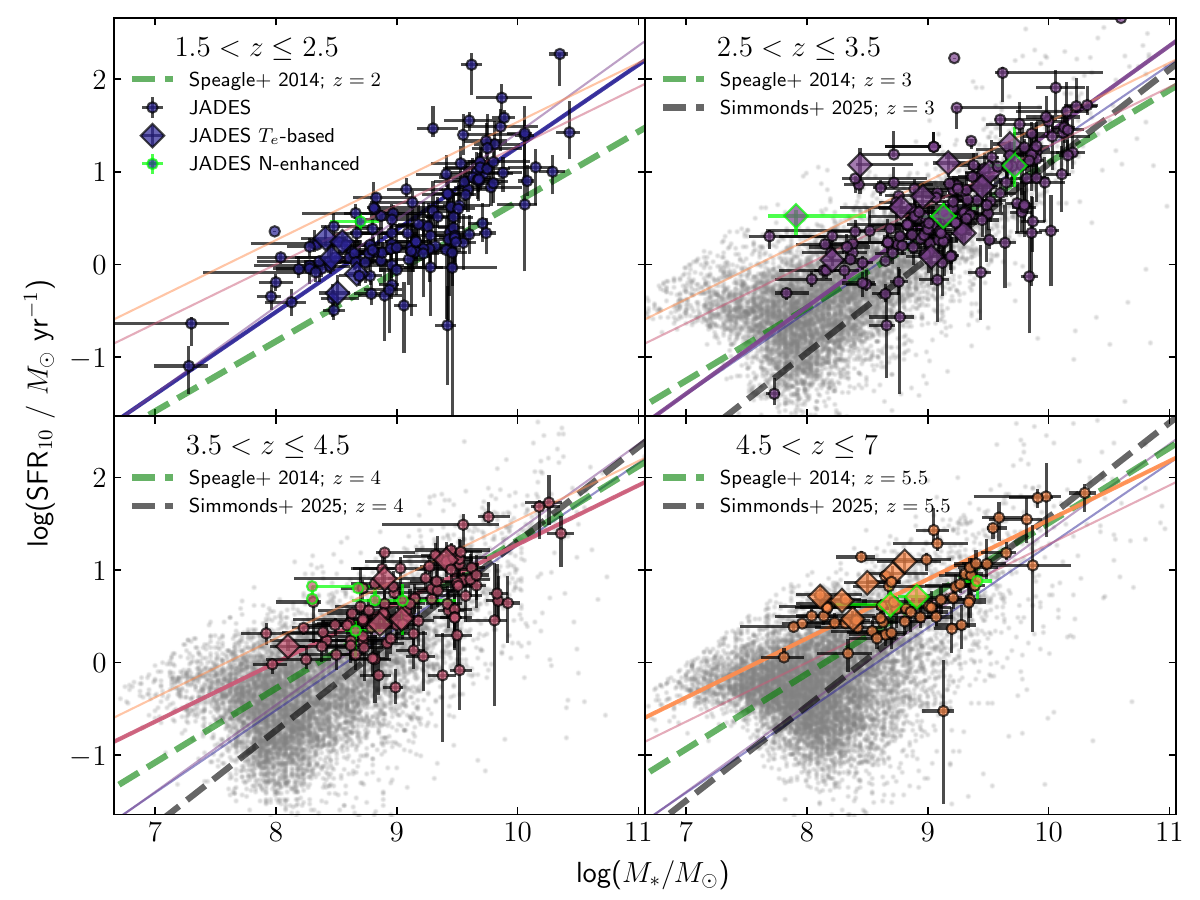}
    \caption{
    Stellar mass vs. star formation rate derived from SED fitting for our JADES sample in bins of redshift. Points with green outlines are nitrogen-enhanced galaxies (see Section~\ref{sub:N_enhanced}). Solid lines are carried across all panels and show the best fit relation to each of the four sub-samples. The green and black dashed lines show the evolving star-forming main sequence fits from \citet{Speagle2014} and \citet{Simmonds2025_SFMS} respectively. The sample from \citet{Simmonds2025_SFMS} are shown as grey points with a $m_{\rm F444W} < 29$ cut applied.}
    \label{fig:SFMS}
\end{figure*}

Figure~\ref{fig:SFMS} shows the $M_*$ and SFR measurements for these nitrogen-enhanced galaxies relative to our full sample and the NIRCam-based sample from \citet{Simmonds2025_SFMS} (see Section~\ref{sub:sed_fitting} for details)\footnote{We adopt star-formation rate values from SED fitting rather than \Ha\ measurements to reflect the `global' SFR since, in many cases, the NIRSpec/MSA shutter does not capture the full extent of the galaxy.}.
In this figure, we can see the effect of our H$\beta$ signal-to-noise requirement -- essentially mimicking a SFR cut -- with very few galaxies below SFR$_{10}\approx1~M_\odot$~yr$^{-1}$ entering our sample. Comparing to the NIRCam-based sample from \citet{Simmonds2025_SFMS} (grey points), we see that above log$(M_*/M_\odot)>9$, the star-forming main sequence (SFMS) is well-sampled, while at lower masses we are biased toward galaxies above the main sequence.
Here we can see our nitrogen-enhanced galaxies tend to lie on or above the SFMS, however there is no obvious trend that emerges, and most galaxies that lie above the SFMS are not nitrogen-enhanced.

The total of 19 nitrogen-enhanced galaxies is drawn from a parent sample of 151 galaxies across our combined sample with $12+\log({\rm O}/{\rm H})<8.0$, equating to a nitrogen-enhanced fraction of 13~\% at this metallicity.
If nitrogen-enhancements tend to occur in highly star-forming systems, our sample should be more complete in selecting for these systems than `nitrogen-normal' systems, in which case this fraction of 13~\% would represent an upper limit.

A recent study by \citet{Cataldi2025_NO} examined \NII-based N/O abundances across a sample of $z>1$ galaxies compiled from the JWST/NIRSpec MARTA program (PID 1879; PI Curti), additional JWST observations obtained from the DAWN JWST Archive (DJA; \citealt{Valentino2025}), and ground-based data sets including MOSDEF \citep{Kriek2015} and KLEVER \citep{Curti2020_KLEVER}. Among their sample, \citet{Cataldi2025_NO} also found that a significant fraction had \logNO $>-1.1$ at low oxygen abundance, but likewise identified very few galaxies with \NII-based abundances exceeding \logNO $>-0.6$.

Comparing to low-redshift samples, we again see evidence for a shift toward more nitrogen-rich systems at higher redshift.
\citet{Bhattacharya2025} searched the DESI data release 1 to find the most extreme N/O galaxies with \lOH $<8.2$ (orange crosses in Figure~\ref{fig:NO_OH_diagram}). \citet{Scholte2026} identified 139 galaxies with $\log({\rm N}/{\rm O})>-0.75$ and \lOH $<8.0$, but these were drawn from a sample of almost 50,000 galaxies with $T_e$-based abundances, and the vast majority of low-metallicity galaxies have N/O below our `N-enhanced' threshold.
While our sample is admittedly small in comparison (40 $T_e$-based abundances, drawn from fewer than six thousand total JADES spectra), if a nitrogen-enhanced fraction of $\sim$13~\% is confirmed in larger samples, this would indicate that the fraction of nitrogen enhancement is higher at $z>1.5$ compared to $z\sim0$.

We note that $z\sim0$ galaxies with \lOH $\lesssim7.3$ -- so called extremely metal poor galaxies (XMPs) -- are seen to exhibit a much larger scatter in N/O, not unlike what is observed in our high-redshift sample \citep{Breneman2025}. 
However, XMPs are at lower metallicity than the bulk of our sample. Only 6 of the JADES galaxies in this work have \lOH $<7.4$. Furthermore, XMPs at $z\sim0$ typically have stellar masses $\log(M_*/M_\odot)\lesssim 6-7$ \citep{Izotov2019, Breneman2025}, whereas 94~\% of our JADES sample has $\log(M_*/M_\odot)\geq 8$. Thus, the link between nitrogen enhancement in XMPs at $z\sim0$ and star-forming galaxies observed with \emph{JWST} at $z>2$ is unclear, though intriguing.

\begin{table*}
    \centering
    \caption{Nitrogen-enhanced galaxies from the JADES strong-line sample, based on our criteria of \logNO $>-1.1$ and \lOH $<8.0$.}
    \renewcommand{\arraystretch}{1.3} 
    \begin{tabular}{cccccc}
    \hline
    Tier & ID & $z_{\rm spec}$ & \lOH & log(\NII/\OII) & \logNO \\
    \hline
goods-n-mediumhst & 920 & 4.885 & $7.75_{-0.23}^{+0.28}$ & $0.15_{-0.05}^{+0.05}$ & $-0.96_{-0.16}^{+0.12}$ \\
goods-s-mediumhst & 58656 & 4.780 & $7.96_{-0.13}^{+0.11}$ & $0.15_{-0.03}^{+0.03}$ & $-0.97_{-0.1}^{+0.08}$ \\
goods-n-mediumhst & 946 & 4.695 & $7.64_{-0.14}^{+0.16}$ & $0.14_{-0.05}^{+0.05}$ & $-0.99_{-0.17}^{+0.12}$ \\
goods-s-mediumjwst & 209839 & 4.468 & $7.85_{-0.17}^{+0.2}$ & $0.22_{-0.08}^{+0.08}$ & $-0.81_{-0.17}^{+0.12}$ \\
goods-s-mediumjwst & 51102 & 4.230 & $7.82_{-0.14}^{+0.15}$ & $0.17_{-0.06}^{+0.06}$ & $-0.91_{-0.18}^{+0.12}$ \\
goods-s-mediumjwst & 203994 & 4.044 & $7.93_{-0.23}^{+0.22}$ & $0.27_{-0.1}^{+0.1}$ & $-0.72_{-0.18}^{+0.12}$ \\
goods-s-mediumjwst & 179198 & 3.831 & $7.91_{-0.13}^{+0.12}$ & $0.17_{-0.02}^{+0.02}$ & $-0.92_{-0.06}^{+0.05}$ \\
goods-s-mediumjwst & 10009453 & 3.705 & $7.73_{-0.14}^{+0.14}$ & $0.22_{-0.05}^{+0.05}$ & $-0.8_{-0.09}^{+0.08}$ \\
goods-s-mediumjwst & 50234 & 3.700 & $7.63_{-0.13}^{+0.14}$ & $0.11_{-0.04}^{+0.04}$ & $-1.09_{-0.16}^{+0.11}$ \\
goods-s-mediumhst & 17349 & 3.491 & $7.62_{-0.12}^{+0.14}$ & $0.19_{-0.06}^{+0.06}$ & $-0.87_{-0.16}^{+0.12}$ \\
goods-s-mediumjwst & 51204 & 3.472 & $7.94_{-0.13}^{+0.12}$ & $0.11_{-0.04}^{+0.04}$ & $-1.09_{-0.19}^{+0.13}$ \\
goods-s-mediumjwst & 63722 & 3.060 & $7.96_{-0.13}^{+0.12}$ & $0.18_{-0.03}^{+0.03}$ & $-0.89_{-0.07}^{+0.06}$ \\
goods-n-mediumhst & 23873 & 2.941 & $7.68_{-0.15}^{+0.18}$ & $0.17_{-0.05}^{+0.05}$ & $-0.93_{-0.13}^{+0.1}$ \\
goods-s-mediumjwst & 282381 & 2.107 & $7.74_{-0.13}^{+0.14}$ & $0.16_{-0.05}^{+0.05}$ & $-0.95_{-0.16}^{+0.11}$ \\
\hline
    \end{tabular}
    \label{tab:highNO_SL}
\end{table*}

\renewcommand{\arraystretch}{1} 

\begin{table}
    \centering
    \caption{Nitrogen-enhanced galaxies from the JADES $T_e$-based sample, based on our criteria of \logNO $>-1.1$ and \lOH $<8.0$. Abundance measurements can be found in Table~\ref{tab:auroral_sample}.}
    \begin{tabular}{ccc}
    \hline
    Tier & ID & $z_{\rm spec}$ \\
    \hline
    goods-n-mediumhst &	607 & 5.1822 \\
    goods-s-mediumjwst & 206035 & 4.7735 \\
    goods-n-mediumjwst & 42939 & 3.4671 \\
    goods-n-mediumhst &	31940 & 3.1295 \\
    goods-n-mediumhst &	24755 & 2.9803 \\
    \end{tabular}
    \label{tab:highNO_Te}
\end{table}

\subsection{Relationship with galaxy properties}
\label{sub:correlation_galaxy_properties}

\begin{figure*}
    \centering
    \includegraphics[width=0.93\linewidth]{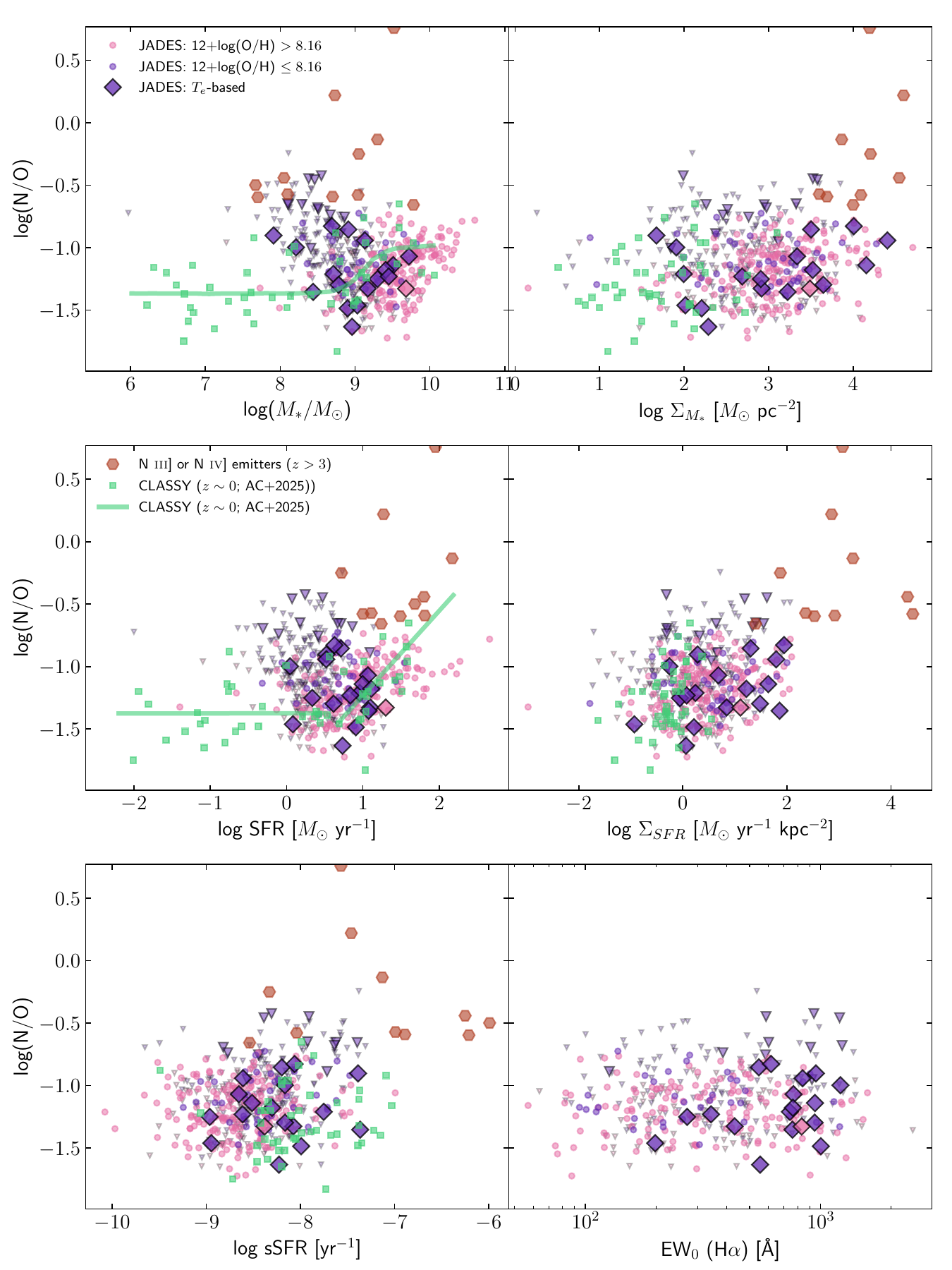}
    \caption{N/O abundances as a function of galaxy properties. From left-to-right, top-to-bottom, these properties are: stellar mass, stellar mass surface density, star formation rate, star formation rate surface density, specific star formation rate, and equivalent width of \Ha.
    In each panel, we show our $T_e$-based measurements (diamonds) alongside strong-line measurements (circles), each divided into \lOH $\leq8.16$ (purple) and \lOH $>8.16$ (pink) sub-samples. Upper limits are denoted by triangles. High-redshift \NIII- and \NIV-emitters are shown as brown hexagons \citep{Martinez2025}, while green squares show $z\sim0$ galaxies from CLASSY \citep{ArellanoCordova2025_CLASSY} for reference.}
    \label{fig:mass_and_SD}
\end{figure*}

\begin{figure*}
    \centering
    \includegraphics[width=\linewidth]{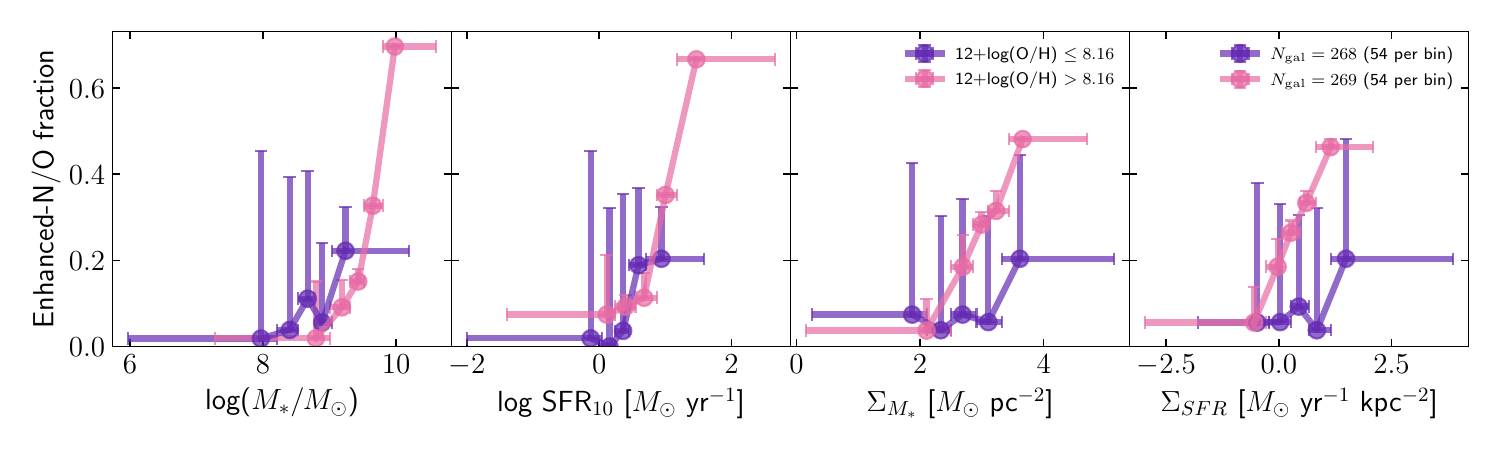}
    \caption{Fraction of galaxies with $\log({\rm N}/{\rm O}) > -1.1$ for each of the \lOH $\leq8.16$ (purple) and \lOH $>8.16$ (pink) sub-samples of our JADES sample in equally-populated bins of (a) stellar mass, (b) star formation rate, (c) stellar mass surface density, and (d) star formation rate surface density.}
    \label{fig:fracNO}
\end{figure*}

We now turn our attention to exploring whether there exists any correlation between N/O and galaxy properties.
Figure~\ref{fig:mass_and_SD} shows N/O abundance ratios of our JADES sample as a function of stellar mass ($M_*$), star formation rate (SFR), the surface density of each of these properties ($\Sigma_{M_*}$ \& $\Sigma_{\rm SFR}$), specific star formation rate (sSFR), and the \Ha\ equivalent width.
We divide our sample into \lOH $\leq8.16$ (purple) and \lOH $>8.16$ (pink), with that cut being set as the median metallicity, to explore how trends may differ in each of these metallicity regimes.
In addition to our JADES $T_e$-based and strong-line sample (diamonds and circles, respectively), we also show a selection of \NIII- and \NIV-emitters compiled from the literature (hexagons) and $z\sim0$ CLASSY galaxies from \citet{ArellanoCordova2025_CLASSY} (squares). 

To explore which properties correlate more strongly with N/O, we calculate Kendall's $\tau$ coefficients, allowing for upper limits as described in \citet{Isobe1986_Kendall} and implemented by \citet{Flury2022_Kendall} and \citet{Herenz2025_Kendall}\footnote{\url{https://github.com/Knusper/kendall}}. 

Considering just the $T_e$-based $z>1$ measurements,
we find the strongest correlation with \SDsfr\ ($\tau_{\Sigma_{\rm SFR}}=0.45$; $p=10^{-4}$), while SFR and \SDmass\ also return clear correlations ($\tau_{\rm SFR}=0.36$; $p=0.002$ and $\tau_{\Sigma_{M_*}}=0.35$, $p=0.004$).
A weaker correlation is observed for sSFR ($\tau_{\rm sSFR}=0.26$; $p=0.03$),
while we find no correlation with $M_*$ ($\tau_{\rm M_*}=-0.04$; $p=0.71$).

This $T_e$-based sample is heavily biased toward galaxies with low metallicity (\lOH $\lesssim8.0$). A great strength of our strong-line analysis is our ability to extend to much higher metallicities. Revisiting these correlations, instead for our full sample of JADES strong-line measurements, we find the correlations with SFR, \SDsfr, and \SDmass\ are much milder; $\tau_{\rm SFR}=0.20$ ($p<10^{-6}$), $\tau_{\Sigma_{\rm SFR}}=0.11$ ($p<10^{-3}$), and $\tau_{\Sigma_{M_*}}=0.14$, $p<10^{-4}$, respectively. On the other hand, the correlation with mass ($\tau_{M_*}=0.20$, $p<10^{-6}$) becomes much more significant. Meanwhile, we do not find any correlation with either sSFR or $EW_0$(\Ha) in our strong-line sample ($\tau_{\rm sSFR}=0.01$, $p=0.71$; $\tau_{EW(H\alpha)}=0.00$, $p=0.86$).

One take-home message arising here is that, considering the full sample, N/O abundances correlate with each of $M_*$, SFR, \SDmass, and \SDsfr. However, the fact that the correlation with $M_*$ emerged much more significantly for the full JADES strong-line sample, which extends the sample to much higher metallicities, suggests the correlation with mass is driven by these most metal rich systems. Indeed, considering only the high-metallicity sample (\lOH $>8.16$), we find the strongest correlation with stellar mass $\tau_{M_*}=0.34$, compared to 0.33, 0.21, and 0.20 for SFR, \SDsfr, and \SDmass, respectively, albeit highly significant ($p<10^{-5}$) in all cases.
This suggests that the nitrogen enrichment in the evolved, chemically-mature portion of our sample is dominated by older stars. This is consistent with the $z\sim0$ picture \citep[e.g.][]{Romano2022}, suggesting secondary enrichment from AGB stars has already taken over as the dominant source of nitrogen for galaxies with $Z\gtrsim 0.3~Z_\odot$ at $z\sim2-7$.
However, the diversity of N/O abundance ratios found among low-metallicity systems correlates more strongly with SFR and \SDsfr, suggesting that enhanced nitrogen abundances at these low metallicities are driven by properties of the young stellar populations during strong starburst events.

In Figure~\ref{fig:fracNO}, we recast this slightly by showing how the fraction of nitrogen-enhanced galaxies in our sample changes in bins of these same galaxy properties. Employing the same \lOH $=8.16$ cut, we divide our high- and low-metallicity samples into five equally-population bins of (a) stellar mass, (b) SFR, (c) \SDmass, and (d) \SDsfr. The low-metallicity sample has total of 268 galaxies and the high-metallicity sample has 269, resulting in $53-54$ galaxies per bin. We calculate the `enhanced-N/O fraction' simply as the number of galaxies in that bin that have a \NII\ detection that puts \logNO $>-1.1$ divided by the total number of galaxies in that bin. 

The \NII\ line is well-detected across much of our sample, especially in our high-metallicity sample and our high stellar mass bins, although we note that for approximately 44~\% of our sample, we only have an upper limit on \NII\ (and, by extension, \logNO).
The majority of these upper limits are sufficiently constraining to be able to confirm \logNO $<-1.1$. However, for 15~\% of the total strong-line sample, we cannot conclusively identify the galaxy as `nitrogen-enhanced' or not, with these being particularly prominent in the low mass and low SFR bins of the low-metallicity sample.
The vertical errorbars in Figure~\ref{fig:fracNO} quantify the contribution of these galaxies which lack a strong constraint on \logNO. The upper extent of the errorbar shows the fraction that would be obtained if half of the `indeterminate' cases proved to fall in the enhanced-N/O category. 

We again find clear evidence for a strong correlation between N/O and stellar mass in the high-metallicity sample.  Almost no nitrogen-enhanced galaxies are found below $\log(M_*/M_\odot)<9.5$, but the fraction rapidly increases above this threshold, with over 60~\% of the galaxies in the highest mass bin exceeding our N/O cut. 

The interpretation of the low-metallicity sample is hampered by the much larger fraction for which we cannot determine whether the galaxy is nitrogen-enhanced. The N-enhanced galaxies that we do identify tend have higher masses and star-formation rates, although it is only in the highest bins that the fraction ever rises above 20~\%. This suggests that while galaxies with $\log({\rm N}/{\rm O})>-1.1$ arise naturally as galaxies grow in mass and metallicity, their existence below \lOH $\lesssim8.16$ tends to be reserved for only systems going through particularly strong starbursts. 
Deeper data pushing the detection limit to lower values of \logNO, would shed more light onto this picture.

\subsection{Relationship with electron density}

\begin{figure}
    \centering
    \includegraphics[width=\columnwidth]{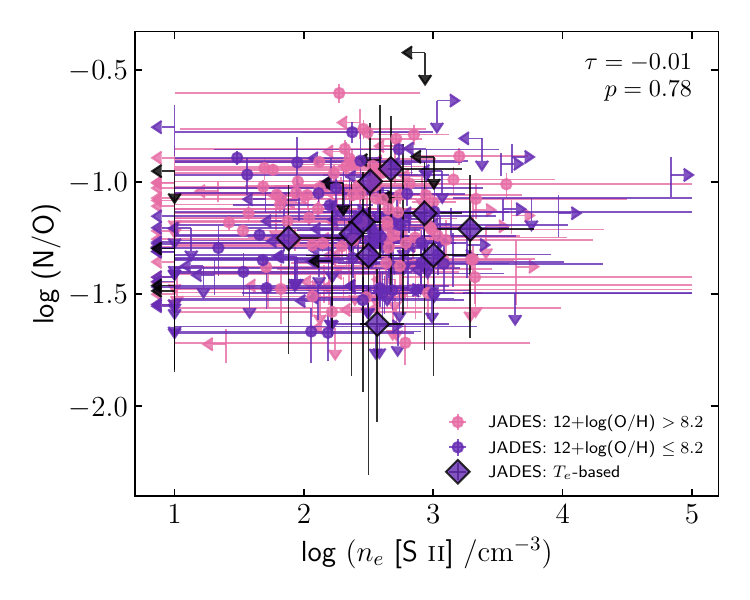}
    \includegraphics[width=\columnwidth]{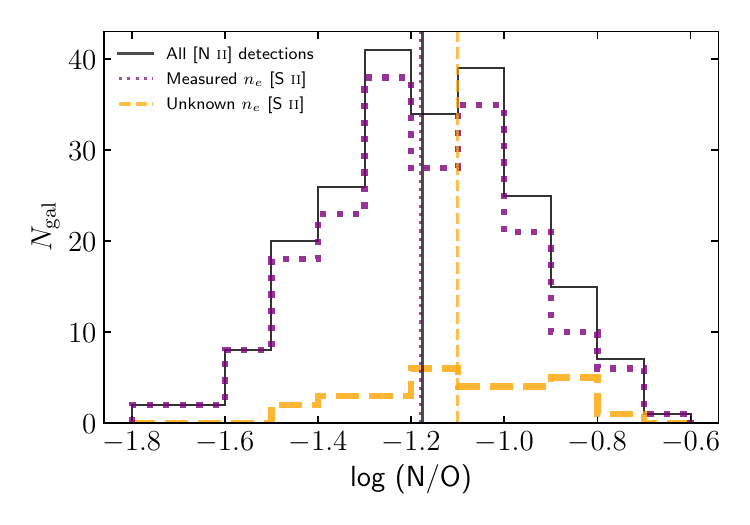}
    \includegraphics[width=\columnwidth]{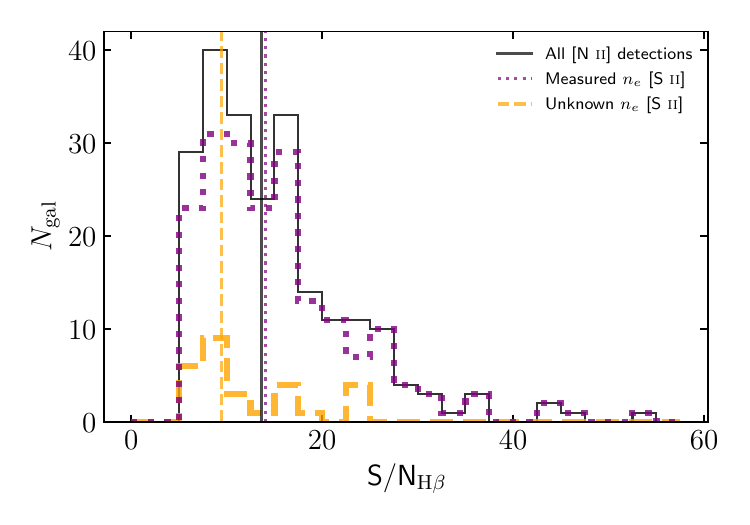}
    \caption{
    \textit{Top:} Relation between \logNO\ and electron density ($n_e$) for all galaxies in our strong-line and $T_e$-based samples for which we can derive $n_e$ constraints from the \SII\ doublet. 
    \textit{Middle:} Histogram of \logNO\ values for all galaxies in the strong-line sample with \NII\ detections (black; i.e. not including upper limits on \logNO). The purple dotted histogram shows the subset of these with an $n_e$ constraint from \SII\ (either measured value or upper- or lower-limit). The orange dashed histogram shows the subset \emph{without} any $n_e$ constraint, due to a non-detection of \SII. Vertical lines show the median values of each of these samples.
    \textit{Bottom:} Histogram of signal-to-noise ratio on the H$\beta$ emission line for the same sub-samples as shown in the middle panel.
    }
    \label{fig:NO_vs_ne}
\end{figure}

\citet{ArellanoCordova2025_CLASSY} found a correlation between electron density from the \SII\ doublet ($n_e$ \SII) and \logNO\ for $z\sim0$ galaxies from the CLASSY survey.
It is particularly pertinent for us to test for a correlation in our strong-line sample since electron density is an important factor in the conversion from \NII/\OII\ to N/O (see Figure~\ref{fig:NO_calib} and discussion in Section~\ref{sub:Te_based_NO}).

Figure~\ref{fig:NO_vs_ne} shows all our JADES galaxies for which we have a constraint on both \logNO\ and $n_e$.
In our $T_e$-based abundance measurements, we derive \logNO\ using the density constraint that we have, removing the systematic uncertainty we might have due to density. For this sample (diamonds), we find no correlation between \logNO\ and $n_e$ ($\tau=0.03$, $p=0.87$).

For our strong-line sample, 268 galaxies have a detection of at least one \SII\ line, allowing us to place a constraint on the $n_e$. These galaxies (pink and purple circles) also show no correlation between \logNO\ and $n_e$ ($\tau=-0.01$, $p=0.78$).
As shown in Figure~\ref{fig:NO_calib}, an increase in density from $n_e=100$~cm$^{-3}$ to $10^4$~cm$^{-3}$ increases the \NII/\OII\ ratio by approximately 0.3 dex at fixed N/O abundance. Inverting this, the same density change can cause \logNO\ to be overestimated by 0.3 dex if not taken into account. This uncertainty is much smaller than the dynamic range of \logNO\ values spanned by our sample. Moreover, the relation fit by \citet{ArellanoCordova2025_CLASSY} implies that \logNO\ changes by $\sim$0.9 dex over a 2 dex range in $n_e$ in their sample. This is larger than the overall spread in \logNO\ in our sample, suggesting that whatever drives this effect within the CLASSY sample is not strongly affecting our JADES sample at high redshift. 

In 171 galaxies in the strong-line sample, we do not have any $n_e$ \SII\ constraint. This is typically due to a non-detection of \SII, but in some cases the \SII\ doublet is not within the spectral coverage. For the majority of these (143/171), this is simply a case of a low signal-to-noise spectrum as these also only have an upper limit on \NII\ (and by extension \logNO).
In the middle panel of Figure~\ref{fig:NO_vs_ne}, we plot the distribution of \logNO\ values obtained for the remaining 28 of these cases, compared to the full sample. We see that this subsample has a slightly higher median value: $-1.10$, rather than $-1.18$ for the full strong-line sample.

One interpretation could be that deeper data would reveal these galaxies to have higher densities. If this were the case, the elevated \NII/\OII\ ratios we observe, which here are being interpreted as elevated N/O abundance ratios, could in fact be due to an elevated $n_e$. This interpretation would imply that this fraction of our sample might be biased high by $\sim0.1$~dex.
Alternatively, this could simply represent the fact that the majority of cases where we have no \SII\ detection are lower signal-to-noise spectra and the fact that we've limited this histogram to detections of \NII\ implicitly biases them to only including galaxies with high N/O, where \NII\ is therefore sufficiently strong so as to be detected.

In the bottom panel of Figure~\ref{fig:NO_vs_ne} we then show the distribution of signal-to-noise on the H$\beta$ line for each of these sub-samples. The sample with \SII\ non-detections is clearly biased to lower S/N (median=$9.5$, rather than 13.7 for the full sample). Furthermore, none of these galaxies have S/N$_{H\beta}>25$. Combined with the fact that actually most of the \SII\ non-detections are also \NII\ non-detections (143/171), we conclude that the effect we're seeing here is primarily a signal-to-noise effect.

That said, we note that the \SII\ doublet is only sensitive at densities below $n_e\lesssim10^4$~cm$^{-3}$ and will underestimate the average density in an inhomogeous system, especially if a significant quantity of gas with $n_e\gtrsim10^4$~cm$^{-3}$ is present \citep{Martinez2025}.
We cannot rule out that this trend may look different if we revisited the relationship between N/O and density using deeper data with access to diagnostics such as C~{\sc iii}]~$\lambda1907/\lambda1909$ or [Ar~{\sc iv}]~$\lambda4714/\lambda4742$ which are sensitive to higher densities.

In summary, we find no evidence that our strong-line N/O measurements are strongly biased by not accounting for $n_e$, due to the lack of any correlation among galaxies that do have measurements of $n_e$, and from the distribution of \logNO\ values derived for galaxies that do not have $n_e$ constraints.
Furthermore, the correlation reported by \citet{ArellanoCordova2025_CLASSY} is sufficiently steep that its presence would outweigh the $n_e$-induced systematic uncertainty. The fact that we find no such correlation in our sample suggests that the nitrogen enrichment mechanisms in our $z>1.5$ sample differ from those in the CLASSY sample.

\section{Discussion} \label{sec:discussion}

\subsection{Why are [N~{\sc ii}]-enhancements milder than N~{\sc iii}]- and N~{\sc iv}]-enhancements?}

Perhaps the most significant finding in this work is that while 13~\% of our \lOH $<8.0$ sample has \logNO $>-1.1$ (`moderately nitrogen-enhanced' hereafter), we do not identify any galaxies across our sample with \NII-based abundances exceeding \logNO\ $>-0.6$, typical of the extreme \NIII\ and \NIV\ emitters identified with \emph{JWST} (`extremely nitrogen-enhanced' hereafter).
There is consensus now that the strong rest-frame UV nitrogen emission found at high-redshift generally does represent highly enhanced N/O \citep[e.g.][]{Cameron2023_GNz11, Senchyna2023, Martinez2025, Zhu2025}.
\citet{Hayes2025} showed that, for their emission line ratios derived from stacked spectra, assuming a high density ($10^6$~cm$^{-3}$) resulted in a higher derived O/H, which may move otherwise discrepant N/O values much closer to the high metallicity end of the $z\sim0$ N/O--O/H relation. However, the multi-phase modeling presented in \citet{Martinez2025} found that the majority of known strong \NIII- and \NIV-emitters remained in the $12+\log({\rm O}/{\rm H})\lesssim 8.0$ regime.

As discussed in Section~\ref{sec:abundances}, assuming hotter low-ionisation temperatures ($T_e$\OII) would increase N/O abundances in our low-metallicity \NII-based sample, but not enough to reconcile them with the \NIII- and \NIV-emitting sample (see Figure~\ref{fig:NO_OH_diagram}). Meanwhile, adopting higher densities, perhaps under the assumption that $n_e$(\SII) is underestimating the average density, would instead decrease the derived N/O abundances.
This rules out that the difference in abundances between \NII-based and UV-based N/O determinations is caused by systematic uncertainties impacting the accuracy of the abundance measurements.

Therefore, two possible explanations for the lack of galaxies with \logNO $> -0.6$ in our \NII-based sample could be: (1) galaxies are chemically stratified, and the most extreme N/O enhancements are only observed in central, highly-ionised zones, or (2) this reflects a sample selection effect, and extreme values of \logNO\ $>-0.6$ are only observed in the most extreme starburst systems which are not adequately represented in our JADES \NII\ sample.

\subsubsection{Chemically stratified galaxies?}

Although many of the \NIII\ and \NIV\ emitters lack spectra with a detection of \NII\footnote{In many cases, this is because they are $z>7$ and \NIIl6583 cannot be observed with \emph{JWST}/NIRSpec, or they lack $R\gtrsim1000$ spectroscopy, and \NIIl6583 cannot be deblended from \Ha. In some cases, it is simply that this line is undetected, due to insufficient depth.}, several examples now exist that bridge this divide.
\citet{Berg2025_WN} measured \NII-based abundance for the lensed \NIV-emitter RXCJ2248 at $z=6.1$, finding $\log({\rm N}/{\rm O})_{{\rm N}^+}=-0.502$. While this is marginally lower than the N/O abundances inferred from the higher ionisation states, it is still highly super-solar, and more nitrogen-enhanced than anything in our JADES sample.
Similarly, the \NII-based abundance observed in the Sunburst Arc -- another lensed galaxy -- was found to be significantly enhanced (\logNO $= -0.65^{+0.16}_{-0.25}$; \citealt{Welch2025}), albeit somewhat lower than that observed from \NIII\ measurements (\logNO $\approx-0.24$; \citealt{Pascale2023}).
One nitrogen-enhanced galaxy in our sample (gsmj\_206035) overlapped with the \NIII- and \NIV-selected sample from \citet{Morel2025_Nemitters}, again suggesting that the high- and low-ionisation abundances can be in good agreement.
On the other hand, \citet{Ji2024} showed that the \NII-based abundance was more than $1.0$~dex lower than the \NIV-based abundance in GS\_3073, a galaxy observed to host a broad-line AGN \citep{Ubler2023}, suggesting a stratified structure with heterogenous chemical enrichment across the different ionisation zones.

Expanding the sample of galaxies which have high signal-to-noise measurements of all of \NII, \NIII, and \NIV\ would clearly shed more light onto what fraction of galaxies exhibit any kind of chemical stratification. Nonetheless, while the evidence here is mixed, it is clear that at least some \NIV-emitting galaxies are not chemically stratified and can have \logNO $>-0.6$ as measured from \NII, which raises the question of why our JADES sample does not identify any galaxies with abundances in this range.

\subsubsection{Impact of sample selection}

We now turn our attention to how our JADES sample selection might contribute to the lack of extremely nitrogen-enhanced galaxies in this work.
The majority of our JADES $1.5<z<5.7$ parent sample comes from a $m_{\rm F444W}$-based pre-selection, approximating a mass-based cut (see \citealt{CurtisLake2025_DR4} and Section~\ref{sec:data}). We then applied an emission line signal-to-noise cut, essentially setting a minimum SFR for our sample, albeit a fairly permissive one, with the star-forming main sequence well sampled down to at least $\log(M_*/M_\odot)\approx9.0$ (see Figure~\ref{fig:SFMS}).
At higher redshifts ($z>5.7$), where most \NIII- and \NIV-emitters are found, spectroscopic sample selections typically rely more on $m_{\rm UV}$-based selections (since the rest-frame optical is progressively less available to \emph{JWST}/NIRCam imaging), which is more akin to a SFR-based selection.

With this in mind, the low fraction of extremely nitrogen-enhanced galaxies compared to higher redshifts could reflect that much more of our sample comes from a mass-based selection, sampling the star-forming main sequence, and that the extremely nitrogen-enhanced galaxies identified at higher redshift give us a biased picture.

Another possibility is that the absence of extremely nitrogen-enhanced galaxies from our sample simply reflects that these extreme emitters are either too highly ionised or have too high $n_e$ to be picked up in our sample.
If the \NIIl6583 and \OIIll3726, 3729 lines are too weak, owing to a low ${\rm N}^+$ and ${\rm O}^+$ abundance (i.e., too highly ionised) or the collisional suppression of these lines (expected at $n_e\gtrsim10^5$~cm$^{-3}$), we might simply miss these galaxies. 
Note that this differs from the chemically stratified scenario -- in this scenario, the low-ionisation (\NII-emitting) gas is still highly N/O enhanced, but we do not detect the low-ionisation lines at the depth of this study.
In the case of RXCJ2248 \citep{Topping2024,Berg2025_WN}, it's clear that shallower spectroscopy would still identify the N/O enhancement based on \NIV, but it is only with exceptionally deep spectroscopy that \NII\ can be accurately measured.

Instead, at moderate spectroscopic depths, \NII-based samples such as this work are representative of `typical' star-forming conditions, where a significant fraction of the emission comes from low-ionisation gas.
Several other studies exploring \NII-based abundances in low metallicity galaxies at $z\gtrsim2$ have similarly identified ample moderately nitrogen-enhanced systems, but very few extremely nitrogen-enhanced objects \citep{Cataldi2025_NO, Zhang2025_WR, Stiavelli2025}. This suggests that the absence of extreme nitrogen-enhancements in our sample is not simply an effect of the way JADES galaxies are selected, but rather reflects that systems with extreme nitrogen-enhancements have very weak low-ionisation lines.

\subsection{What sources drive nitrogen enhancement at low metallicity?}

We now turn our attention to how these findings relate to the mechanisms that drive nitrogen-enhancement at high-redshift.
Identifying the dominant sources of nitrogen enrichment requires first establishing whether nitrogen-enhancements represent enrichment from previous generations (and that we are witnessing stars forming from nitrogen-enhanced gas), or whether they reflect enrichment from the current generation of stars, with young massive stars directly enriching the \HII regions.

Our finding that the most extreme nitrogen-enhancements are limited to the most extreme starbursts supports the latter scenario. If nitrogen-enhancement arose from properties of previous generations of stars or from dilution, outflows, and gas-mixing effects 
then we would expect to see such enhancements across a diverse range of star-formation conditions, implying a stronger representation in this sample.

Instead, the fact that \NII-based studies do not seem to identify extreme N/O-enhancements supports the scenario in which the enrichment is driven by the young stars that are also responsible for powering the emission.
This then raises the question of what, if any, connection there is between the extreme nitrogen-enhancements seen in UV-based studies, and the mild nitrogen-enhancements seen in the \NII-based studies, such as this work.

Many explanations have been put forward that explain the observed nitrogen enhancement in the context of prompt enrichment from massive stars ($\gtrsim50~M_\odot$).
Enrichment from Wolf-Rayet (WR) stars, expected to have progenitor masses $\gtrsim40~M_\odot$ at SMC metallicity and below \citep{Shenar2020}, is believed to be an important source of nitrogen at low metallicity \citep[][]{LimongiChieffi2018} and has been invoked in many models of nitrogen-enhanced systems \citep{Bekki2023, KobayashiFerrara2023, Zhang2025_WR}. 
Conventional WR models overproduce carbon relative to what is observed in extremely nitrogen-enhanced galaxies, prompting some studies to instead invoke winds of rotating massive stars without WR enrichment \citep[e.g.][]{Tapia2024}, VMS winds \citep{Vink2023}, or SMS winds \citep[e.g.][]{Charbonnel2023, NageleUmeda2023}. On the other hand, \citet{Berg2025_WN} argued that abundances can be reconciled by extrapolating the observed increase in WN/WC ratio from the Milky Way to the SMC down to even lower metallicity, an argument supported by recent observations of low metallicity stars that may be evolving directly from WN to WO stars, skipping the WC phase \citep{Sander2025_WN_WO}.

The relative lack of extremely \NII-enhanced systems found here could indicate that our moderately nitrogen-enhanced galaxies are an extension of these massive-star-based scenarios toward slightly less extreme star formation rates.
If milder star-formation events are not always capable of fully populating the upper regions of the stellar IMF ($\gtrsim 50 - 200$~M$_\odot$, as variously invoked aforementioned studies) it might naturally follow that these systems do not attain the extreme nitrogen-enhancements seen for more intense starbursts. 
Moreover, the existence of a significant population of moderately nitrogen-enhanced objects suggests that extremely nitrogen-enhanced galaxies are not a discrete population, but rather the upper-most extension of a broad distribution. This may favour explanations invoking enrichment from a distributed population of objects (e.g. many $\sim50-200~M_\odot$ stars) over solutions involving a single monolithic source (e.g. a $>1000~M_\odot$ SMS).

However, an alternative explanation is that these moderately nitrogen-enhanced systems might reflect the remnants of extremely nitrogen-enhanced galaxies after a short-lived phase. 
If a softening radiation field resulting from an aging population (leading to an increased \NII/\NIV\ ratio), is mirrored by a decreasing N/O abundances due to increased oxygen enrichment from CCSNe, the result could be a galaxy that appears closer to the star-forming main sequence and maintains an abundance pattern similar to the moderately nitrogen-enhanced galaxies identified in this study.

Ultimately, more detailed studies of nitrogen-enhanced galaxies are needed to accurately determine the properties of the sources responsible for this enrichment.
One promising approach is to directly search for stellar features in the spectra of these galaxies.
From deep $R\sim1000$ spectroscopy, \citet{Berg2025_WN} identified Wolf-Rayet wind features in the spectrum of RXCJ2248. Meanwhile \citet{Zhang2025_WR} also found tentative evidence for Wolf-Rayet features in a moderately nitrogen-enhanced system. This provides compelling evidence that massive stars in this brief evolutionary phase are at least part of the solution.
It is also worth noting that the wind features of  Wolf-Rayet stars may not be as prominent at low metallicity, and the absence of Wolf-Rayet features in an integrated spectrum does not necessarily confirm their absence \citep{GonzalezTora2025}.

Given these massive stars impact on very local scales, detailed spatially resolved data will also be very insightful. Gravitational lensing is, of course, very appealing in this endeavour. 
Many studies have explored enrichment in the Sunburst Arc on small scales \citep{Pascale2023, Choe2025, RiveraThorsen2025, Welch2025}.
While $z>1.5$ galaxies in blank fields cannot be studied on parsec spatial scales, it is worth noting that many of the galaxies in this study show extended and/or clumpy morphologies. 
An example of this is gnmh\_31940, which has two clumps which, given the slit alignment, are resolvable along the spatial direction (Figure~\ref{fig:spec_auroral}, bottom row). The global spectrum resulted in a N-rich abundance pattern of \logNO $=-0.90^{+0.62}_{-0.35}$, but extracting 1D spectra for each clump separately, we tentatively obtain an even higher N/O abundance in the central clump (\logNO $\approx -0.72$; see Appendix~\ref{app:spatial_extraction}). This further highlights that nitrogen enhancement likely arises from processes happening on very small scales, and detailed studies with high spatial resolution will be an important component of establishing a complete picture.
Revisiting correlations between nitrogen abundances and physical properties on a spatially-resolved scale is beyond the scope of this paper, but may reveal important new insights.


\section{Conclusions} \label{sec:conclusion}

We have presented a systematic analysis of nitrogen abundances based on measurements of the \NIIl6583 emission line in galaxies at redshifts $1.5<z<7.0$ from the JADES survey.
Arising from singly-ionised nitrogen, this line is sensitive to nitrogen abundances across galaxies with a wider range of ISM conditions than the rest-frame UV \NIII\ and \NIV\ lines that have been widely studied at high redshift.

After removing AGN and employing a cut of S/N$_{H\beta}>5$, we assembled a sample of \Nall\ galaxies, of which \NallTe ~had $>3\sigma$ detections of the \OIIIl4363 auroral line, affording temperature-based abundance measurements.
We find that our $T_e$-based N/O measurements have a significant dependence on the adopted calibration between the high-ionisation $T_e$\OIII\ and the low-ionisation $T_e$\OII, with the final N/O abundance changing by more than 0.14 dex depending on which relation is adopted.
Poorly constrained $n_e$ can give rise to over 0.3 dex of systematic uncertainty, however for over half of our sample we are able to place adequate constraints on $n_e$\SII\ to alleviate this uncertainty. Furthermore, we find no correlation between electron density and derived N/O, suggesting our measurements are not systematically biased by density variations.

Comparing our $T_e$-based abundances to strong-line calibrations from the literature, we find that our sample is in good agreement with the calibrations of \citet{PerezMontero2009} and \citet{Cataldi2025_NO}, favouring a steeper relation than is derived from higher metallicity samples in \citet{HaydenPawson2022} or \citet{Florido2022}. 

Our \NII-based nitrogen abundance measurements are sensitive to much lower N/O values than \NIII- and \NIV-based measurements, with individual measurements from our $T_e$-based sample probing as low as \logNO $\approx-1.63$, below typical N/O abundances in local metal-poor dwarf galaxies \citep[e.g.][]{Berg2019_CNO_Dwarf}, allowing us to probe the primary plateau at high-redshift.
Under conservative assumptions around $T_e$- and $n_e$- structure, we find that our low-metallicity (i.e., \lOH $<8.2$) $T_e$-based sample
has an average N/O value of \logNO $=-1.26\pm0.03$,
at least 0.1 dex higher than values found at $z\approx0$ \citep[e.g.][]{Scholte2026}.
Furthermore, we identify six galaxies for which we measure \logNO $>-1.1$, a threshold which is very rarely exceeded in the same O/H range at $z\approx0$ \citep{Bhattacharya2025}.
Applying strong-line measurements to the remainder of our sample, we find an additional 14 candidate galaxies with \lOH $<8.0$ and  \logNO $>-1.1$, suggesting that as many as 13~\% of star-forming galaxies at $1.5<z<7$ may be `nitrogen-enhanced'.

Despite the presence of many `moderately nitrogen-enhanced' (\logNO $>-1.1$) galaxies in our sample, we find a notable absence of \NII-based abundances exceeding \logNO $>-0.6$, values which are typical of the $z>3$ \NIII- and \NIV-emitters.
This suggests that these most extreme N/O ratios are only reached during the most extreme starburst phases, and not during mild star-formation activity, which our sample is much more heavily weighted towards. 
Indeed, we find N/O abundance among low-metallicity systems at $z>1.5$ to correlate with \SDsfr, SFR, and \SDmass, further highlighting that the strongest nitrogen enhancements are found in the most extreme systems.

This favours an interpretation in which the nitrogen enrichment in low-metallicity systems observed at high-redshift is dominated by young massive stars from the present generation of star formation, rather than scenarios in which a star cluster is formed from nitrogen-rich gas.
If nitrogen-enhanced galaxies represent a brief phase of a galaxy's evolution, the lower typical nitrogen enhancements from our \NII-based sample could reflect the fact that, in milder bursts of star formation, the impact from these massive stars is reduced.
Alternatively, our \NII-based sample might tend to pick up milder systems that are the remnants of extreme \NIII- and \NIV-emitters as they are in the process of evolved down to more typical N/O abundances.

Meanwhile, for the high metallicity portion of our sample (\lOH $>8.0$), we find that N/O correlates strongly with $M_*$.
This is in line with the chemical evolution picture at $z\sim0$, suggesting that, even at $z\sim1.5-7$, nitrogen enrichment of more massive, chemically-mature systems is dominated by secondary nitrogen enrichment from intermediate-mass AGB stars.

It is clear now that the extreme nitrogen abundances observed in \NIII- and \NIV-emitting galaxies do not represent a discrete population of unusual systems, but rather are the extreme end of a distribution of nitrogen abundances observed in low metallicity systems at high redshift.
Better characterising the properties of the stars responsible for this enrichment and how they might impact other galaxy properties, such as feedback budget, ionising photon output, and mass-to-light ratios, will be greatly aided by detailed analyses of deep spectroscopy capable of spatially and spectrally resolving key features of massive stars.

\section*{Acknowledgements}

The authors wish to thank Danielle Berg and Karla Arellano-C{\'o}rdova for helpful discussions in relation to this work.
This work is based on observations made with the NASA/ESA/CSA James Webb Space Telescope. The data were obtained from the Mikulski Archive for Space Telescopes at the Space Telescope Science Institute, which is operated by the Association of Universities for Research in Astronomy, Inc., under NASA contract NAS 5-03127 for JWST.
AJC \& JW gratefully acknowledges support from the Cosmic Dawn Center through the DAWN Fellowship. The Cosmic Dawn Center (DAWN) is funded by the Danish National Research Foundation under grant No. 140.
AJC, AJB, JC \& AS acknowledge funding from the "FirstGalaxies" Advanced Grant from the European Research Council (ERC) under the European Union’s Horizon 2020 research and innovation programme (Grant agreement No. 789056).
CC, ZJ, BDJ, BER \& CNAW acknowledge support from the JWST/NIRCam Science Team contract to the University of Arizona, NAS5-02105.
CC, BER \& JAAT acknowledge support from JWST Program 3215. Support for program 3215 was provided by NASA through a grant from the Space Telescope Science Institute, which is operated by the Association of Universities for Research in Astronomy, Inc., under NASA contract NAS 5-03127.
JAAT also acknowledges support from the Simons Foundation.
CS, XJ \& JS acknowledge support from the Science and Technology Facilities Council (STFC).
CS, XJ \& JS acknowledge support by the ERC through Advanced Grant 695671 “QUENCH”.
CS \& XJ acknowledge support by the UKRI Frontier Research grant RISEandFALL.
SC acknowledges support by European Union’s HE ERC Starting Grant No. 101040227 - WINGS.
ECL acknowledges support of an STFC Webb Fellowship (ST/W001438/1).
Some funding for this research was provided by the Johns Hopkins University, Institute for Data Intensive Engineering and Science (IDIES).
MSS acknowledges support by the Science and Technology Facilities Council (STFC) grant ST/V506709/1.
ST acknowledges support by the Royal Society Research Grant G125142.
H\"U acknowledges funding by the European Union (ERC APEX, 101164796). Views and opinions expressed are however those of the authors only and do not necessarily reflect those of the European Union or the European Research Council Executive Agency. Neither the European Union nor the granting authority can be held responsible for them.
The research of CCW is supported by NOIRLab, which is managed by the Association of Universities for Research in Astronomy (AURA) under a cooperative agreement with the National Science Foundation.

\section*{Software}

In addition to software packages referenced directly in the text, this work made use of the following Python packages:
Astropy \citep{Astropy},
CMasher \citep{cmasher},
LMFIT \citep{lmfit},
Matplotlib \citep{Matplotlib},
NumPy \citep{Numpy},
PyNeb \citep{Luridiana2015_pyneb}, and
SciPy \citep{Scipy}.

\section*{Data Availability}

This study primarily makes use of data from JADES \citep{Eisenstein2023_JADES}. The spectroscopic data are available for direct download\footnote{\url{https://jades-survey.github.io/scientists/data.html}} or by querying the JADES database\footnote{\url{https://jades.herts.ac.uk/search/}} and have been described in detail in \citet{Bunker2023_DR}, \citet{DEugenio2025_DR3}, \citet{CurtisLake2025_DR4}, and \citet{Scholtz2025_DR4}. The imaging has been described in \citet{Rieke2023_DR1}, 
Johnson et al. (2026) and Robertson et al. (2026). 



\bibliographystyle{mnras}
\bibliography{abunbib} 




\appendix

\section{Spatial decomposition of goods-n-mediumhst 31940}
\label{app:spatial_extraction}

\begin{figure}
    \centering
    \includegraphics[width=0.5\textwidth]{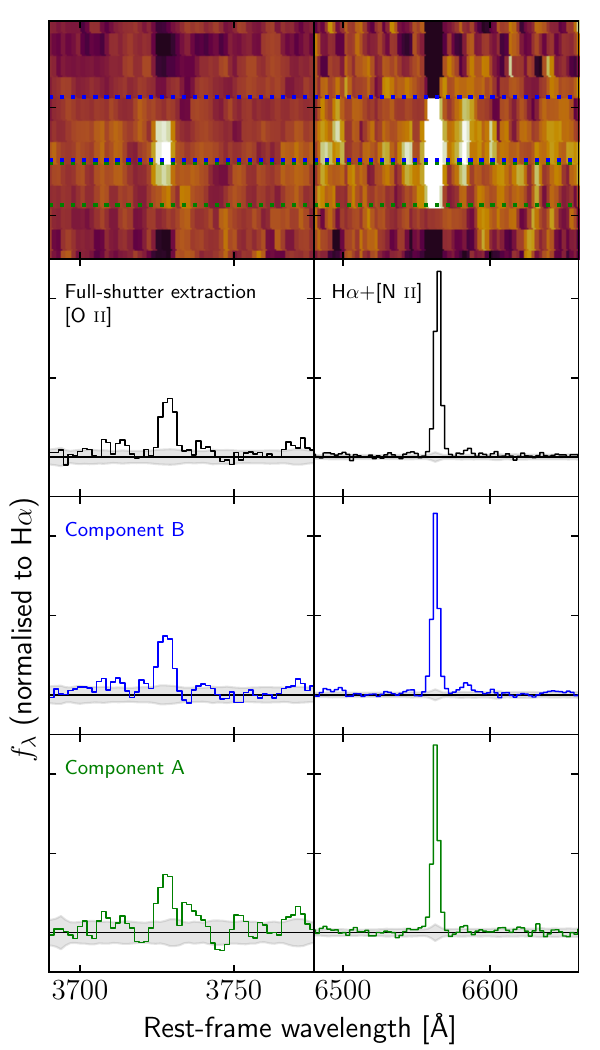}
    \caption{Re-extraction of the spectra for each component for the system observed as goods-n-mediumhst\_39140. The imaging thumbnail of this system can be found in the bottom row of Figure~\ref{fig:spec_auroral}.
    The top row shows the 2D spectra for gnmh\_31940, zoomed in on the areas around \OIIll3726,3729 and \Ha+\NIIl6583, showing the extractions of component B as the area between the blue dotted lines and component A as the area between the green dotted lines. The full shutter extraction encompasses the union of these two areas.
    Lower panels show the 1D spectra for the full shutter extraction (black), component B (blue), and component A (green), with each spectrum normalised to the \Ha\ flux. We see that all three spectra have very similar \OII/\Ha\ ratios, but component B has a particularly strong \NII\ feature, corresponding to a $\sim0.2$ dex higher N/O abundance.
    }
    \label{app_fig:special_extraction}
\end{figure}

From imaging of gnmh\_31940 (Figure~\ref{fig:spec_auroral}, bottom panel), it is clear that the morphology comprises two separate components. The alignment of the slit allows for the spectra of these two clumps to be separated in the 2D spectrum, where we find that they are at the same redshift.

The top panels in Figure~\ref{app_fig:special_extraction} show our extraction of the spectra of the two components of gnmh\_31940 from the 2D spectrum. The extracted 1D spectra are shown in the subsequent panels, with the full-shutter extracted spectrum shown in black, `Component A' (extracted from the lowermost two rows of spatial pixels in the object shutter) shown in green, and `Component B' (extracted from the uppermost three rows of spatial pixels in the object shutter) shown in blue. We then fit the emission lines in the same way described in Section~\ref{sub:line_fitting}.

The 1D spectra shown have been renormalised to the \Ha\ flux and we find that the \OII/\Ha\ flux ratios are very similar across the three 1D spectra. However, the \NII/\OII\ ratio is much higher in the spectrum extracted for Component B, corresponding to a nitrogen-to-oxygen abundance ratio of $\log({\rm N}/{\rm O})\approx-0.72$, around 0.2 dex higher than the value of $\log({\rm N}/{\rm O})=-0.9_{-0.35}^{+0.62}$ reported for the full-shutter extraction in Table~\ref{tab:auroral_sample}. We find the \NII\ line is undetected in the spectrum of Component A.

While a more detailed analysis is beyond the scope of this paper, this preliminary finding suggests that nitrogen abundances might vary significantly on small spatial scales, suggesting that the processes driving nitrogen-enhancement operate on small physical scales.

\section{Nitrogen-enhanced galaxies}
\label{app:n_enhanced}

\begin{figure*}
    \centering
\includegraphics[width=0.2475\textwidth]{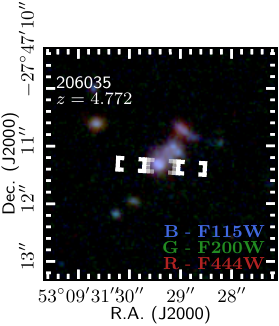}
\includegraphics[width=0.495\textwidth]{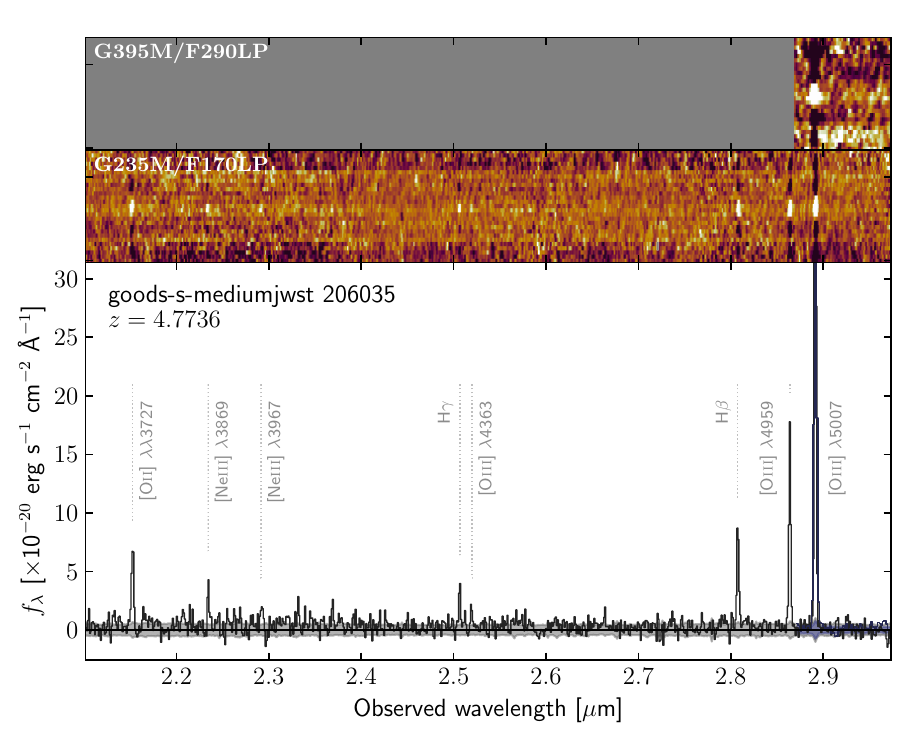}
\includegraphics[width=0.2475\textwidth]{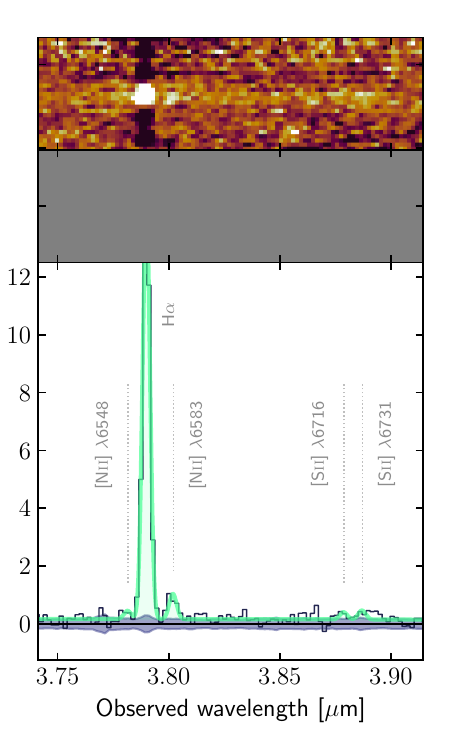}
\includegraphics[width=0.2475\textwidth]{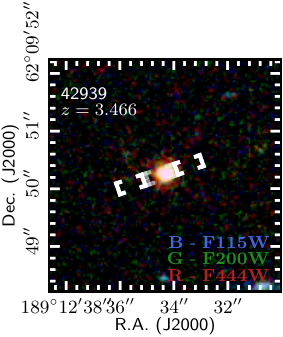}
\includegraphics[width=0.495\textwidth]{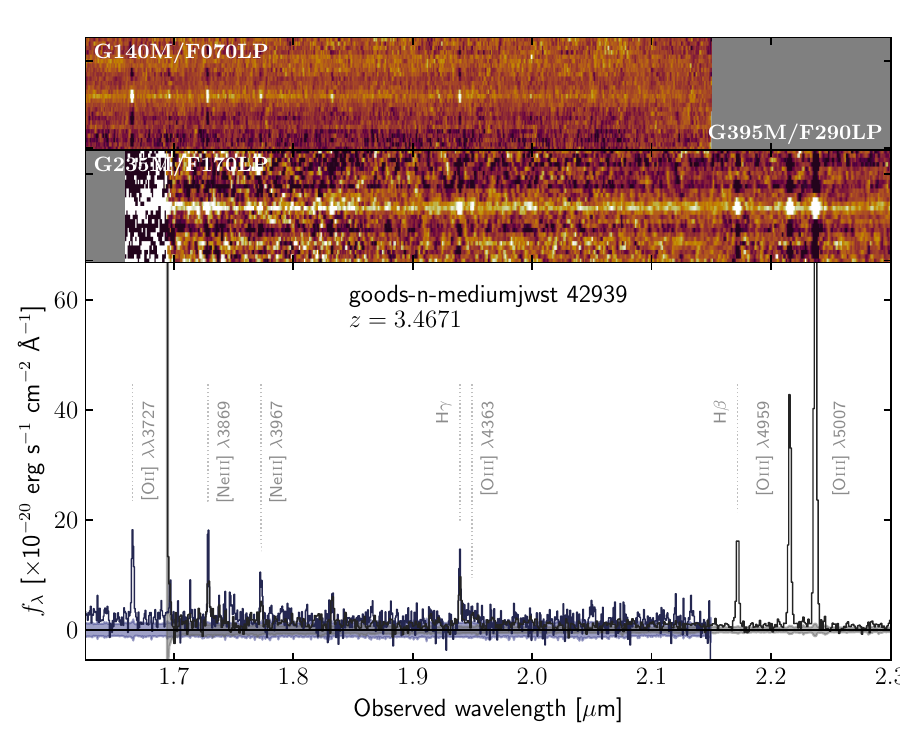}
\includegraphics[width=0.2475\textwidth]{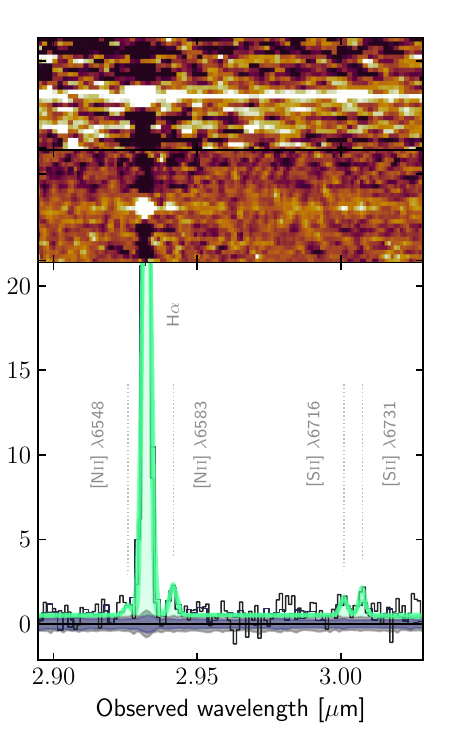}
    \caption{As for Figure~\ref{fig:spec_auroral}, showing two additional nitrogen-enhanced galaxies from the $T_e$-based sample. The upper 2D panel of the second row (gnmj\_42939) is shared between G140M/F070LP (left, covering \OIIIl5007) and G395M/F290LP (right, covering \Ha).}
    \label{app_fig:highNO_Te}
\end{figure*}

\begin{figure*}
    \centering
\includegraphics[width=0.2475\textwidth]{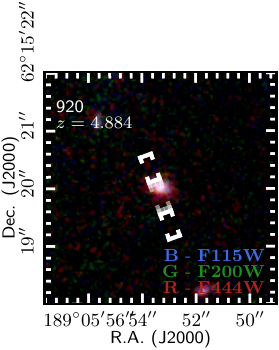}
\includegraphics[width=0.495\textwidth]{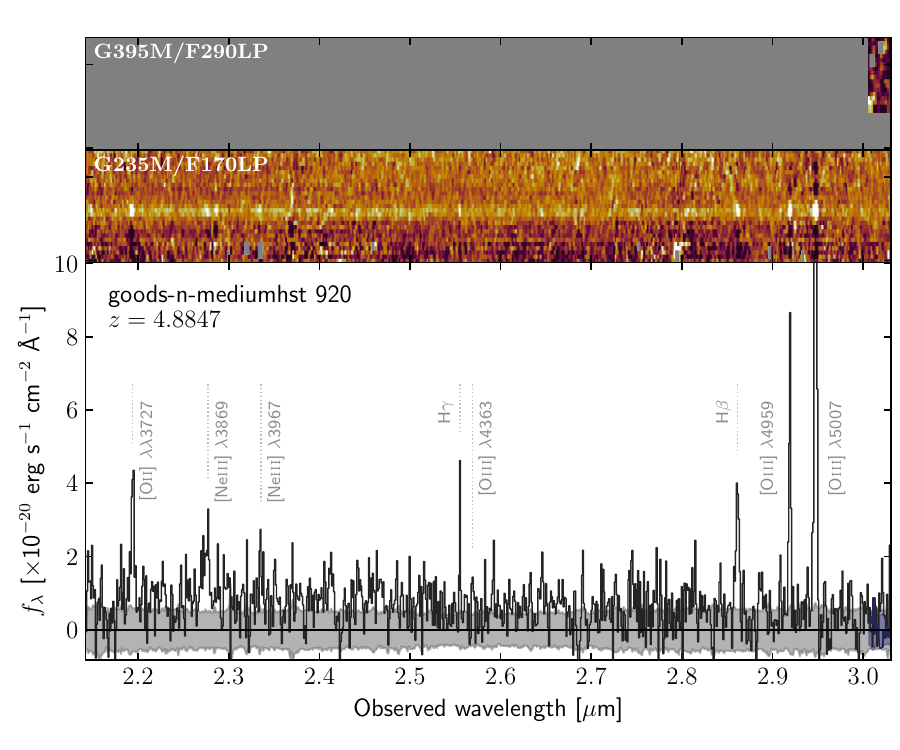}
\includegraphics[width=0.2475\textwidth]{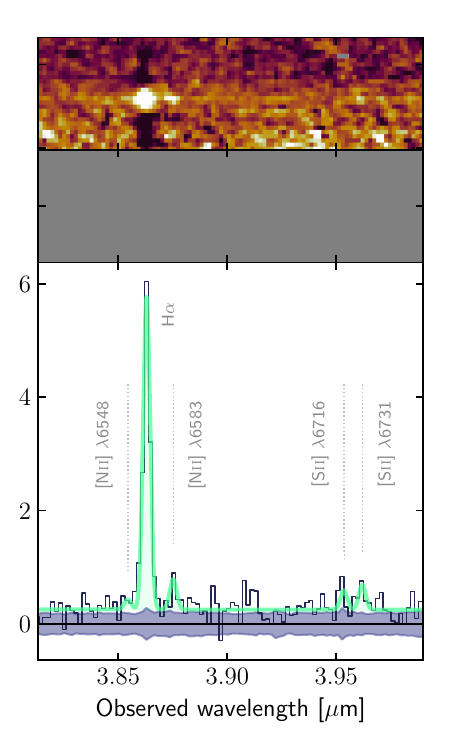}
\includegraphics[width=0.2475\textwidth]{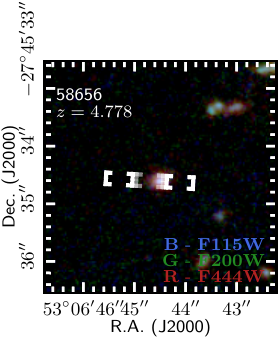}
\includegraphics[width=0.495\textwidth]{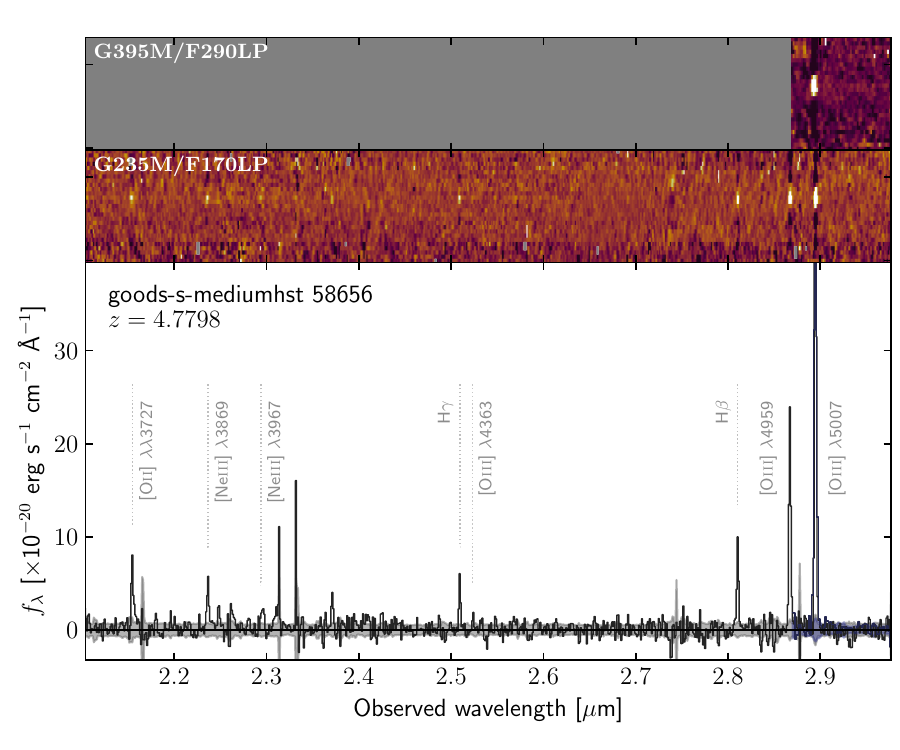}
\includegraphics[width=0.2475\textwidth]{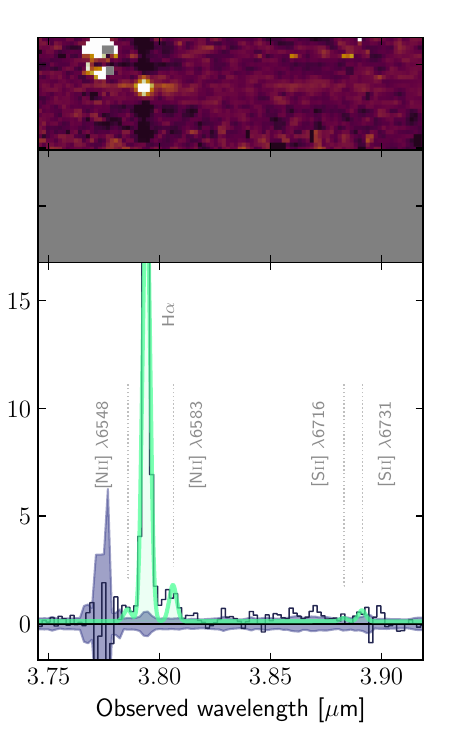}
\includegraphics[width=0.2475\textwidth]{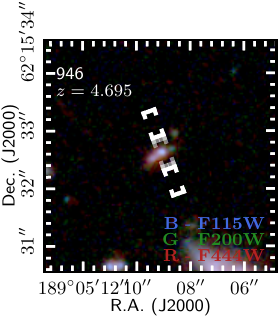}
\includegraphics[width=0.495\textwidth]{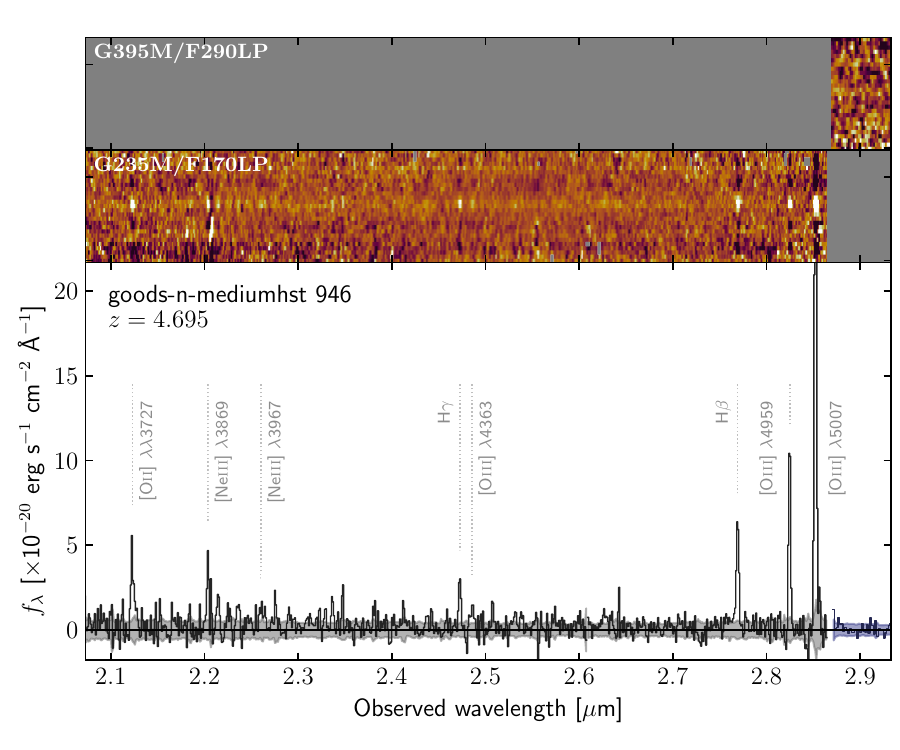}
\includegraphics[width=0.2475\textwidth]{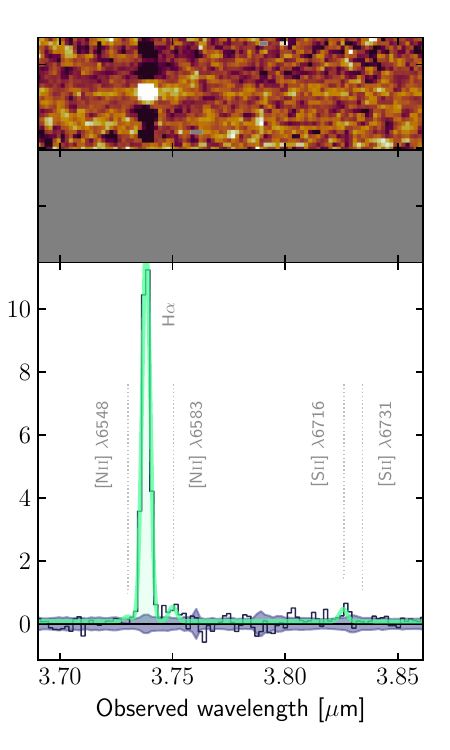}
    \caption{As for Figure~\ref{app_fig:highNO_Te}, showing three nitrogen-enhanced galaxies from the strong-line sample.}
    \label{app_fig:highNO_strongline1}
\end{figure*}

\begin{figure*}
    \centering
\includegraphics[width=0.2475\textwidth]{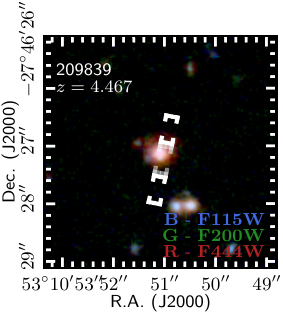}
\includegraphics[width=0.495\textwidth]{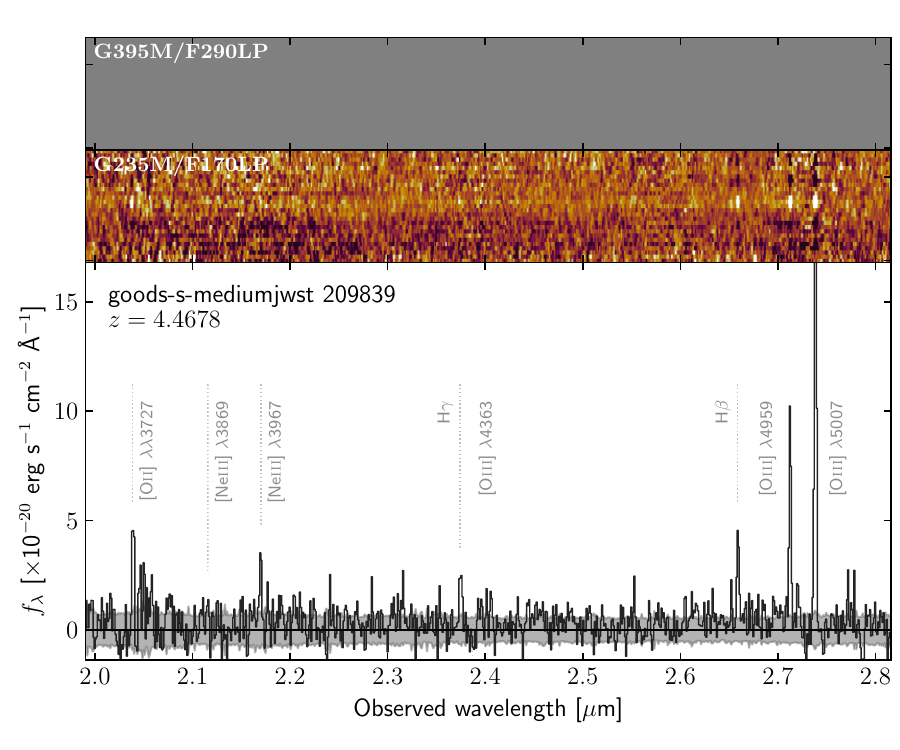}
\includegraphics[width=0.2475\textwidth]{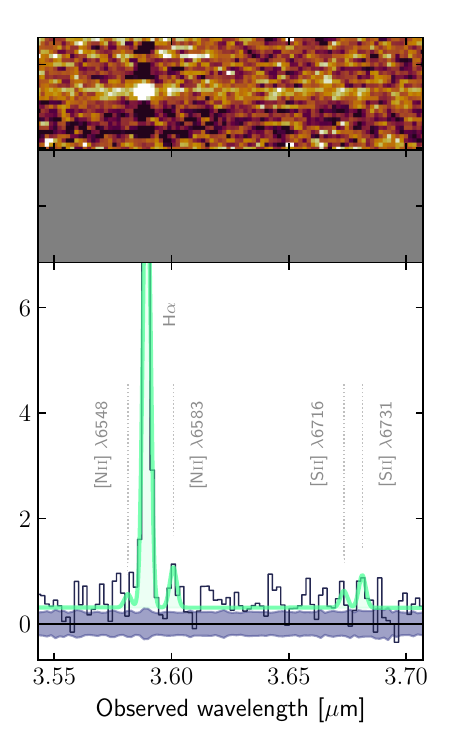}
\includegraphics[width=0.2475\textwidth]{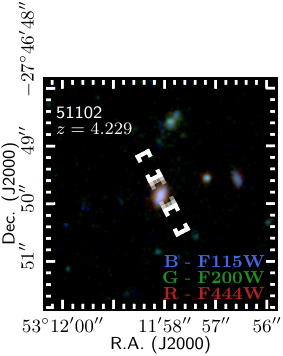}
\includegraphics[width=0.495\textwidth]{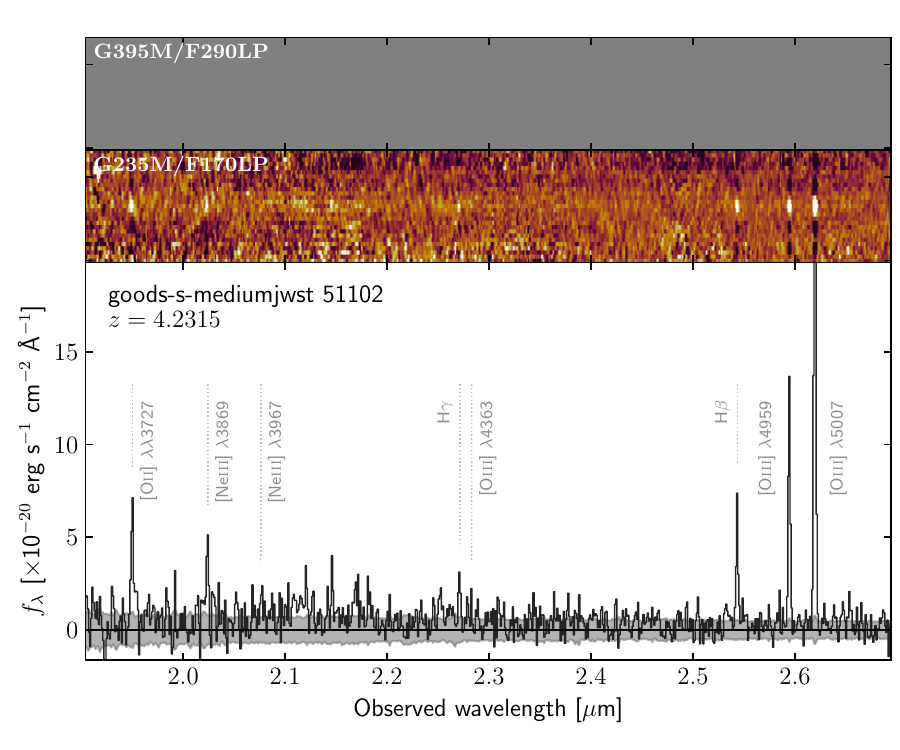}
\includegraphics[width=0.2475\textwidth]{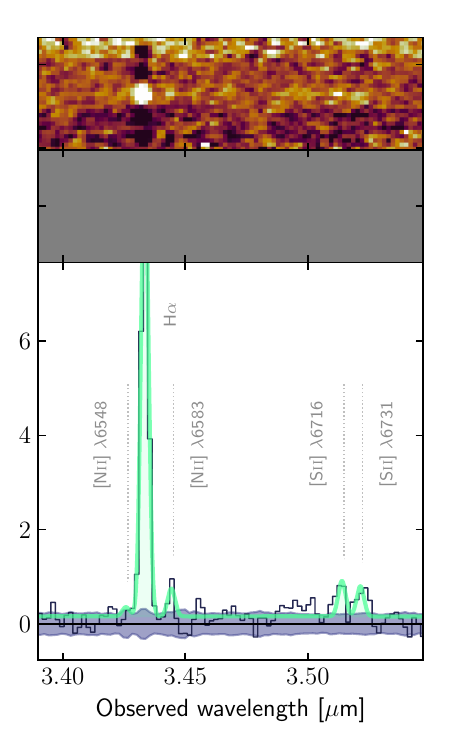}
\includegraphics[width=0.2475\textwidth]{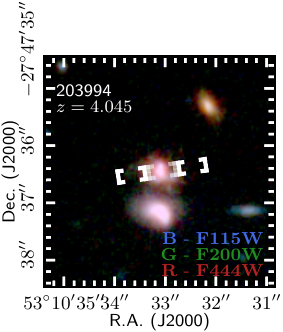}
\includegraphics[width=0.495\textwidth]{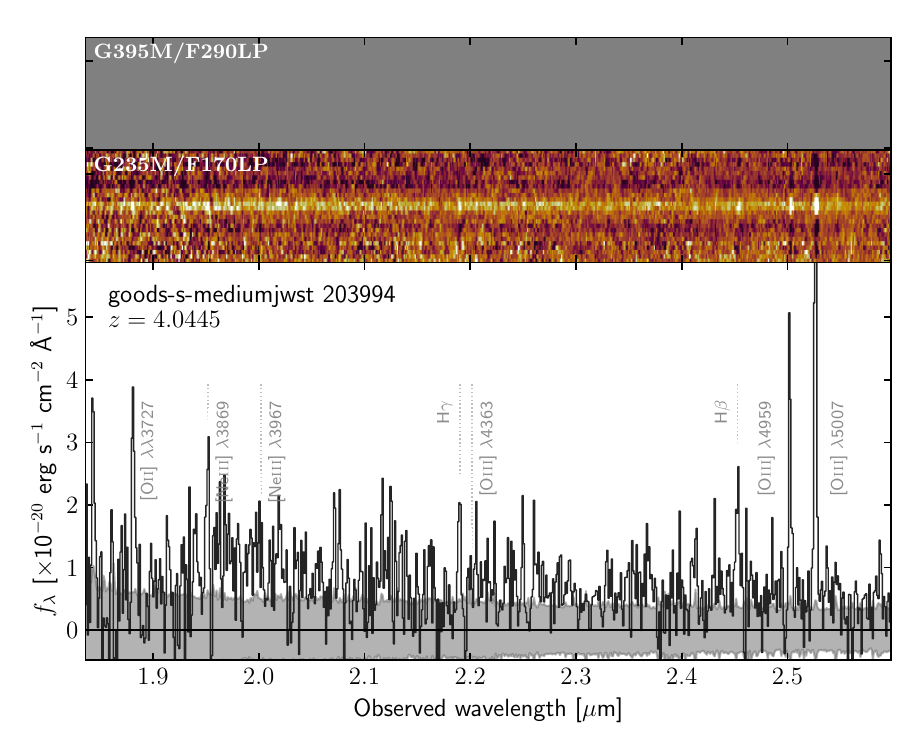}
\includegraphics[width=0.2475\textwidth]{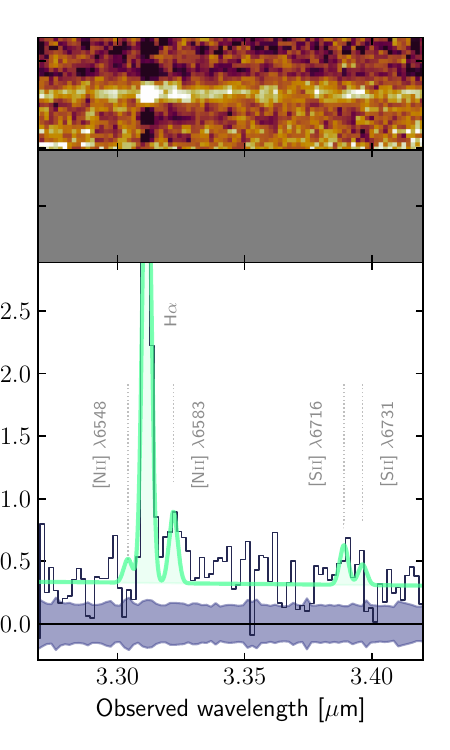}

    \caption{As for Figure~\ref{app_fig:highNO_Te}, showing a further three nitrogen-enhanced galaxies from the strong-line sample.}
    \label{app_fig:highNO_strongline2}
\end{figure*}

\begin{figure*}
    \centering
\includegraphics[width=0.2475\textwidth]{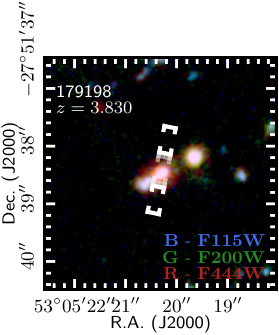}
\includegraphics[width=0.495\textwidth]{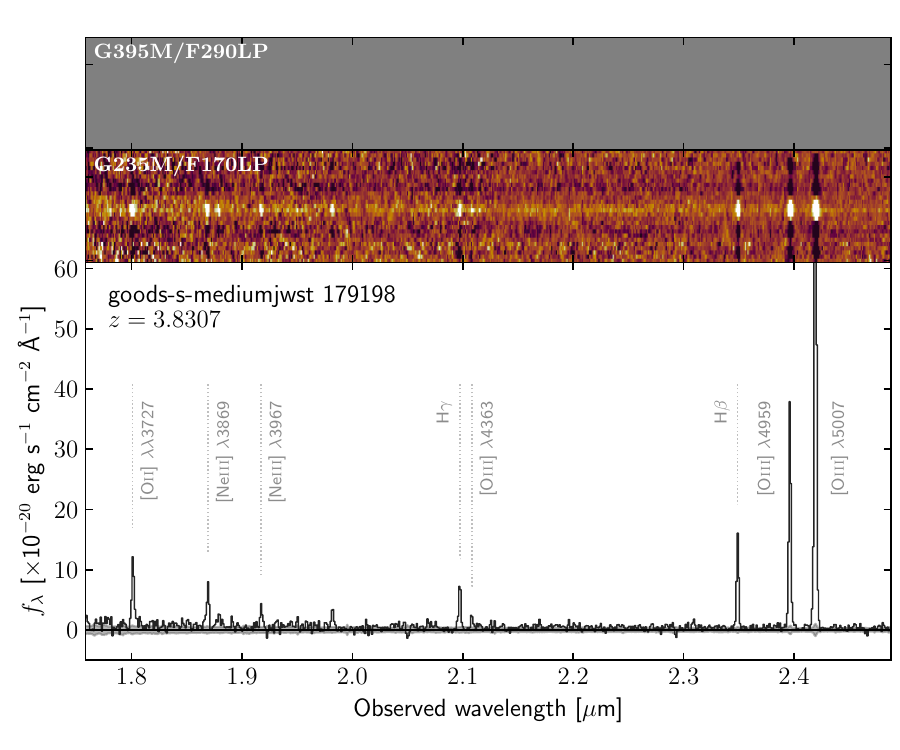}
\includegraphics[width=0.2475\textwidth]{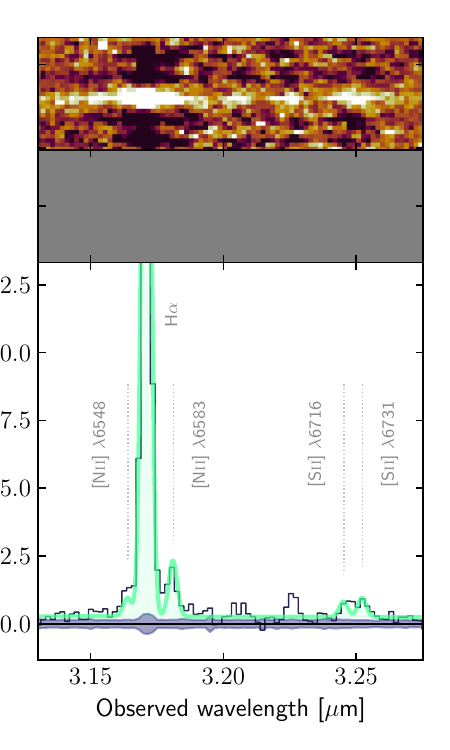}
\includegraphics[width=0.2475\textwidth]{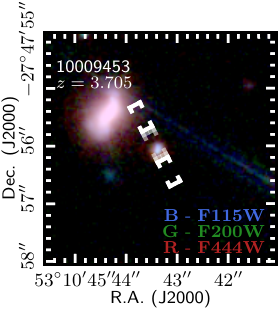}
\includegraphics[width=0.495\textwidth]{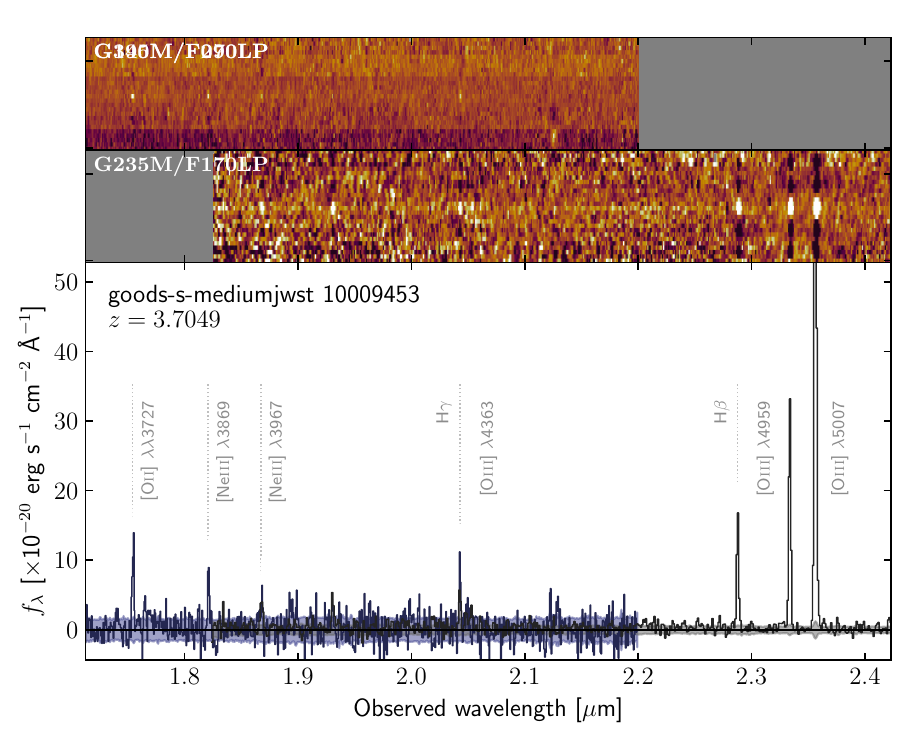}
\includegraphics[width=0.2475\textwidth]{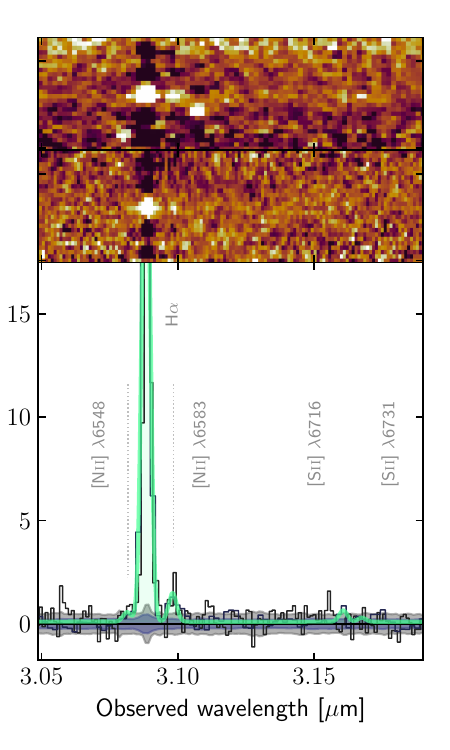}
\includegraphics[width=0.2475\textwidth]{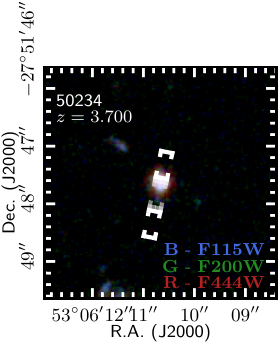}
\includegraphics[width=0.495\textwidth]{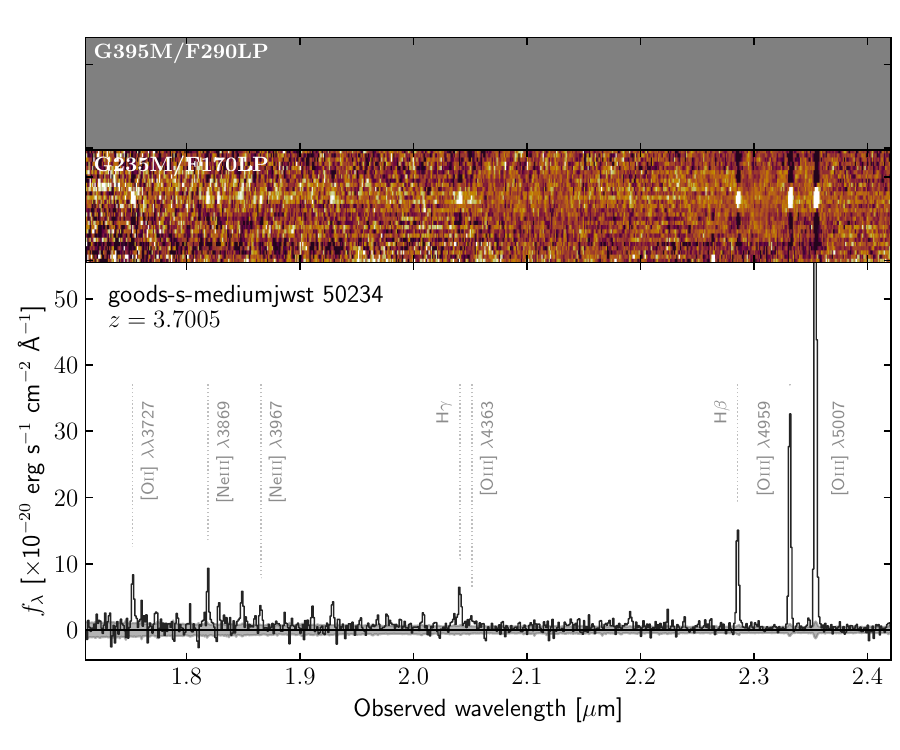}
\includegraphics[width=0.2475\textwidth]{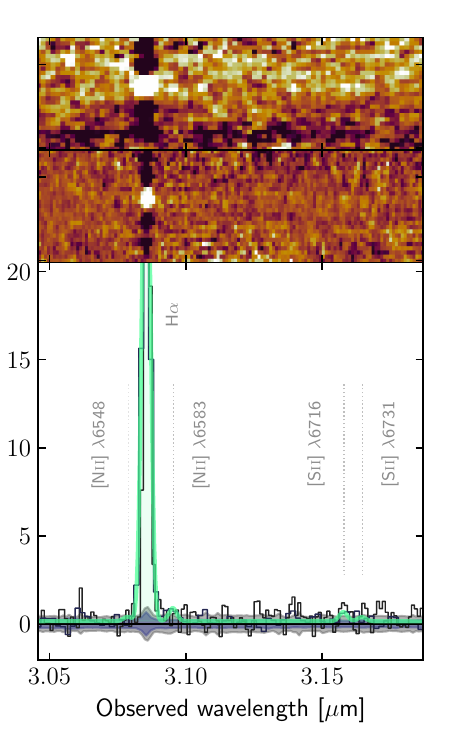}

    \caption{As for Figure~\ref{app_fig:highNO_Te}, showing a further three nitrogen-enhanced galaxies from the strong-line sample.}
    \label{app_fig:highNO_strongline3}
\end{figure*}

\begin{figure*}
    \centering
\includegraphics[width=0.2475\textwidth]{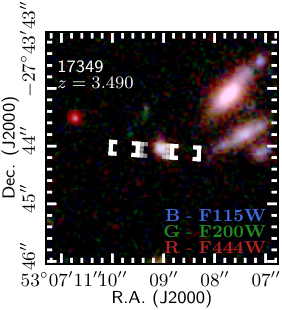}
\includegraphics[width=0.495\textwidth]{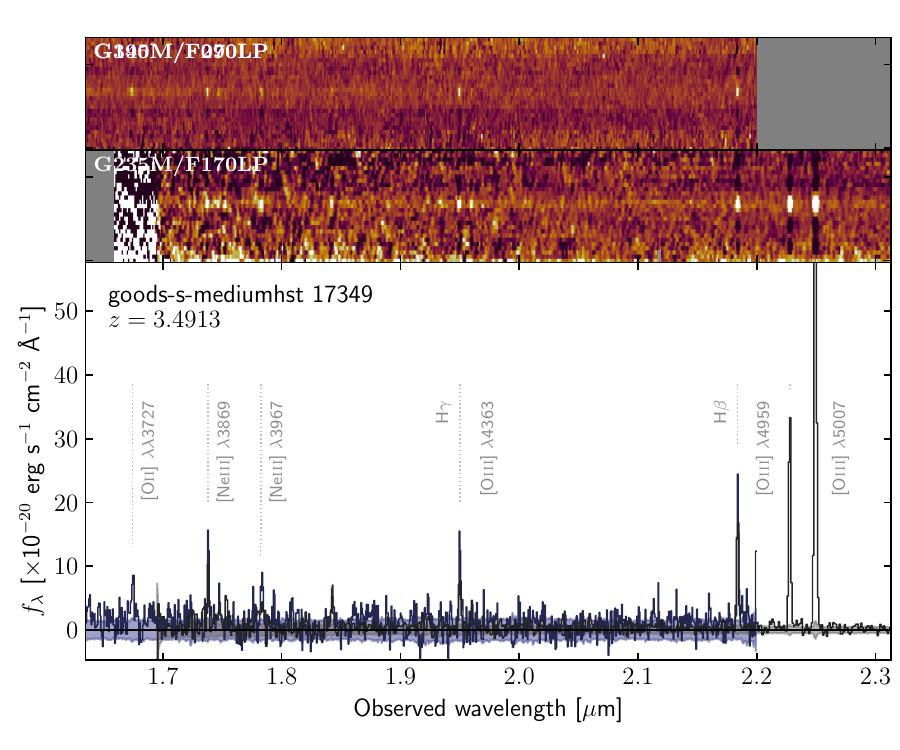}
\includegraphics[width=0.2475\textwidth]{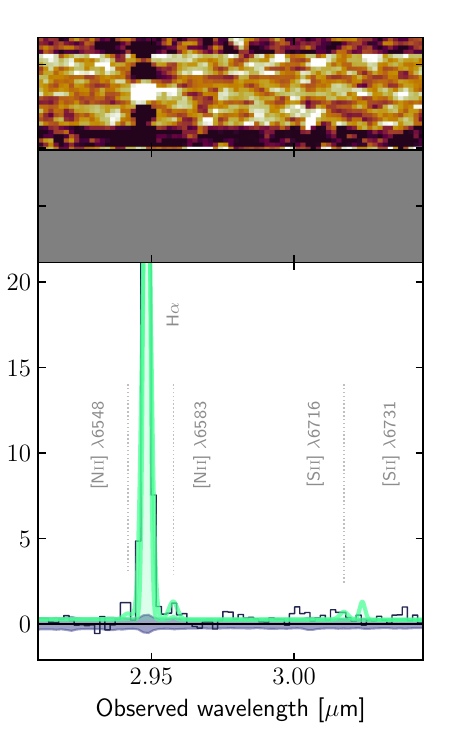}
\includegraphics[width=0.2475\textwidth]{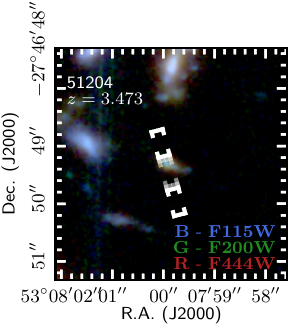}
\includegraphics[width=0.495\textwidth]{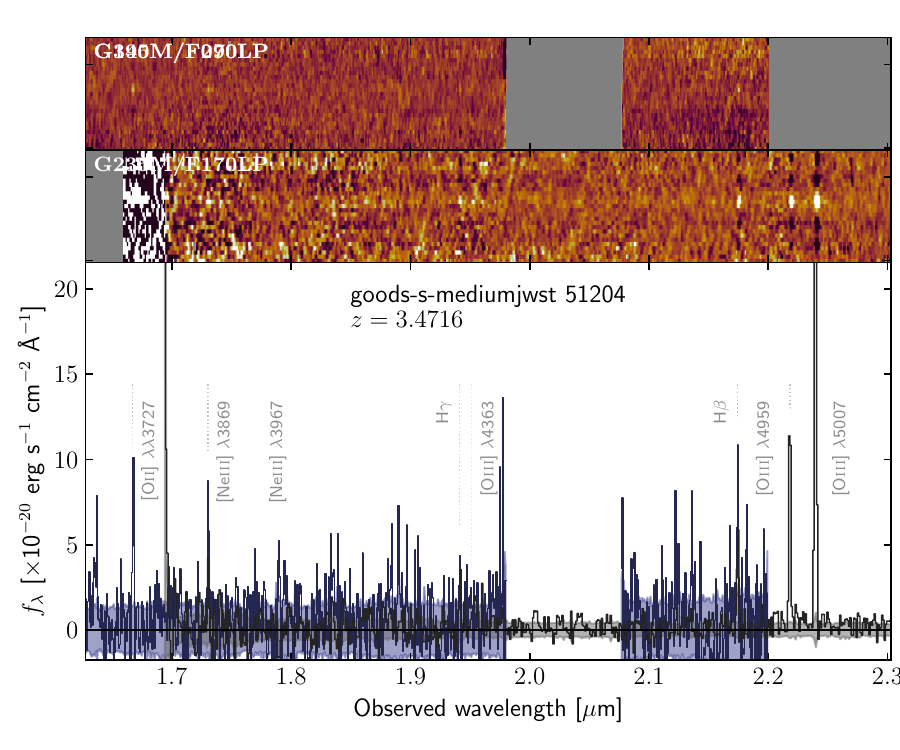}
\includegraphics[width=0.2475\textwidth]{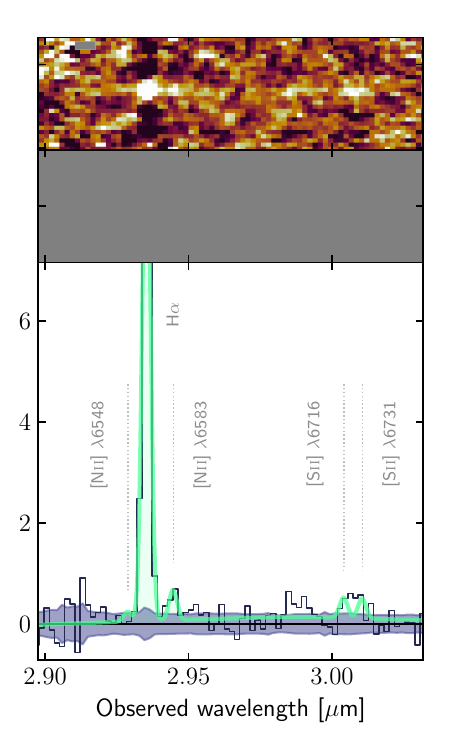}
\includegraphics[width=0.2475\textwidth]{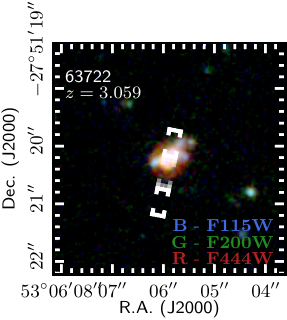}
\includegraphics[width=0.495\textwidth]{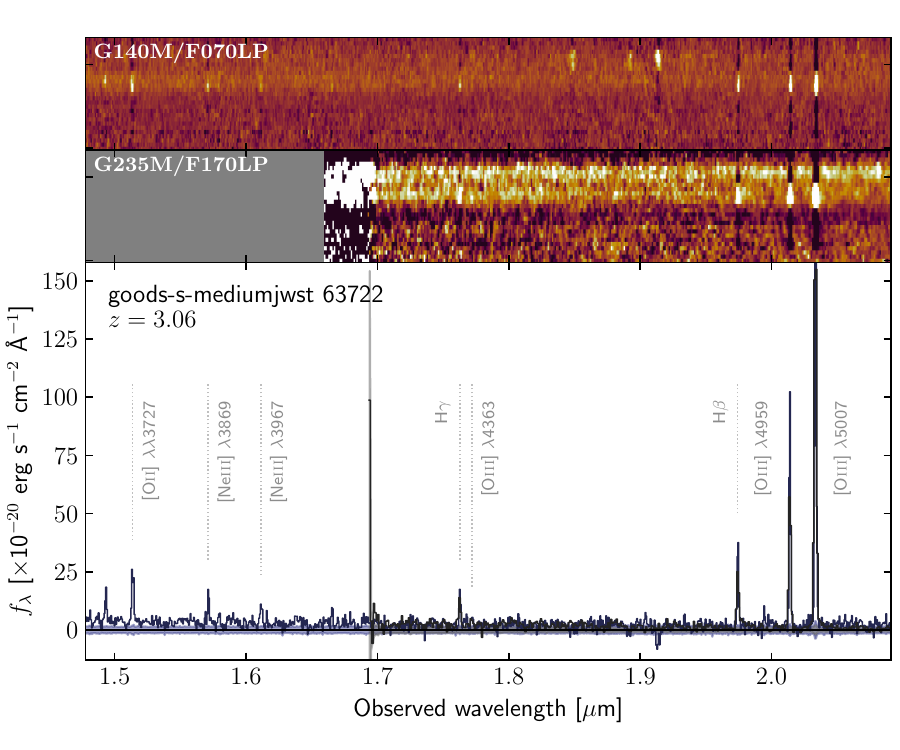}
\includegraphics[width=0.2475\textwidth]{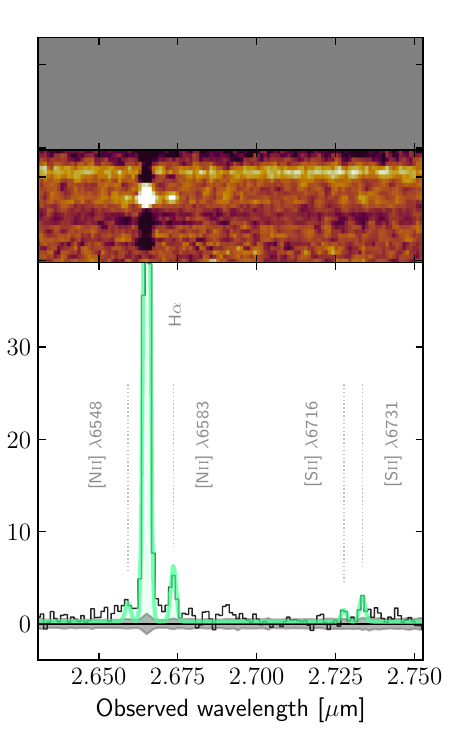}
    \caption{As for Figure~\ref{app_fig:highNO_Te}, showing a further three nitrogen-enhanced galaxies from the strong-line sample.}
    \label{app_fig:highNO_strongline4}
\end{figure*}

\begin{figure*}
    \centering
\includegraphics[width=0.2475\textwidth]{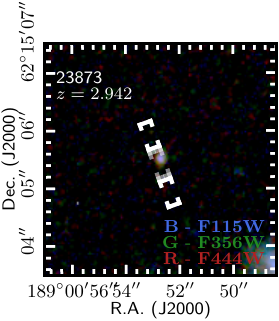}
\includegraphics[width=0.495\textwidth]{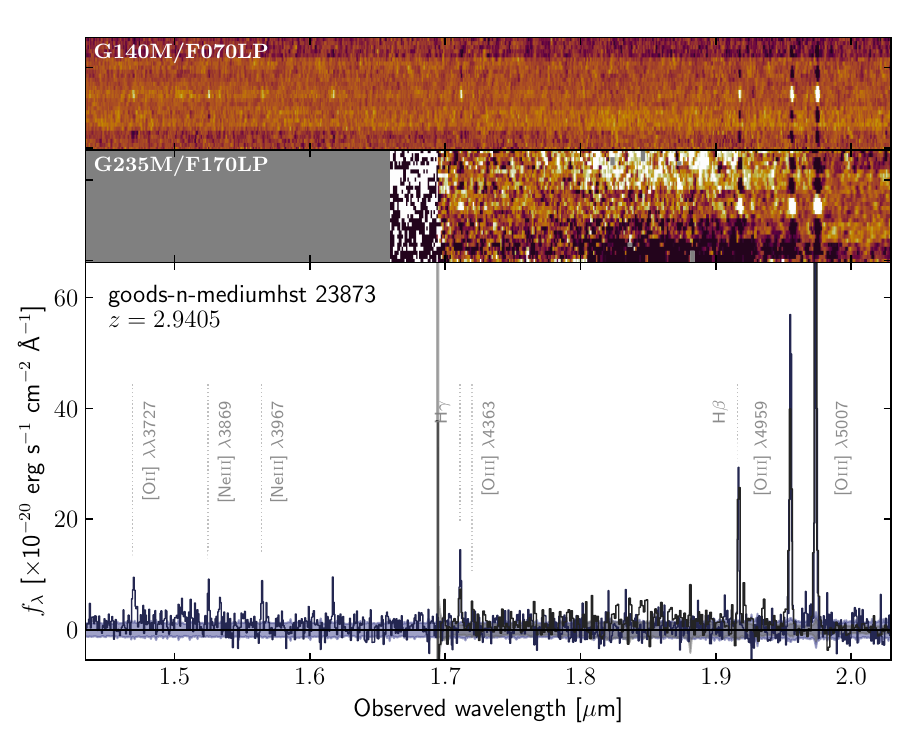}
\includegraphics[width=0.2475\textwidth]{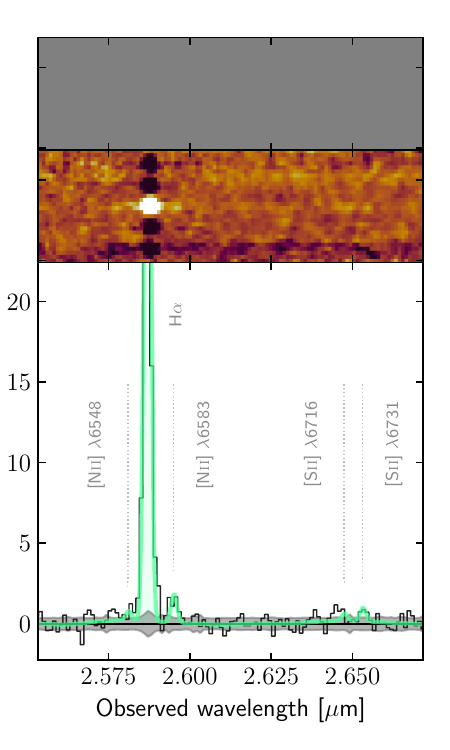}
\includegraphics[width=0.2475\textwidth]{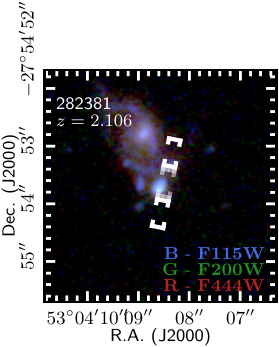}
\includegraphics[width=0.495\textwidth]{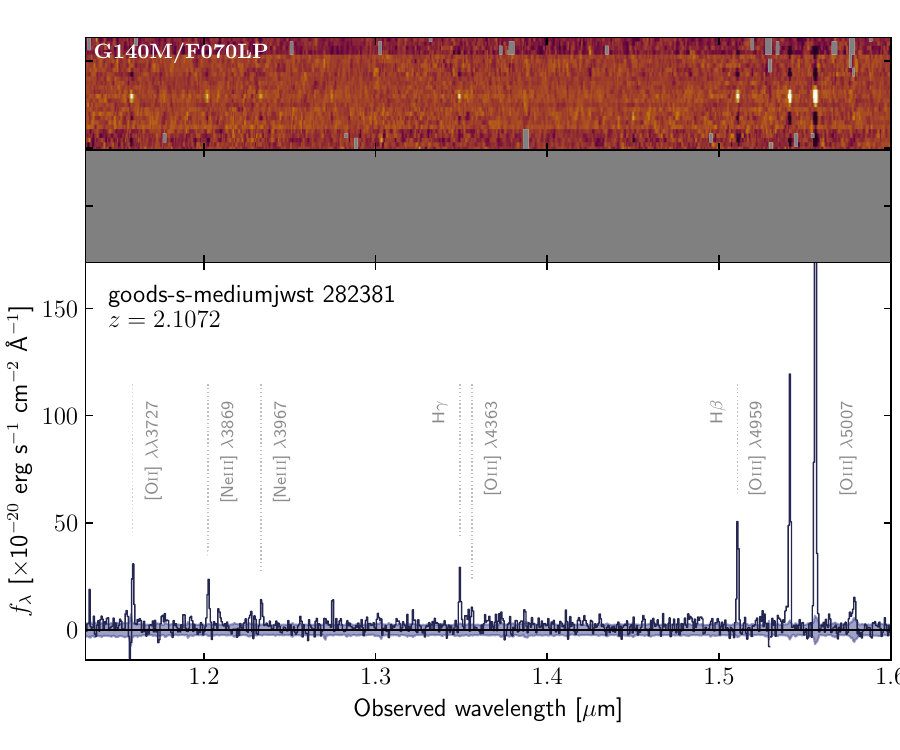}
\includegraphics[width=0.2475\textwidth]{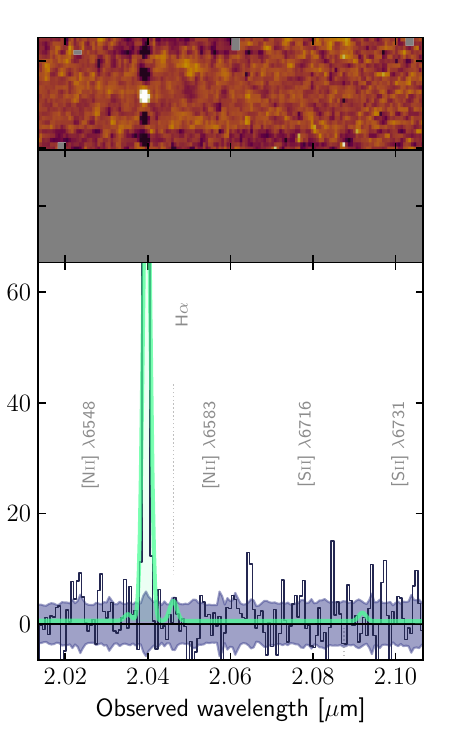}
    \caption{As for Figure~\ref{app_fig:highNO_Te}, showing a further two nitrogen-enhanced galaxies from the strong-line sample.}
    \label{app_fig:highNO_strongline5}
\end{figure*}

Figure~\ref{app_fig:highNO_Te} shows thumbnails and spectra for the other two nitrogen-enhanced galaxies from the $T_e$-based sample (with the other three being shown in Figure~\ref{fig:spec_auroral}).
Spectra and thumbnails of all nitrogen-enhanced galaxies from the strong-line sample are shown in Figures~\ref{app_fig:highNO_strongline1}~--~\ref{app_fig:highNO_strongline5}.

\section{Table of strong-line abundance measurements}
\label{app:emline_tables}

We present in Table~C.1 our strong-line abundance measurements and associated galaxy properties measurements for galaxies with robust properties derived from SED and morphological fitting. In Table~C.2 we report additional strong-line measurements for the galaxies without these photometrically-derived properties. 

\clearpage
\onecolumn

\centering
\tablefirsthead{
\multicolumn{9}{l}{
Table C.1. Strong-line oxygen and nitrogen abundances for the \NmassSL\ galaxies in our strong-line sample with galaxy properties derived from}\\
\multicolumn{9}{l}{
SED and morphological fitting. Details on the derivation of these quantitites can be found in Sections~\ref{sec:data}~\&~\ref{sec:abundances}.}
\\
\toprule Tier & ID & $z_{\rm spec}$ & 12+log(O/H) & log(N/O) & log($M_*/M_\odot$) & SFR & \SDmass\ & \SDsfr\ \\ \midrule}
\tablehead{%
 Table C.1. \textit{(continued)} \\
\toprule
Tier & ID & $z_{\rm spec}$ & 12+log(O/H) & log(N/O) & log($M_*/M_\odot$) & SFR & \SDmass\ & \SDsfr\  \\ \midrule}
\tabletail{%
\midrule }
\tablelasttail{%
\bottomrule}
\begin{supertabular}{ccccccccc}
\shrinkheight{2 cm}
goods-n-mediumhst & 643 & 5.3458 & $8.02^{+0.1}_{-0.12}$ & $<-1.37$ & $8.99^{+0.2}_{-0.52}$ & $1.11^{+0.07}_{-0.12}$ & $2.6^{+0.2}_{-0.52}$ & $0.72^{+0.07}_{-0.12}$ \\
goods-n-mediumhst & 721 & 5.1778 & $7.47^{+0.11}_{-0.09}$ & $<-0.49$ & $7.96^{+0.25}_{-0.17}$ & $0.42^{+0.01}_{-0.03}$ & $3.6^{+0.25}_{-0.17}$ & $2.06^{+0.01}_{-0.03}$ \\
goods-n-mediumhst & 761 & 5.5449 & $7.59^{+0.16}_{-0.13}$ & $<-0.53$ & $8.42^{+0.2}_{-0.35}$ & $0.38^{+0.09}_{-0.07}$ & $2.09^{+0.2}_{-0.35}$ & $0.05^{+0.09}_{-0.07}$ \\
goods-n-mediumhst & 772 & 3.847 & $7.96^{+0.12}_{-0.13}$ & $-1.24^{+0.09}_{-0.11}$ & $9.29^{+0.04}_{-0.06}$ & $0.69^{+0.05}_{-0.11}$ & $3.36^{+0.04}_{-0.06}$ & $0.75^{+0.05}_{-0.11}$ \\
goods-n-mediumhst & 804 & 3.6614 & $8.41^{+0.07}_{-0.09}$ & $-0.85^{+0.04}_{-0.04}$ & $9.92^{+0.1}_{-0.1}$ & $0.64^{+0.3}_{-0.43}$ & $3.47^{+0.1}_{-0.1}$ & $0.19^{+0.3}_{-0.43}$ \\
goods-n-mediumhst & 853 & 3.4723 & $8.3^{+0.08}_{-0.11}$ & $-0.89^{+0.04}_{-0.04}$ & $9.9^{+0.08}_{-0.08}$ & $0.93^{+0.27}_{-0.26}$ & $3.66^{+0.08}_{-0.08}$ & $0.69^{+0.27}_{-0.26}$ \\
goods-n-mediumhst & 896 & 6.7585 & $8.1^{+0.1}_{-0.12}$ & $-0.97^{+0.08}_{-0.1}$ & $9.82^{+0.13}_{-0.31}$ & $1.55^{+0.23}_{-0.26}$ & $4.28^{+0.13}_{-0.31}$ & $2.0^{+0.23}_{-0.26}$ \\
goods-n-mediumhst & 902 & 4.0612 & $8.08^{+0.13}_{-0.16}$ & $<-1.2$ & $8.82^{+0.41}_{-0.2}$ & $0.85^{+0.06}_{-0.04}$ & $3.13^{+0.41}_{-0.2}$ & $1.17^{+0.06}_{-0.04}$ \\
\end{supertabular}%

\pagebreak
\tablefirsthead{
\multicolumn{5}{l}{
Table C.2. Strong-line oxygen and nitrogen abundances for the 95 galaxies in }\\
\multicolumn{5}{l}{
our strong-line sample without galaxy properties derived from SED and}\\
\multicolumn{5}{l}{morphological fitting. }
\\
\toprule Tier & ID & $z_{\rm spec}$ & 12+log(O/H) & log(N/O) \\ \midrule}
\tablehead{%
 Table C.2. \textit{(continued)} \\
\toprule
Tier & ID & $z_{\rm spec}$ & 12+log(O/H) & log(N/O)  \\ \midrule}
\tabletail{%
\midrule }
\tablelasttail{%
\bottomrule}
\begin{supertabular}{ccccc}
goods-n-mediumhst & 910 & 4.4144 & $8.39^{+0.08}_{-0.1}$ & $<-1.31$ \\
goods-n-mediumhst & 946 & 4.6948 & $7.64^{+0.16}_{-0.14}$ & $-0.99^{+0.12}_{-0.17}$ \\
goods-n-mediumhst & 956 & 5.4225 & $7.82^{+0.13}_{-0.13}$ & $<-1.15$ \\
goods-n-mediumhst & 964 & 5.5992 & $8.1^{+0.11}_{-0.13}$ & $<-1.01$ \\
goods-n-mediumhst & 971 & 4.4221 & $8.25^{+0.1}_{-0.13}$ & $-1.29^{+0.11}_{-0.16}$ \\
goods-n-mediumhst & 1940 & 3.6602 & $8.19^{+0.08}_{-0.1}$ & $-1.29^{+0.05}_{-0.06}$ \\
goods-n-mediumhst & 2890 & 3.3335 & $8.03^{+0.12}_{-0.13}$ & $<-0.89$ \\
goods-n-mediumhst & 22517 & 2.5343 & $8.01^{+0.11}_{-0.12}$ & $<-1.16$ \\
\end{supertabular}%


\bsp	
\label{lastpage}
\end{document}